\documentclass[11pt]{article}
\usepackage{graphicx}
\usepackage{amsmath,amsfonts,amssymb,amsthm,mathtools}
\usepackage{relsize}
\usepackage{fullpage}
\usepackage{bigints}
\usepackage{float}
\usepackage{subfigure}
\usepackage{captcont}
\usepackage{lmodern} 			
\usepackage[T1]{fontenc}		
\usepackage{gensymb}
\usepackage{textcomp}
\usepackage{lineno}		
\usepackage{color}
\usepackage{verbatim}
\usepackage{comment}
\usepackage{hyperref}		
\usepackage{array}

\graphicspath{{./../PureComponents/CO2/} {./../PureComponents/Water/} {./../Mixtures/CO2-Water/PatelEubank1988/90CO2/} {./../Mixtures/CO2-Water/PatelEubank1988/50CO2/} {./../Mixtures/CO2-Water/Koscheletal2006/} {./../Duan/pressureRange/} {./../}}

\begin{document}

\title{The CPA Equation of State and an Activity Coefficient Model for Accurate Molar Enthalpy Calculations of Mixtures with Carbon Dioxide and Water/Brine}
\author{Philip C.\ Myint$^{1,2,}$\thanks{\texttt{myint1@llnl.gov}} , Yue Hao$^{2,}$\thanks{\texttt{hao1@llnl.gov}} , and Abbas Firoozabadi$^{1,3,}$\thanks{\texttt{abbas.firoozabadi@yale.edu}}\\
\\ $^1$Department of Chemical and Environmental Engineering, Yale University \\
$^2$Lawrence Livermore National Laboratory, Livermore, CA \\
$^3$Reservoir Engineering Research Institute, Palo Alto, CA}
\date{\today}
\maketitle

\begin{abstract} 
Thermodynamic property calculations of mixtures containing carbon dioxide (CO$_2$) and water, including brines, are essential in theoretical models of many natural and industrial processes. The properties of greatest practical interest are density, solubility, and enthalpy. Many models for density and solubility calculations have been presented in the literature, but there exists only one study, by Spycher and Pruess, that has compared theoretical molar enthalpy predictions with experimental data~\cite{Spycher2011}. In this report, we recommend two different models for enthalpy calculations: the CPA equation of state by Li and Firoozabadi~\cite{Li2009}, and the CO$_2$ activity coefficient model by Duan and Sun~\cite{Duan2003}. We show that the CPA equation of state, which has been demonstrated to provide good agreement with density and solubility data, also accurately calculates molar enthalpies of pure CO$_2$, pure water, and both CO$_2$-rich and aqueous (H$_2$O-rich) mixtures of the two species. It is applicable to a wider range of conditions than the Spycher and Pruess model. In aqueous sodium chloride (NaCl) mixtures, we show that Duan and Sun's model yields accurate results for the partial molar enthalpy of CO$_2$. It can be combined with another model for the brine enthalpy to calculate the molar enthalpy of H$_2$O-CO$_2$-NaCl mixtures. We conclude by explaining how the CPA equation of state may be modified to further improve agreement with experiments. This generalized CPA is the basis of our future work on this topic.
\end{abstract}

\section{Introduction}
\label{sec:intro}
Mixtures of carbon dioxide (CO$_2$) with water, including brines, are important in a number of industries, such as food products, combustion systems, cosmetics, and petrochemicals. They also play a major role in Earth science applications like geological carbon sequestration, in which CO$_2$ is injected and stored in saline aquifers located deep in the Earth's subsurface~\cite{Spycher2003, Myint2013a, Smith2013a, Hao2013, Elenius2014, Slim2014, Loodts2014}. In all of these applications, accurate models of thermodynamic properties are essential. The models must often be reliable over a wide range of temperatures and pressures. This is the case, for example, in the work of the National Risk Assessment Partnership (NRAP), which is a collaborative effort between several United States Department of Energy national laboratories~\cite{NRAP}. One of NRAP's main missions is to predict the risk of CO$_2$ leakage from storage reservoirs (deep saline aquifers) to overlying shallow aquifers, which are potential sources of drinking water. The models employed by NRAP need to compute thermodynamic properties in the range of conditions between those of the shallow aquifers, which are roughly between 290--300~K and 3--20~bar, and those of the CO$_2$ storage reservoirs, which are between 323--423~K and 100--500~bar, depending on the subsurface depth.

Thermodynamic properties of greatest practical interest can be classified into three categories: volumetric (e.g., density), phase behavior (e.g., solubility), and thermal (e.g., enthalpy). For mixtures with CO$_2$ and water, there are several different models that can be used to compute the density and/or solubility over a wide range of conditions. Enthalpy models, however, are much more rare. This may be a reflection of the fact that enthalpy measurements are far less common than density or solubility measurements. To the best of our knowledge, there has been only one reference that has compared theoretical molar enthalpy (or specific enthalpy) predictions with experimental data~\cite{Spycher2011}. This study by Spycher and Pruess has been limited to pure components and to mixtures of CO$_2$ with pure water, not brines. Mixtures with pure water may be of direct interest to the industrial applications listed above. For carbon sequestration, fresh water is sometimes used as a surrogate for the brine in which CO$_2$ is dissolved. Nevertheless, in some cases it may be desirable to represent the saline environment of the aquifers with true brine properties.

The main goal of this report is to identify models that can accurately calculate the enthalpy of mixtures containing CO$_2$ and water. For mixtures with pure water (not electrolytic solutions), we promote use of the cubic-plus-association (CPA) equation of state (EOS) developed by Li and Firoozabadi~\cite{Li2009}. The CPA equation of state is applicable to pure components as well as to aqueous mixtures containing substances such as CO$_2$, hydrogen sulfide (H$_2$S) and hydrocarbons. For these mixtures, it has been demonstrated to provide good agreement with experimental density and solubility data, but no comparisons have been made with enthalpy data. In fact, simultaneous prediction of phase equilibria and enthalpy is considered one of the remaining challenges for CPA equations of state~\cite{Kontogeorgis2006a}. In fluids where water is not present, such as in hydrocarbon-CO$_2$ mixtures, the CPA reduces to the widely-used Peng-Robinson equation of state. The CPA has been successfully implemented in codes that simulate three-phase (oil, water, gas) flows~\cite{Moortgat2012}. For aqueous mixtures of CO$_2$ dissolved in brines, we recommend using the Duan and Sun activity coefficient model to compute the partial molar enthalpy of CO$_2$~\cite{Duan2003}. We consider brines composed only of water and sodium chloride (NaCl), which is the predominant salt found in deep saline aquifers.

The report is organized as follows. We present molar enthalpy calculations from the CPA equation of state for pure components and CO$_2$-H$_2$O mixtures in Sections~\ref{sec:pure} and~\ref{sec:mixture}, respectively. We show that it is applicable to a wider range of conditions than the Spycher and Pruess model~\cite{Spycher2011}. In an effort to make the CPA EOS accessible and readily usable, extensive details-- much more than presented in the original research paper~\cite{Li2009}-- are included in Appendices~\ref{sec:thermo} and~\ref{sec:CPA}. Enthalpy calculations of H$_2$O-CO$_2$-NaCl mixtures using the Duan and Sun model are presented in Section~\ref{sec:aqueous} and Appendix~\ref{sec:duan}. Although CPA accurately predicts molar enthalpies, it exhibits relatively poor agreement with experimental excess enthalpy data. Section~\ref{sec:excess} together with~\ref{sec:Helmholtz}--\ref{sec:pressureCPA} explain how the Li and Firoozabadi version of the CPA equation of state may be generalized to improve the agreement with excess enthalpy data. This generalized model is the starting point of our future work that we outline in Section~\ref{sec:future}.

\section{Enthalpy of pure components}
\label{sec:pure}

\subsection{CO$_2$}
\label{sec:CO2}

The molar enthalpy~$h$ is given by~(\ref{eq:molarEnthalpy}), which for a pure component fluid simplifies to

\begin{equation}
\label{eq:molarEnthalpyPure}
h(T, P)  = -RT^2 \left. \left( \frac{\partial \ln \phi(T, P)}{\partial T} \right) \right|_{P,\mathbf{n}} + h^\textrm{ig}.
\end{equation}
The fugacity coefficient~$\phi$ of the pure component can be obtained from~(\ref{eq:lnphiZLi}), which is the expression for the fugacity coefficient from Li and Firoozabadi's version of the CPA EOS. Since CO$_2$ is not self-associating (hydrogen bonds are unable to form between CO$_2$ molecules), this expression simplifies tremendously because the association contribution to~$\ln \phi$, which involves the site fractions~$\chi_i$, is zero. For these non-self-associating substances, Li and Firoozabadi's CPA EOS is equivalent to the Peng-Robinson EOS. The molar enthalpy~$h^\textrm{ig}$ of CO$_2$ in the ideal gas state is found by integrating, with respect to temperature, the correlation for the constant pressure heat capacity presented in Appendix C of the textbook by Smith, Van Ness, and Abbott~\cite{Smith2004}:

\begin{equation}
\label{eq:enthalpyIG}
h^\textrm{ig} = R \left(a T + b \frac{T^2}{2} - \frac{c}{T} \right), 
\end{equation}
where~$a = 5.457$, $b = 1.045 \times 10^{-3}$~K$^{-1}$, and~$c = -1.157 \times 10^{5}$~K$^{2}$. For pure components, we compare theoretical calculations to experimental data from the National Institute of Standards and Technology (NIST)~\cite{NIST}. In order for the calculations to be compatible with the data, the same reference point for the enthalpy must be chosen. NIST presents their enthalpy data using a reference point where the internal energy of liquid water along the vapor-liquid (VLE) equilibrium curve at 273.16 K is assigned to be zero. With this choice of reference point, NIST reports the molar enthalpy of CO$_2$ at 273.16 K and 34.861 bar (the CO$_2$ vapor pressure at 273.16~K) to be 8.804~kJ/mole. Therefore, the molar enthalpy at any other temperature~$T$ and pressure~$P$ is

\begin{equation}
\label{eq:enthalpyCO2}
h(T, P)  = -RT^2 \left. \left( \frac{\partial \ln \phi(T, P)}{\partial T} \right) \right|_{P,\mathbf{n}} + h^\textrm{ig}(T) - h(T = 273.16\mathrm{~K}, P = 34.861\mathrm{~bar}) + 8.804,
\end{equation}
where~$h(T = 273.16\mathrm{~K}, P = 34.861\mathrm{~bar})$ is the right-hand side of~(\ref{eq:molarEnthalpyPure}) evaluated at the indicated conditions. Computing the enthalpy in this way circumvents the issue that the reference point for the CPA EOS, which is used to calculate the fugacity coefficient, may not be the same as that for the correlation in~(\ref{eq:enthalpyIG}).

\begin{figure}
\centering
(a)\includegraphics[width=75mm,height=60mm]{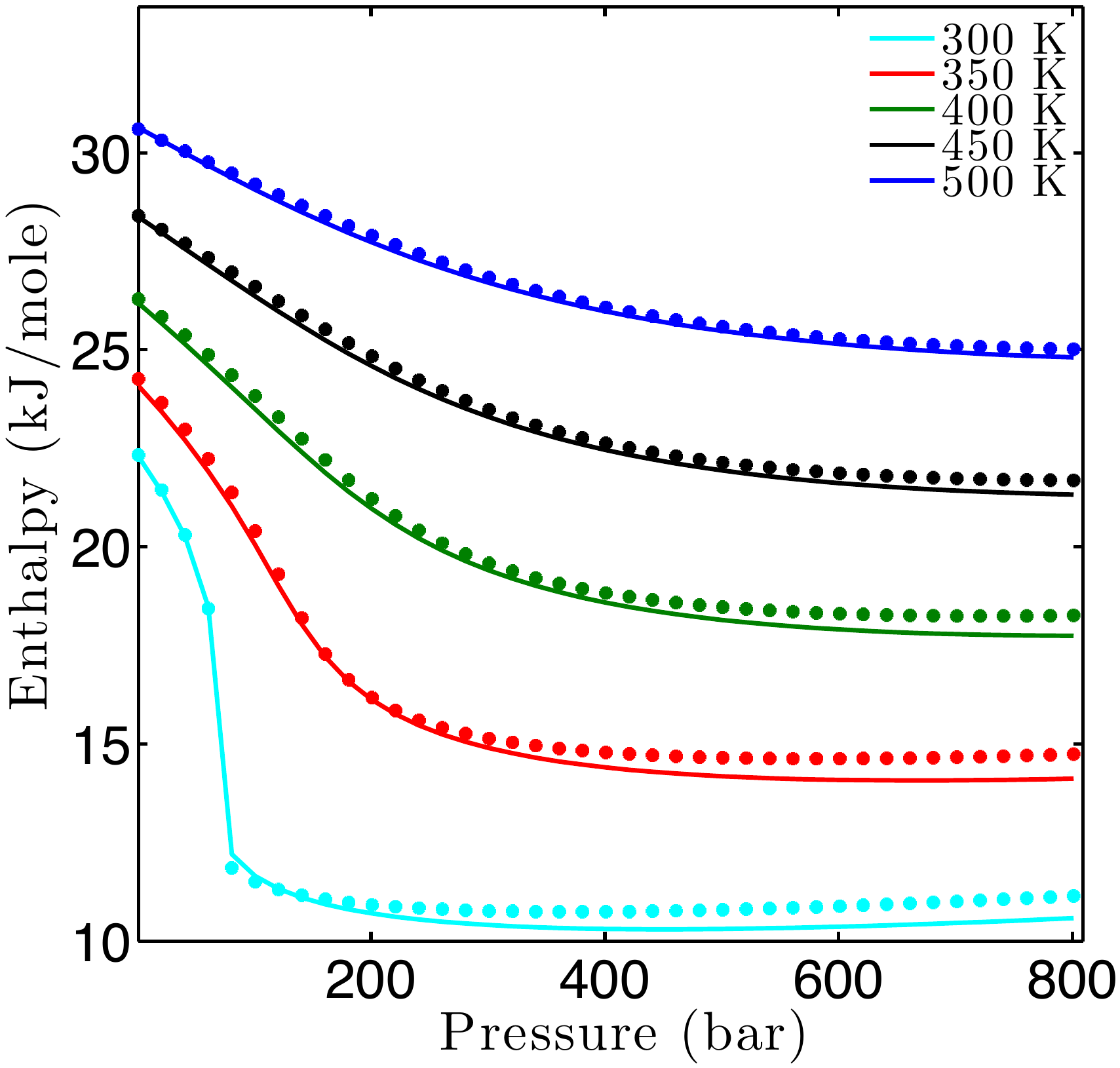} \quad
(b)\includegraphics[width=75mm,height=60mm]{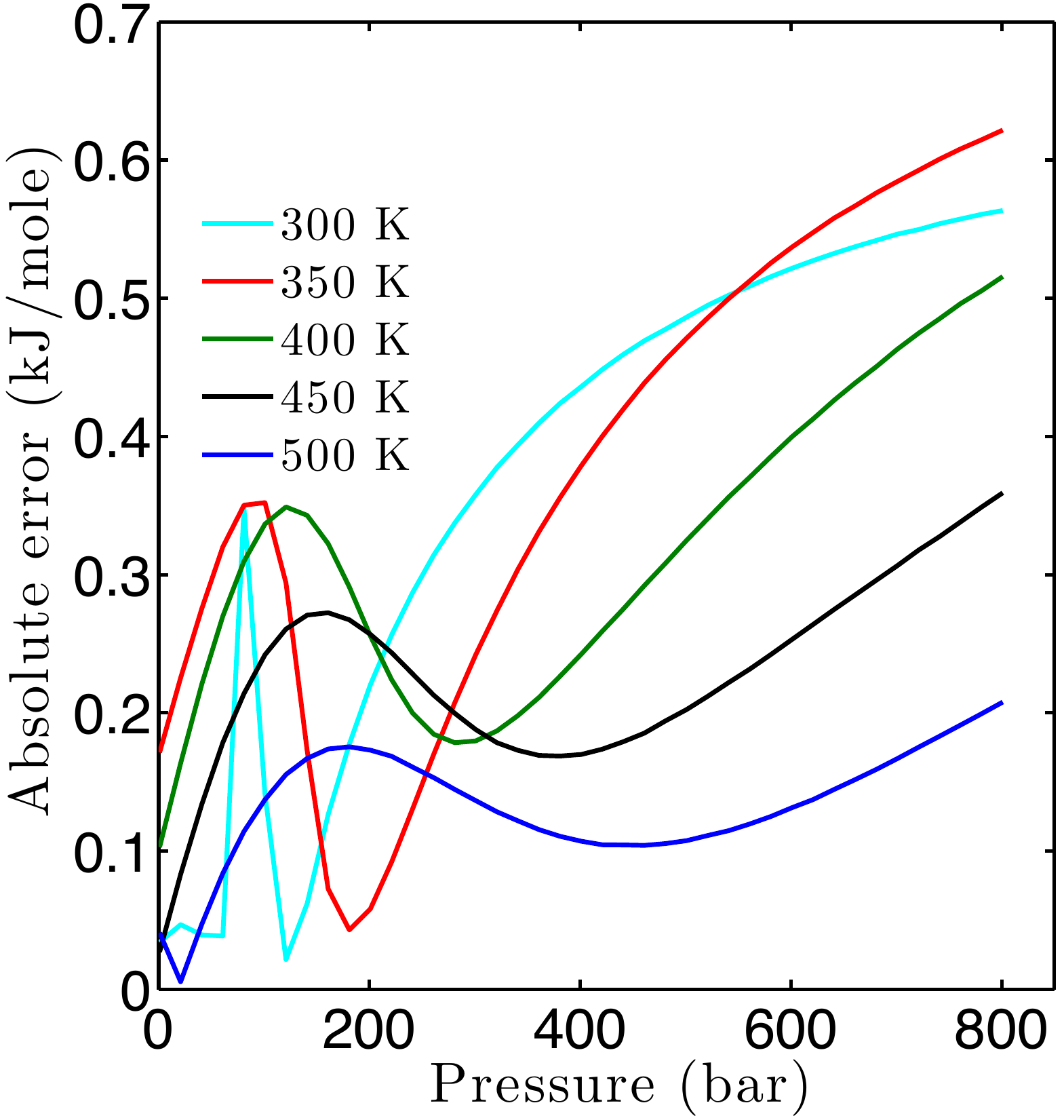} \\
\caption{CPA EOS-predicted enthalpies of pure CO$_2$ vs.\ pressure along five isotherms. Theoretical calculations are depicted in solid lines and are calculated from~(\ref{eq:enthalpyCO2}). Experimental data (dots) are obtained from NIST~\cite{NIST}. The absolute error of the predictions in (a) is shown in (b).}
\label{fig:enthalpyCO2}
\end{figure}

\begin{figure}
(a)\includegraphics[width=75mm,height=60mm]{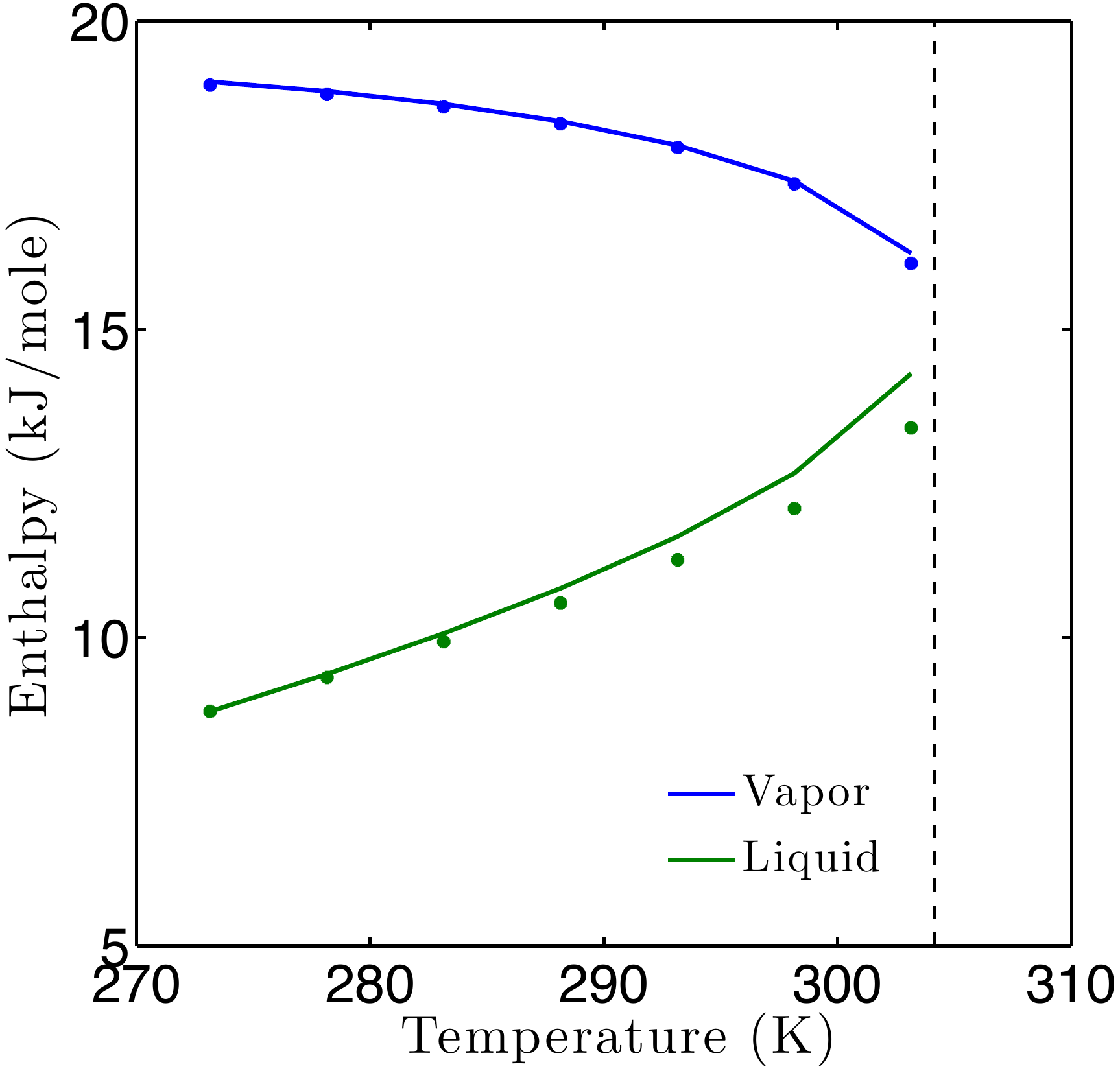} \quad
(b)\includegraphics[width=75mm,height=60mm]{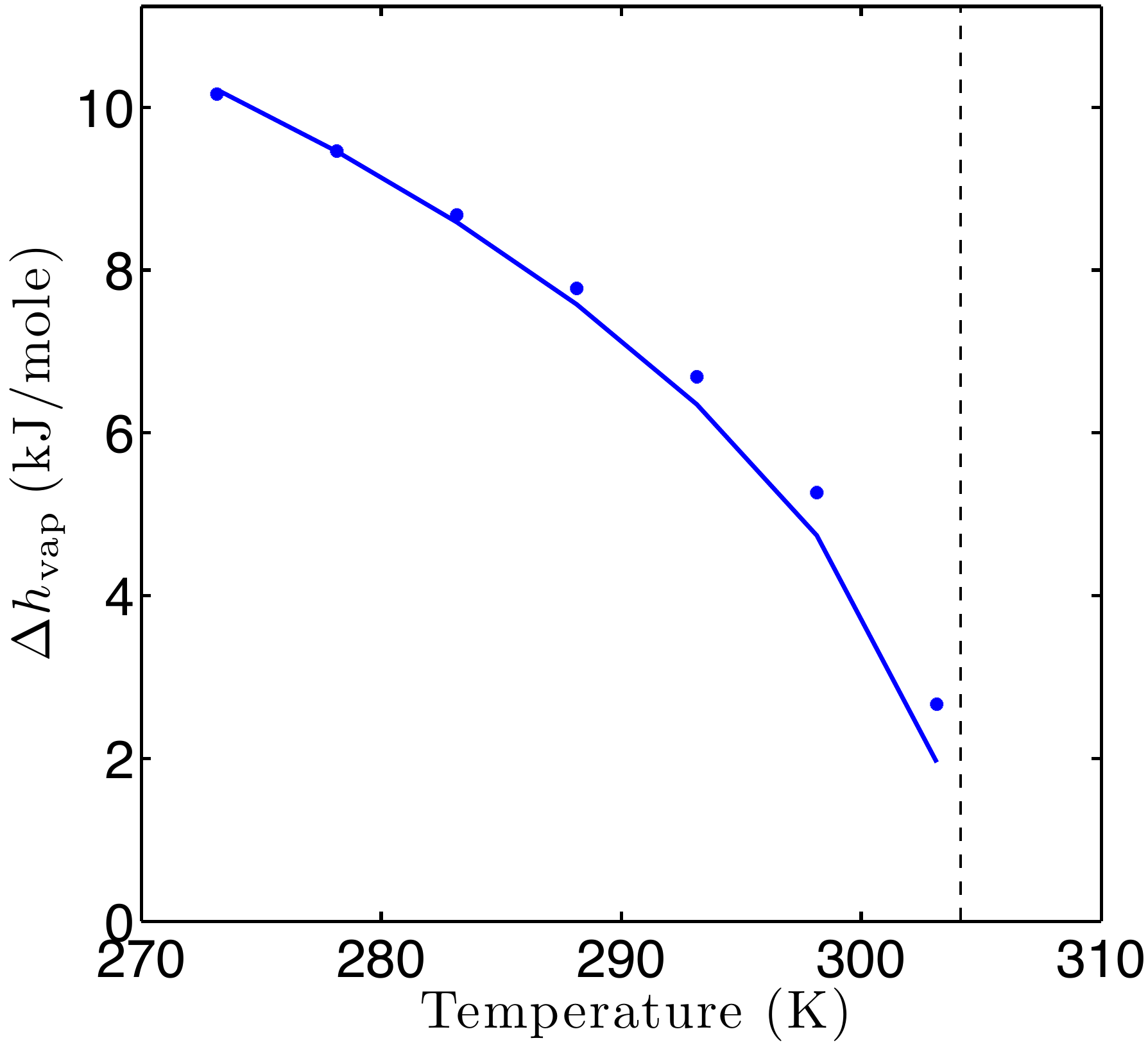} 
\caption{CPA EOS-predicted molar enthalpies of pure CO$_2$ (solid lines) vs.\ temperature along the vapor-liquid equilibrium (VLE) curve. Experimental data (dots) are obtained from NIST~\cite{NIST}.~The dashed vertical line crosses the abscissa at 304.14 K, which is the critical temperature of CO$_2$. (a) Vapor-phase and liquid-phase enthalpies along the VLE curve. (b) presents the molar enthalpy of vaporization $\Delta h_\textrm{vap}$, which is the difference between the vapor and liquid enthalpies in (a).}
\label{fig:HvapCO2}
\end{figure}

The CPA EOS (or more precisely, the Peng-Robinson EOS) produces accurate CO$_2$ enthalpy predictions over a wide range of temperature and pressure conditions (Figure~\ref{fig:enthalpyCO2}), including those stated in Section~\ref{sec:intro} for NRAP. The maximum error is less than 0.60~kJ/mole in virtually the entire range considered. Although not shown here, we have found that calculations for isotherms above 500~K follow the trend in Figure~\ref{fig:enthalpyCO2} of increasingly better agreement with data at higher temperatures. The results are comparable to those from Spycher and Pruess, where the accuracy is stated to be within a few percent~\cite{Spycher2011}. The parameters of the EOS have been fitted to experimental vapor pressure measurements, so that the vapor pressure predictions closely match the data, even near the critical temperature. Except for near the critical temperature, the CPA EOS can also accurately predict the molar enthalpy of vaporization $\Delta h_\textrm{vap}$ [Figure~\ref{fig:HvapCO2}(b)], which is defined as the difference in the vapor-phase and liquid-phase enthalpies shown in Figure~\ref{fig:HvapCO2}(a). The enthalpy of vaporization rapidly approaches zero near the critical point; most of the error near this point comes from the liquid-phase enthalpy predictions. Due to thermodynamic singularities that arise, it is not uncommon for equations of state to have larger errors near the critical point. 

\subsection{H$_2$O}
\label{sec:water}

The molar enthalpy of pure water is calculated from~(\ref{eq:molarEnthalpyPure}). In this case, the association contribution to the fugacity coefficient in~(\ref{eq:lnphiZLi}) is non-zero so that the CPA EOS yields different results from the Peng-Robinson EOS on which it is based. The molar enthalpy~$h^\textrm{ig}$ of water as an ideal gas can be calculated using~(\ref{eq:enthalpyIG}), with~$a = 3.470$, $b = 1.450 \times 10^{-3}$~K$^{-1}$, and~$c = 0.121 \times 10^{5}$~K$^{2}$~\cite{Smith2004}. An alternative and slightly more accurate correlation for~$h^\textrm{ig}$ has been presented by Cooper~\cite{Cooper1982}. Using the same reference point mentioned in the previous section (internal energy of liquid water along the VLE curve at 273.16~K), we calculate the enthalpy according to

\begin{equation}
\label{eq:enthalpyH2O}
h(T, P)  = -RT^2 \left. \left( \frac{\partial \ln \phi(T, P)}{\partial T} \right) \right|_{P,\mathbf{n}} + h^\textrm{ig}(T) - h(T = 273.16\mathrm{~K}, P = 0.006117\mathrm{~bar}),
\end{equation}
where~0.006117~bar is the vapor pressure of water at 273.16~K. Equation~(\ref{eq:enthalpyH2O}) should also contain a term that represents the difference between the enthalpy and the internal energy of water at $(T = 273.16\mathrm{~K}, P = 0.006117\mathrm{~bar})$, but this term is so small that we neglect it.

\begin{figure}
\centering
(a)\includegraphics[width=75mm,height=60mm]{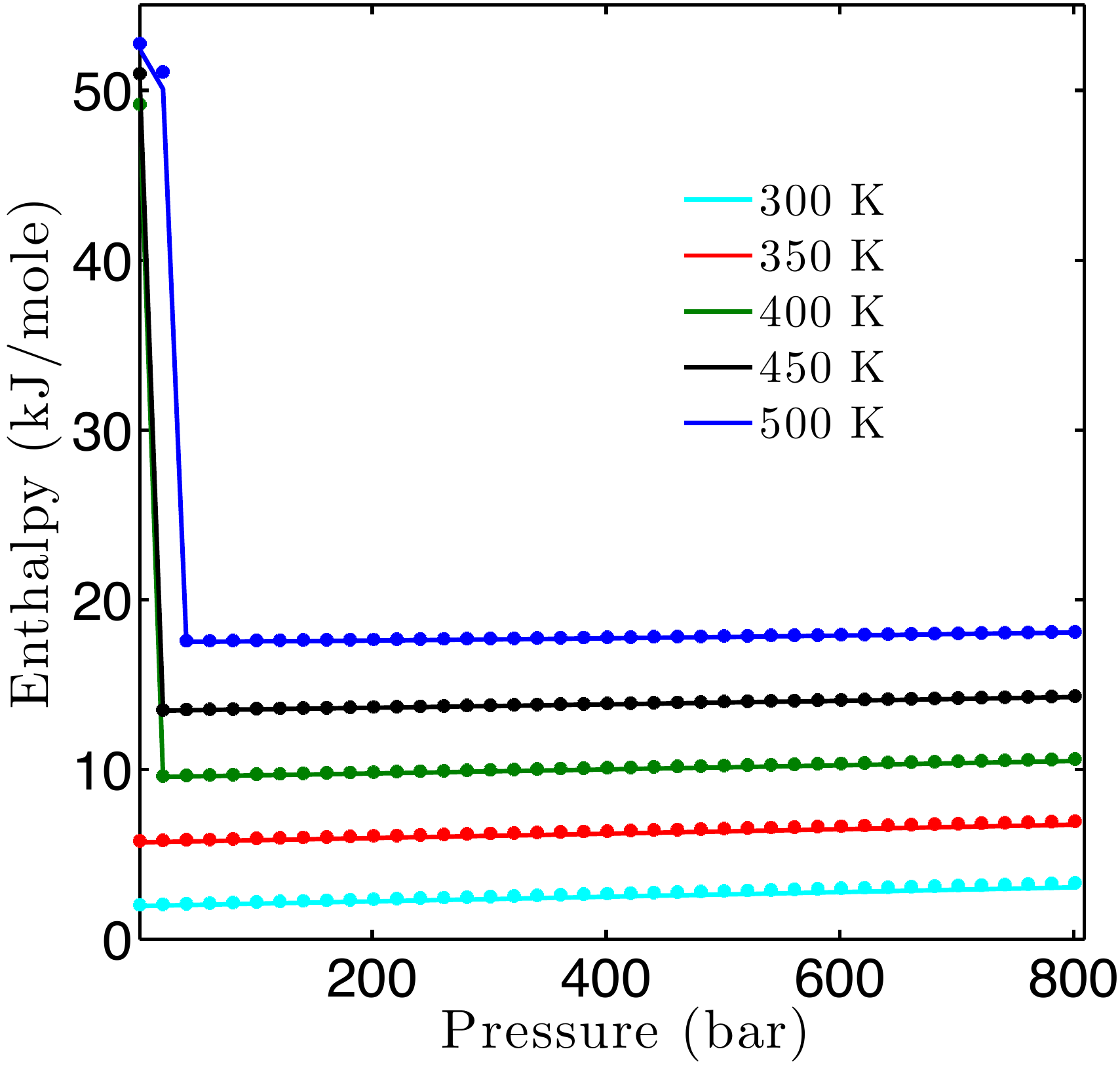} \quad
(b)\includegraphics[width=75mm,height=60mm]{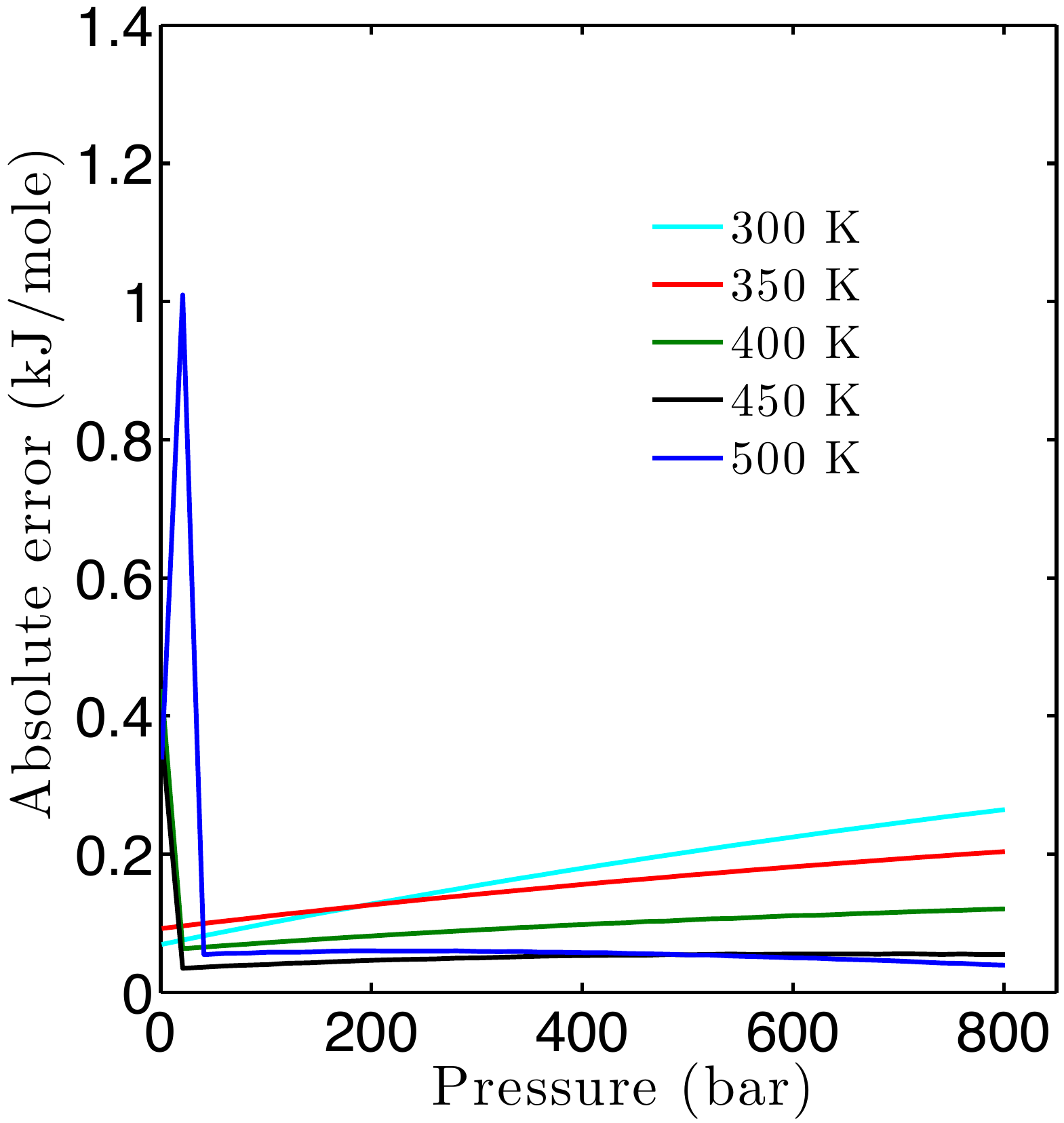} \\
\caption{CPA EOS-predicted enthalpies of pure H$_2$O vs.\ pressure along five isotherms. Theoretical calculations are depicted in solid lines and are calculated from~(\ref{eq:enthalpyH2O}). Experimental data (dots) are obtained from NIST~\cite{NIST}. The absolute error of the predictions in (a) is shown in (b). The 1.0~kJ/mole error for the 500~K isotherm is associated with the vapor formed during the phase change at that temperature (see Figure~\ref{fig:HvapH2O}).}
\label{fig:enthalpyH2O}
\end{figure}

\begin{figure}
\centering
(a)\includegraphics[width=75mm,height=60mm]{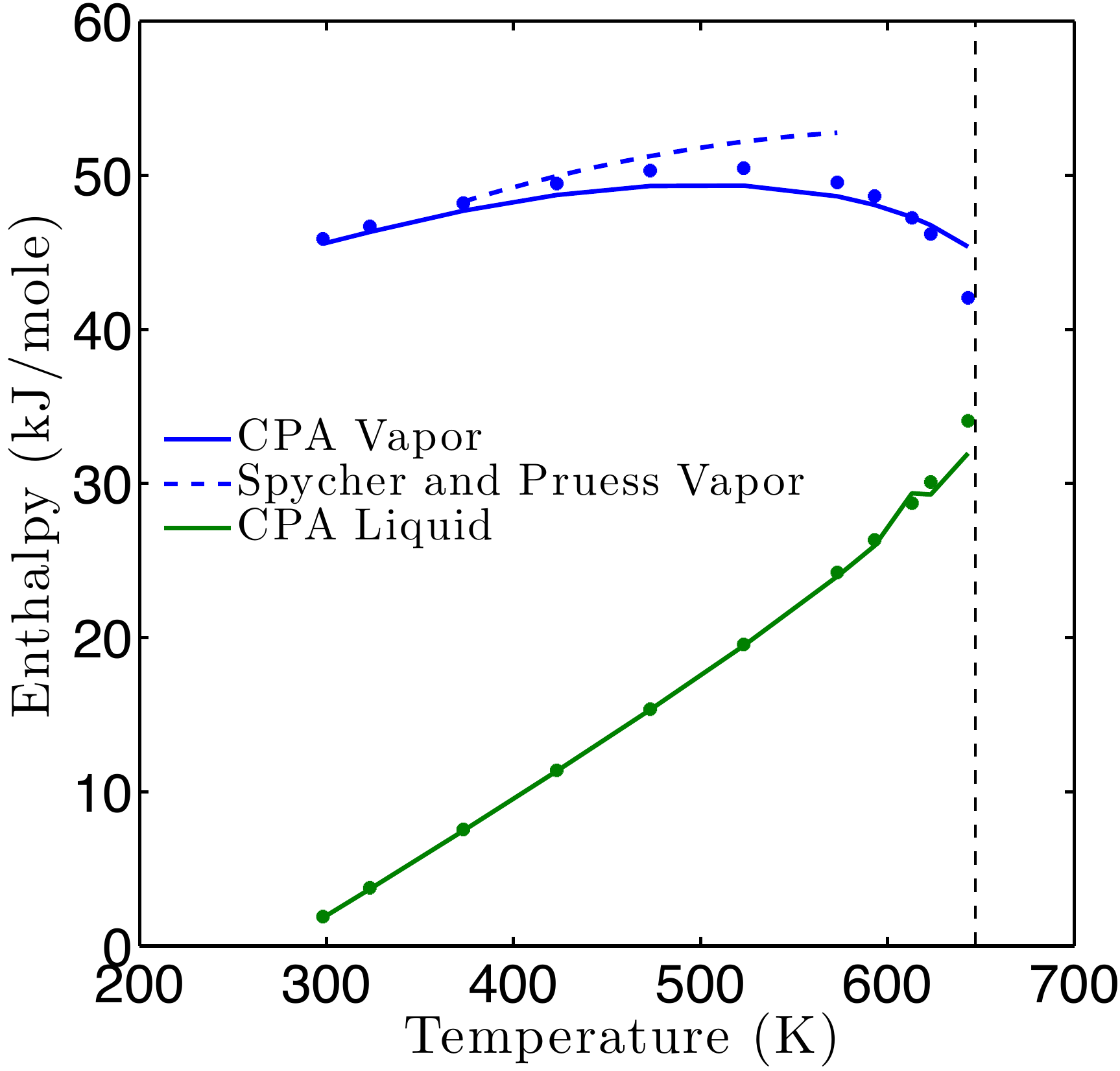} \quad
(b)\includegraphics[width=75mm,height=60mm]{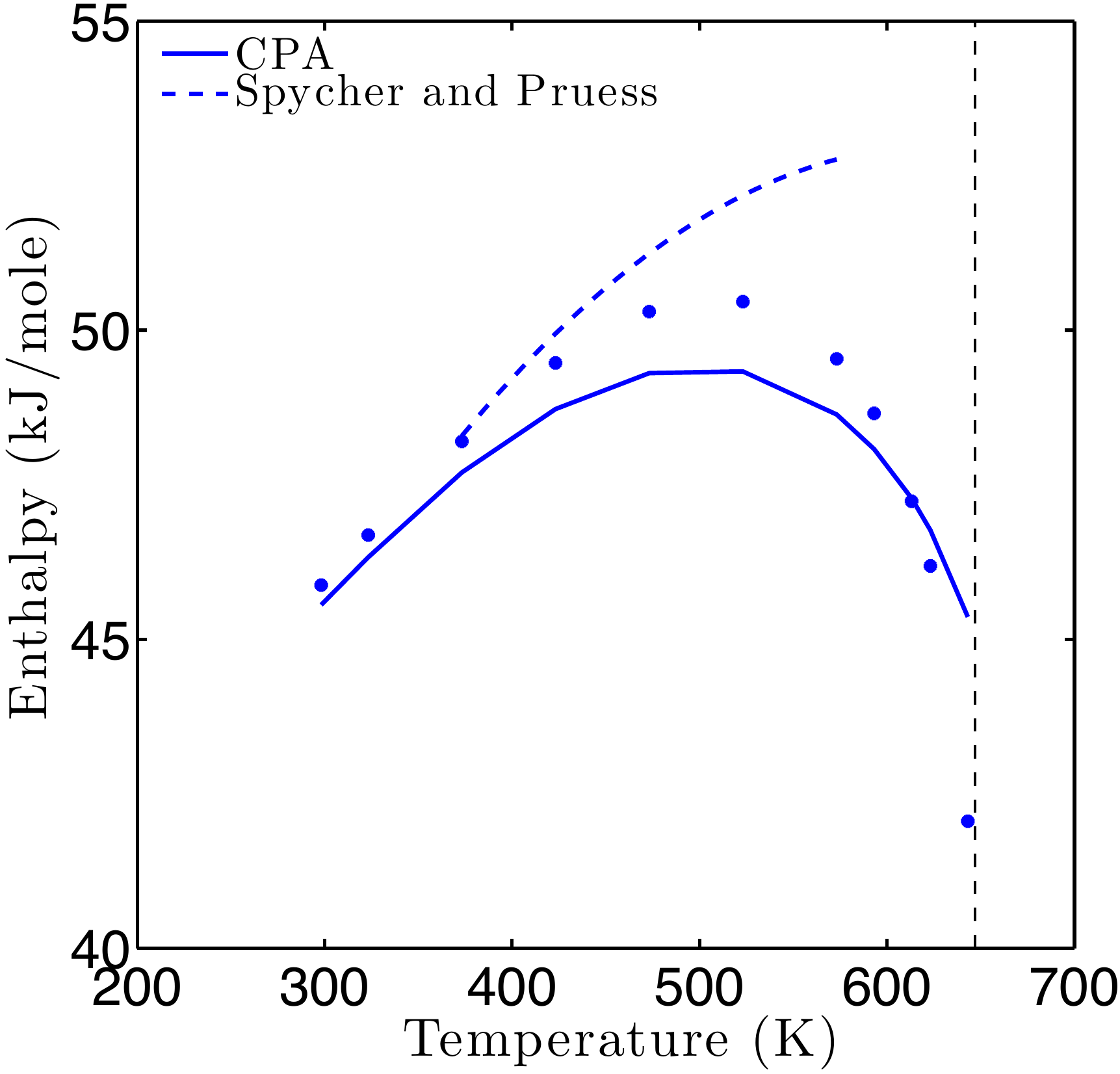} \\ 
\vspace{10pt} (c)\includegraphics[width=75mm,height=60mm]{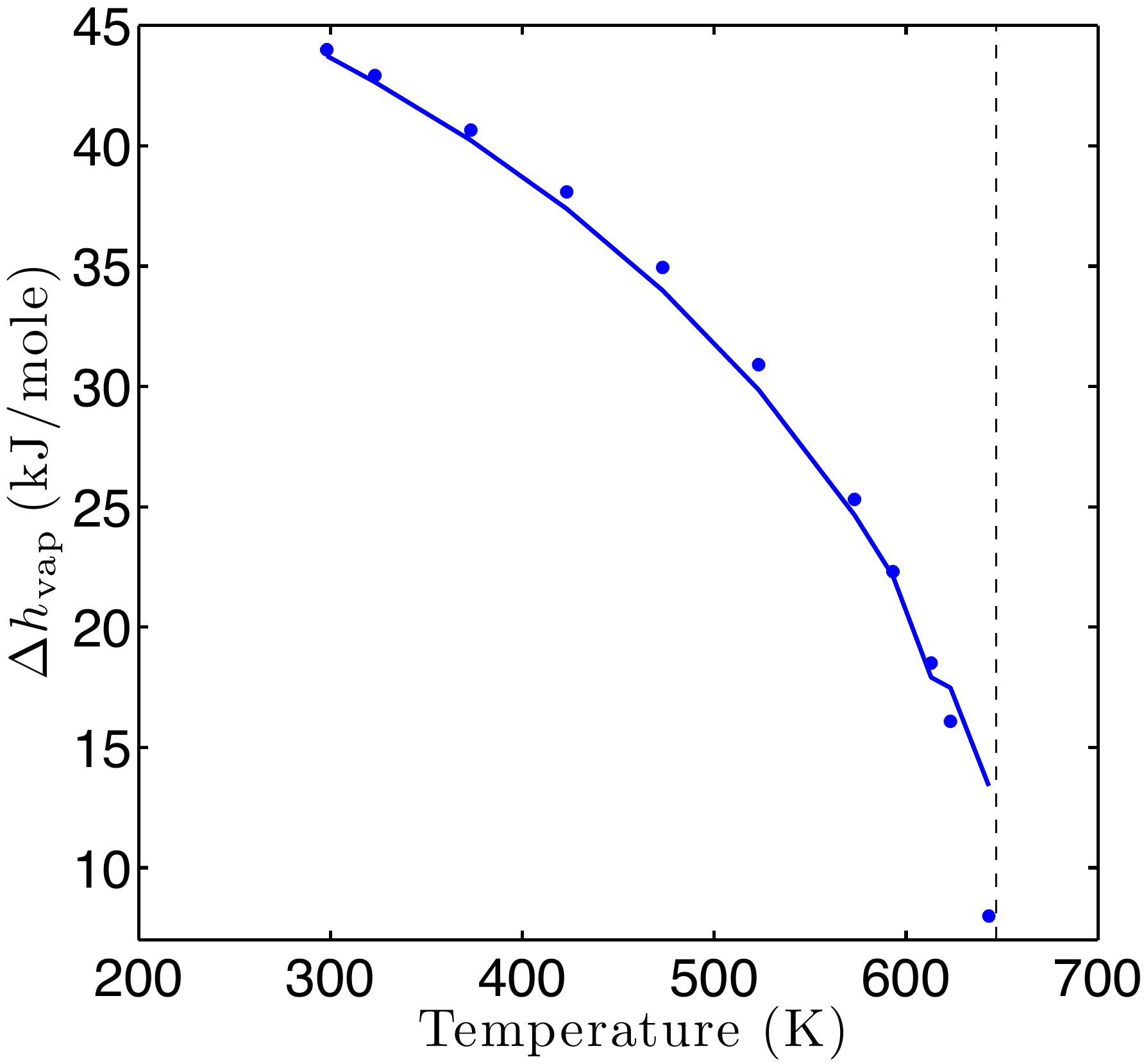}
\caption{CPA EOS-predicted molar enthalpies of pure H$_2$O vs.\ temperature along the VLE curve. Results from Spycher and Pruess~\cite{Spycher2011} are included for comparison. Experimental data (dots) are obtained from NIST~\cite{NIST}. The dashed line vertical crosses the abscissa at 647.1 K, which is the critical temperature of H$_2$O. A magnified view of the vapor-phase results in (a) is shown in (b). Predictions of the molar enthalpy of vaporization is compared with experimental data in (c).}
\label{fig:HvapH2O}
\end{figure}

Results for five isotherms are shown in Figures~\ref{fig:enthalpyH2O}(a) and~\ref{fig:enthalpyH2O}(b), while the molar enthalpy of vaporization is shown in Figure~\ref{fig:enthalpyH2O}(c). Similar to the case for CO$_2$, the the EOS parameters for water (such as~$\epsilon$ and~$\kappa$ described in Section~\ref{sec:FdepartAssoc}) are fit to vapor pressure data. Predictions of~$\Delta h_\textrm{vap}$ are accurate to within a few percent as long as the temperature is not close to the critical temperature. Spycher and Pruess have obtained good agreement with experimental data for pure water, except for the enthalpy of the vapor along certain regions of the VLE curve~\cite{Spycher2011}. A comparison of their model's predictions with those of the CPA EOS along the VLE curve is shown in Figure~\ref{fig:HvapH2O}. The accuracy of the two models is similar in the range of temperatures between 373--550~K (less than 2.5~\% maximum error for both). Above 550~K, the CPA EOS yields noticeably better results, although it fails to reproduce the sharp decrease in the vapor-phase enthalpy near the critical temperature. More importantly, however, Spycher and Pruess state that their model cannot calculate the enthalpy of water below 373~K. This could have practical implications for the applications described in Section~\ref{sec:intro}. For example, in numerical simulations of CO$_2$ sequestration leakage, Spycher and Pruess' model cannot be used to calculate the enthalpy in the aqueous environments of the shallow aquifers and nearby leakage pathways.

\section{Enthalpy of mixtures}
\label{sec:mixture}

Mixtures of CO$_2$ and water can in general exist as two different phases, a CO$_2$-rich phase and an aqueous (H$_2$O-rich) phase. We treat the CO$_2$-rich phase as being composed only of CO$_2$ and water, while the aqueous phase may have dissolved sodium chloride.

\subsection{CO$_2$-rich phase (CO$_2$ + H$_2$O)}
\label{sec:CO2-rich}

At high ($T,P$) conditions, the CO$_2$-rich phase may be composed almost entirely of CO$_2$ that exists in a relatively dense supercritical state. The solubility of water in this form of CO$_2$ is less than 2~mole~\%~\cite{Li2009}, and often times it is even much less than that. Since the mixture is almost pure CO$_2$, the Peng-Robinson EOS is a very good approximation to the CPA EOS, and we have already demonstrated in Section~\ref{sec:CO2} that it can accurately model the behavior of CO$_2$ over a wide range of ($T,P$), including supercritical conditions. However, at sufficiently low pressures for a given temperature, CO$_2$ may exist as a gaseous substance. When water vapor is exposed to CO$_2$ at these conditions, the resulting gaseous mixture may have very high concentrations of water, since gases are completely miscible in each other. These mixtures find applications in technologies such as oxy-fuel combustion, which is a method for carbon capture involving the formation of mixtures rich in CO$_2$ and water vapor~\cite{Firoozabadi2010}.

\begin{figure}
(a)\includegraphics[width=75mm,height=60mm]{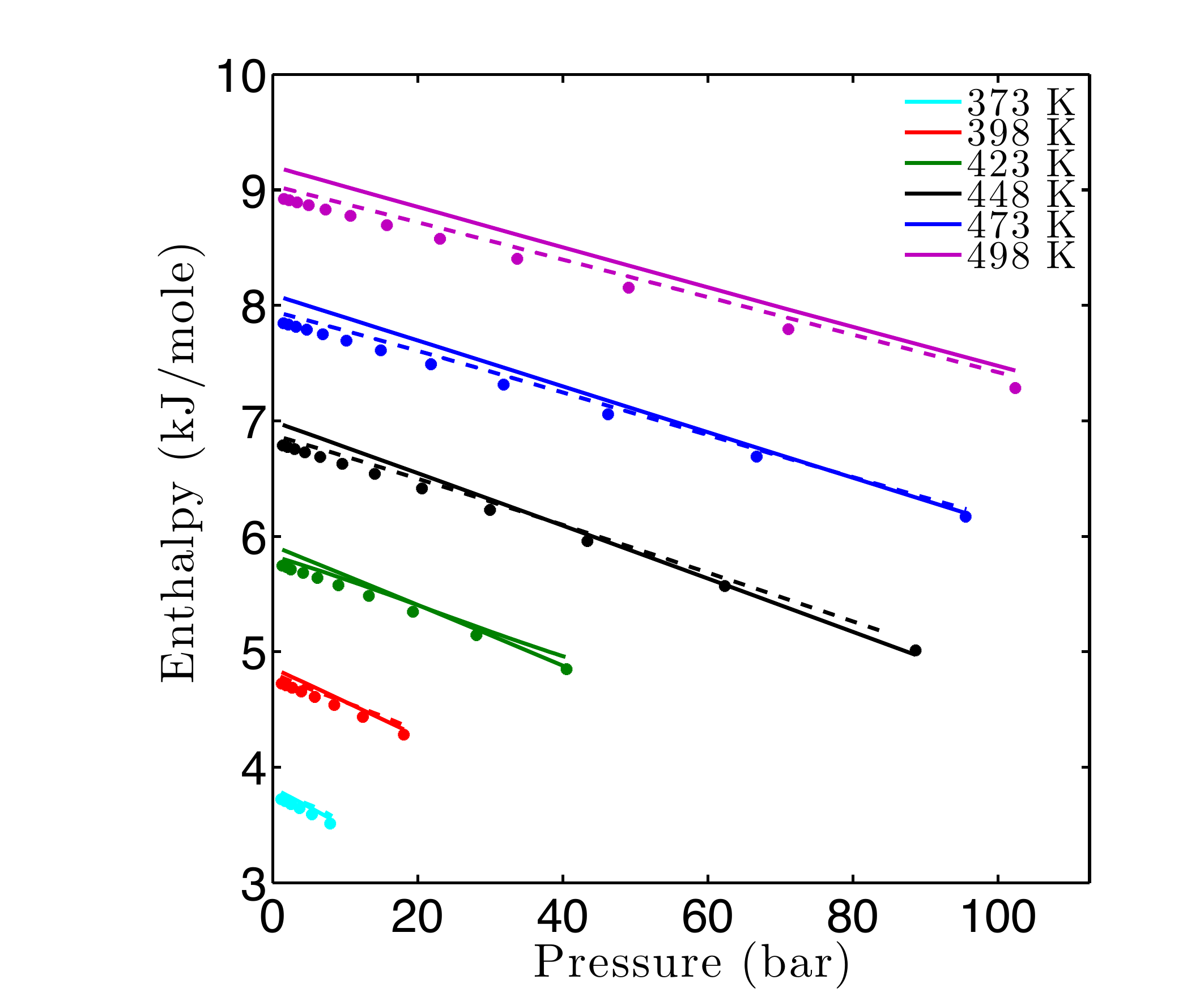} \quad
(b)\includegraphics[width=75mm,height=60mm]{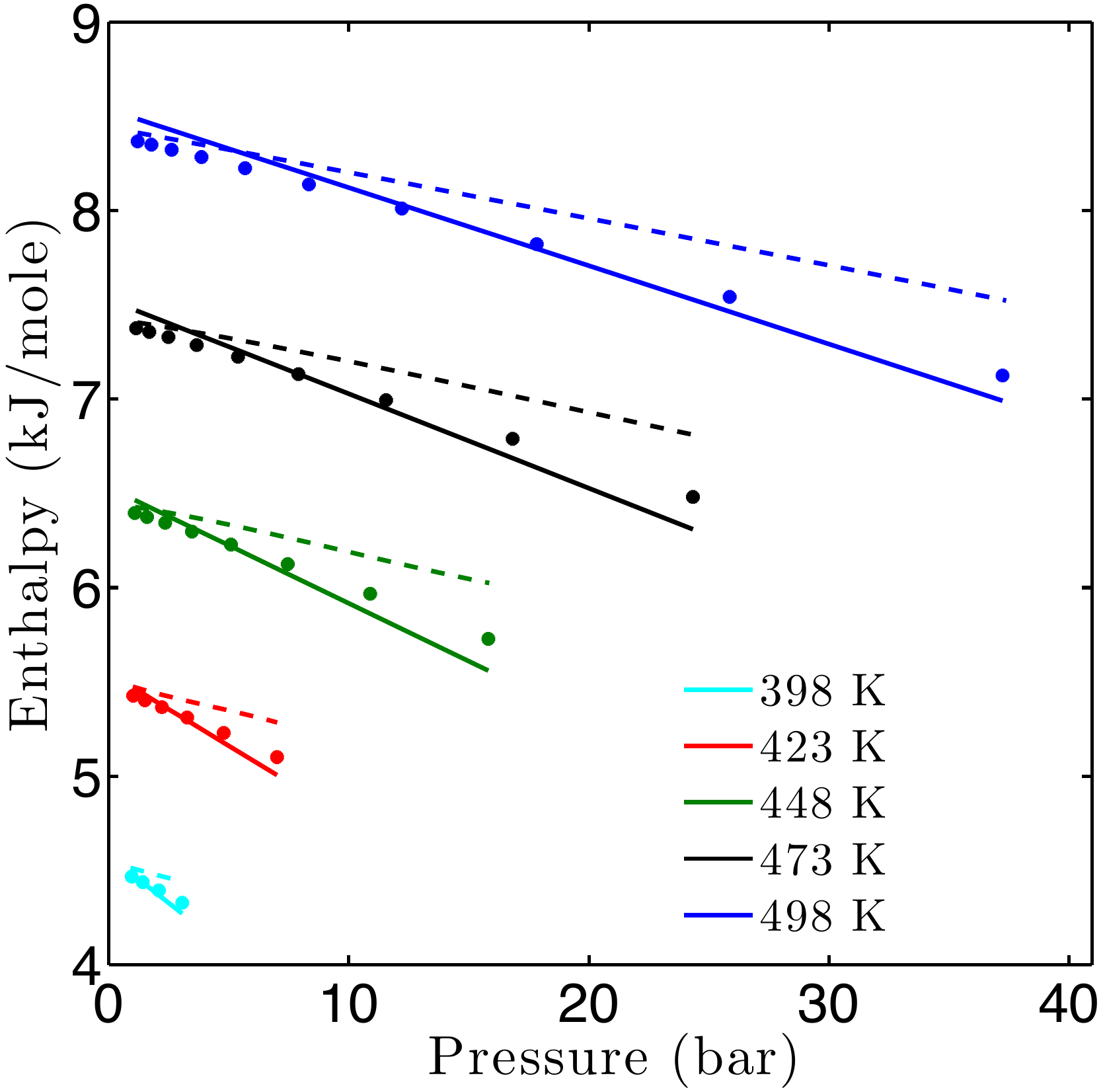} 
\caption{CPA EOS-predicted molar enthalpies of gaseous CO$_2$-H$_2$O mixtures (solid lines) along different isotherms vs.\ pressure. Experimental data (dots) are obtained from Patel and Eubank~\cite{Patel1988}. Results from Spycher and Pruess (dashed lines)~\cite{Spycher2011} are included for comparison. (a)~9:1 mole ratio of CO$_2$ to  H$_2$O; (b)~1:1 mole ratio of CO$_2$ to H$_2$O.}
\label{fig:enthalpyCO2rich}
\end{figure}

Experimental enthalpy data for CO$_2$-H$_2$O mixtures, which are much less common than data for the pure components, have been presented by Patel and Eubank~\cite{Patel1988}. They have studied CO$_2$-rich gaseous mixtures between 323~K and 498~K where the water vapor mole fraction can be as high as 50~\%. The same set of data is also considered by Spycher and Pruess~\cite{Spycher2011}. Patel and Eubank choose the reference to be the enthalpy of the corresponding ideal gas mixture at $T = 273.16\mathrm{~K}$. With this reference, we calculate the molar enthalpy~$h(T, P, \mathbf{z})$ of the real fluid mixture from the CPA EOS according to

\begin{equation*}
h(T, P, \mathbf{z}) = -RT^2 \sum_{i=1}^2 z_i \left. \left( \frac{\partial \ln \phi_i(T, P, \mathbf{n})}{\partial T} \right) \right|_{P,\mathbf{n}} + h^\textrm{ig}(T) - h^\textrm{ig}(T = 273.16\mathrm{~K}),
\end{equation*}
where~$h^\textrm{ig}(T) = \sum_{i=1}^2 z_i h_i^\textrm{ig}(T)$,~$z_i$ is the mole fraction of component~$i$, and the pure component ideal gas enthalpies~$h_i^\textrm{ig}(T)$ are calculated as described in Section~\ref{sec:pure}. Enthalpy results for several isotherms of CO$_2$-H$_2$O mixtures at two different compositions are shown in Figure~\ref{fig:enthalpyCO2rich}. Spycher and Pruess' model is more accurate at low pressures. As the pressure increases, the CPA EOS becomes more accurate. The point at which this switch occurs depends on the temperature and the composition; the CPA EOS tends to be more accurate at higher concentrations of water and at lower temperatures. In Figure~\ref{fig:enthalpyCO2rich}(a), the maximum error for the CPA EOS is 0.26~kJ/mole. In Figure~\ref{fig:enthalpyCO2rich}(b), the maximum error is 0.17~kJ/mole. Spycher and Pruess state that their maximum error is about 0.40~kJ/mole (13~kJ/kg). Besides the fact that the CPA EOS can simultaneously predict phase equilibria behavior, density, and enthalpy, its main advantage is that it can calculate the enthalpy of pure water or mixtures involving water below 373~K. Although not shown in Figure~\ref{fig:enthalpyCO2rich}(a), Patel and Eubank present a few enthalpy data points at 348~K for the 90~\%~CO$_2$ mixture. The maximum error of the CPA EOS for these points is less than 0.03~kJ/mole.

\subsection{Aqueous phase (H$_2$O + CO$_2$ + NaCl)}
\label{sec:aqueous}

\begin{figure}
\centering
(a)\includegraphics[width=75mm,height=60mm]{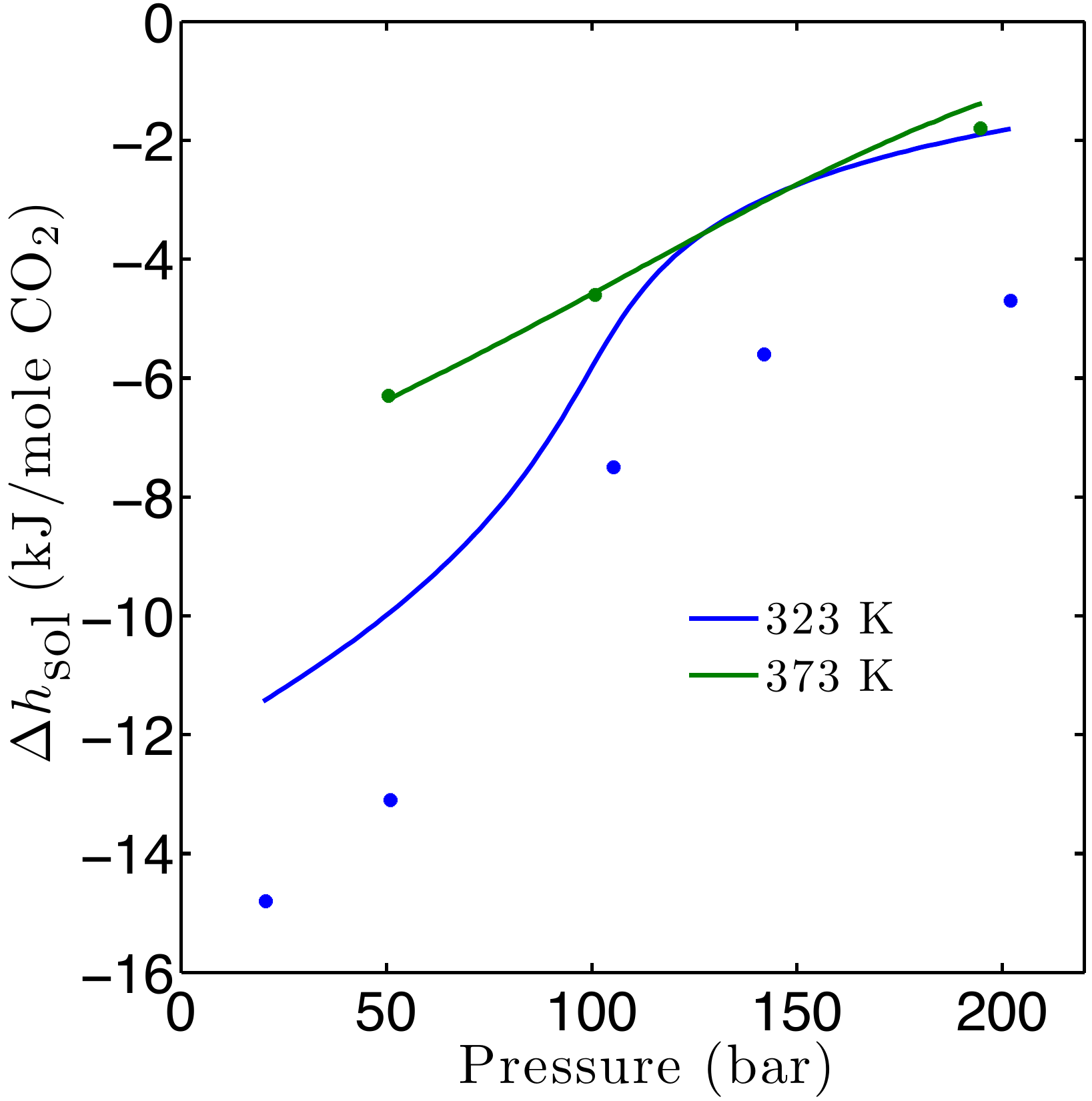} \quad
(b)\includegraphics[width=75mm,height=60mm]{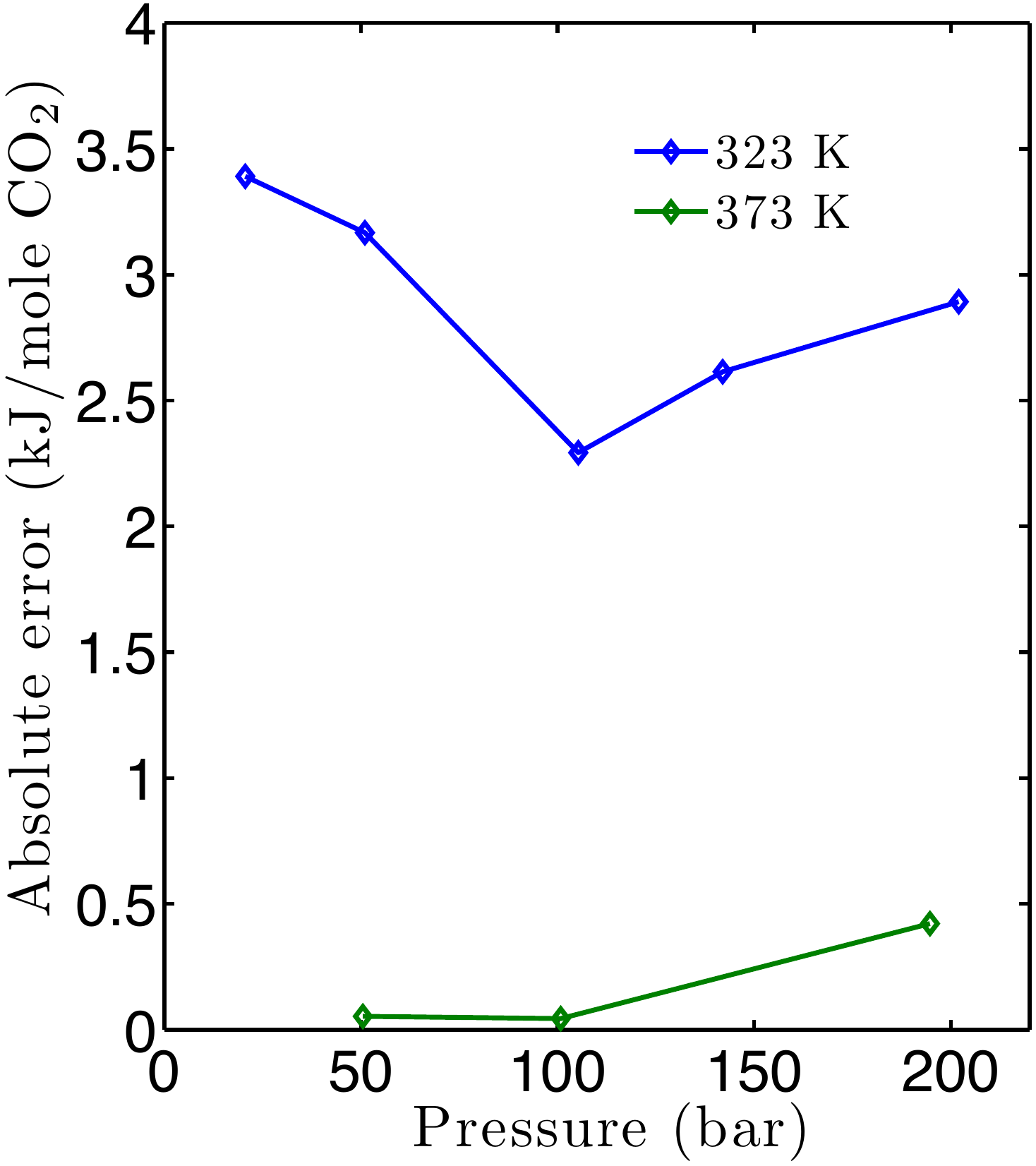} \\
\caption{(a) CPA EOS-predicted CO$_2$ molar enthalpy of solution in pure water vs.\ pressure at two different temperatures. Experimental data are obtained from Koschel~\textit{et al}.~\cite{Koschel2006}. (b) shows the absolute error of the predictions. Despite the large error in~$\Delta h_\mathrm{sol}$ at~323~K, the EOS can still accurately compute the molar enthalpy~$h(T, P, \mathbf{z})$ of H$_2$O-CO$_2$ mixtures, as explained in the text.}
\label{fig:deltaHsolCPA}
\end{figure}

To the best of our knowledge, there are no reported molar enthalpy measurements for aqueous mixtures with CO$_2$. Measurements of the CO$_2$ molar enthalpy of solution~$\Delta h_\mathrm{sol}$ are available, however. This quantity is defined as

\begin{equation}
\label{eq:deltaHsol}
\Delta h_\mathrm{sol} = \bar{H}_{\mathrm{CO}_2}(T, P, z_{\mathrm{CO}_2} \rightarrow 0) - h_{\mathrm{CO}_2}(T, P).
\end{equation}
That is,~$\Delta h_\mathrm{sol}$ is the difference between the partial molar enthalpy of CO$_2$ in an infinitely dilute aqueous solvent and the molar enthalpy of pure CO$_2$. Koschel~\textit{et al}.~\cite{Koschel2006} have presented~$\Delta h_\mathrm{sol}$ data at 323~K and 373~K in three different aqueous solvents: pure water, 1~molal~($m$) NaCl, and 3~$m$~NaCl. As explained in Section~\ref{sec:Helmholtz}, the CPA EOS is limited to aqueous-phase mixtures containing only pure water. For such mixtures, Figure~\ref{fig:deltaHsolCPA} compares~$\Delta h_\mathrm{sol}$ calculations from the CPA EOS with experimental data from Koschel~\textit{et al}. The agreement at 323~K is poor, while that at 373~K is relatively good. Since the EOS accurately calculates the molar enthalpy of pure CO$_2$, most of the error in~$\Delta h_\mathrm{sol}$ comes from calculation of the infinite dilution partial molar enthalpy. 

Despite the large error in~$\Delta h_\mathrm{sol}$, the CPA EOS can still provide accurate predictions of the molar enthalpy~$h(T, P, \mathbf{z})$ of the mixture. This can be reasoned as follows. From~(\ref{eq:molarEbar}), the molar enthalpy of H$_2$O-CO$_2$ mixtures is

\begin{equation*}
h(T, P, \mathbf{z}) = z_{\mathrm{CO}_2} \bar{H}_{\mathrm{CO}_2} + z_{\mathrm{H}_2\mathrm{O}} \bar{H}_{\mathrm{H}_2\mathrm{O}}.
\end{equation*}
For the temperatures and pressures considered in this report (say,~$T < 500$~K and~$P < 500$~bar), the CO$_2$ solubility in pure water is less than 3.5~mole~\%~\cite{Li2009}. Suppose that the error in the calculation of the partial molar enthalpy~$\bar{H}_{\mathrm{CO}_2}$ is 4.0~kJ/mole~CO$_2$, which is a value greater than the maximum error of 3.5~kJ/mole~CO$_2$ in Figure~\ref{fig:deltaHsolCPA}. Then the error in the product~$z_{\mathrm{CO}_2} \bar{H}_{\mathrm{CO}_2}$, assuming that~$z_{\mathrm{CO}_2} = 0.035$, is 0.14~kJ/mole. The partial molar enthalpy~$\bar{H}_{\mathrm{H}_2\mathrm{O}}$ of water is similar to the molar enthalpy~$h_{\mathrm{H}_2\mathrm{O}}$ of the pure component because the aqueous mixture is predominantly water. For example at 323~K, the CPA EOS predicts that the difference between the two quantities is less than 0.006~kJ/mole~H$_2$O even for CO$_2$-saturated water at 500~bar. It is therefore reasonable to expect the error in~$\bar{H}_{\mathrm{H}_2\mathrm{O}}$ to be similar to that for~$h_{\mathrm{H}_2\mathrm{O}}$ shown in Figure~\ref{fig:enthalpyH2O}, where the maximum error is 0.25~kJ/mole~H$_2$O in the liquid phase. Thus, the error in~$h(T, P, \mathbf{z})$ over a wide range of conditions will likely be less than 0.40~kJ/mole ($\approx 0.14$~kJ/mole~$+~0.25$~kJ/mole), which is rather small considering the expected magnitude of~$h(T, P, \mathbf{z})$. Nonetheless, we propose a generalized version of the Li and Firoozabadi CPA EOS in Section~\ref{sec:excess}, and explain how it may provide better agreement with the excess enthalpy, which is related to~$\Delta h_\mathrm{sol}$. Another alternative is the Duan and Sun CO$_2$ activity coefficient model~\cite{Duan2003}, which not only agrees better with~$\Delta h_\mathrm{sol}$ measurements in pure water, but also provides an accurate way to calculate the CO$_2$ partial molar enthalpy (and subsequently~$\Delta h_\mathrm{sol}$) in brine solutions. Their study is the focus of the rest of this section.

\begin{figure}
\centering
(a)\includegraphics[width=75mm,height=60mm]{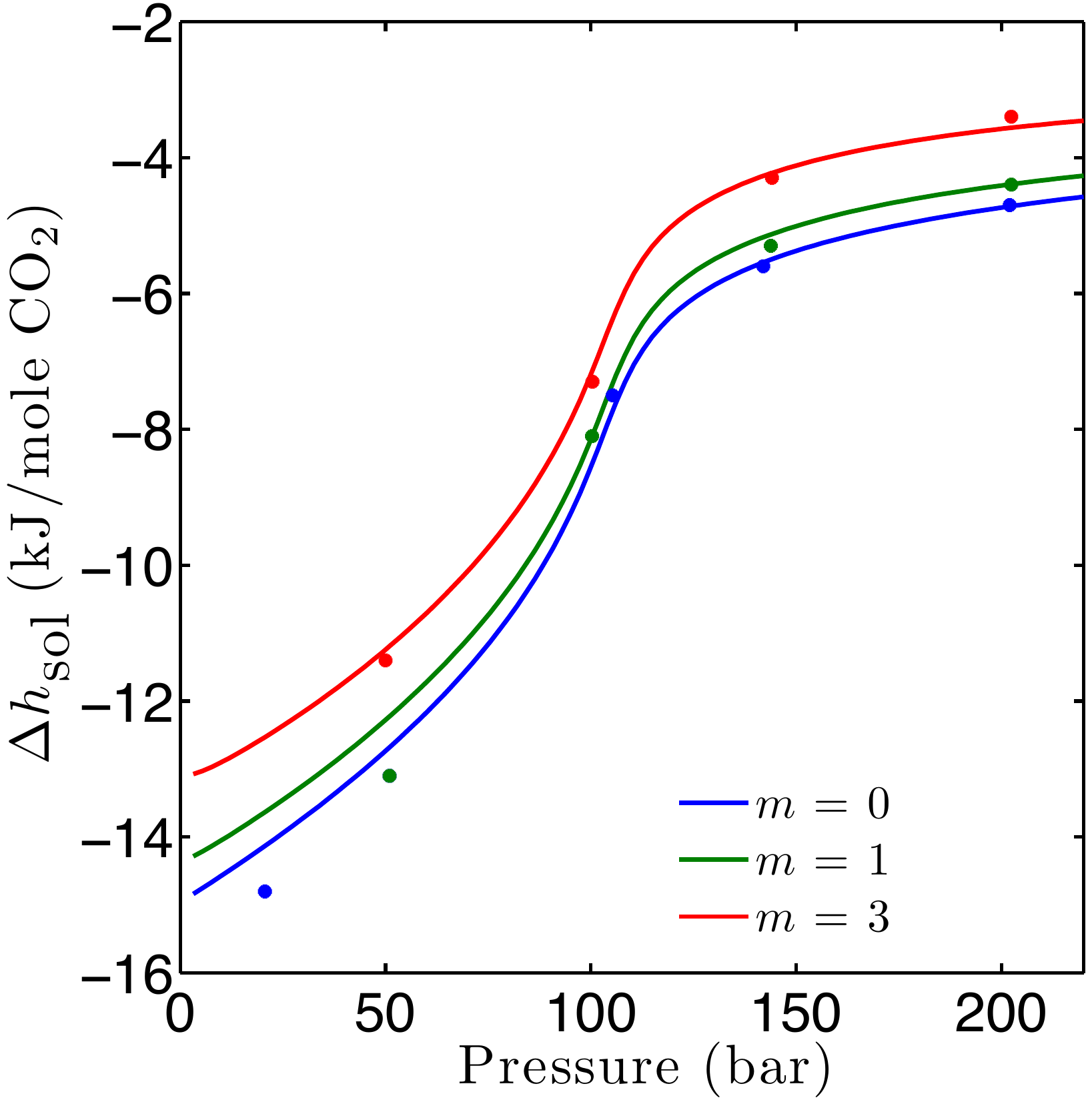} \quad
(b)\includegraphics[width=75mm,height=60mm]{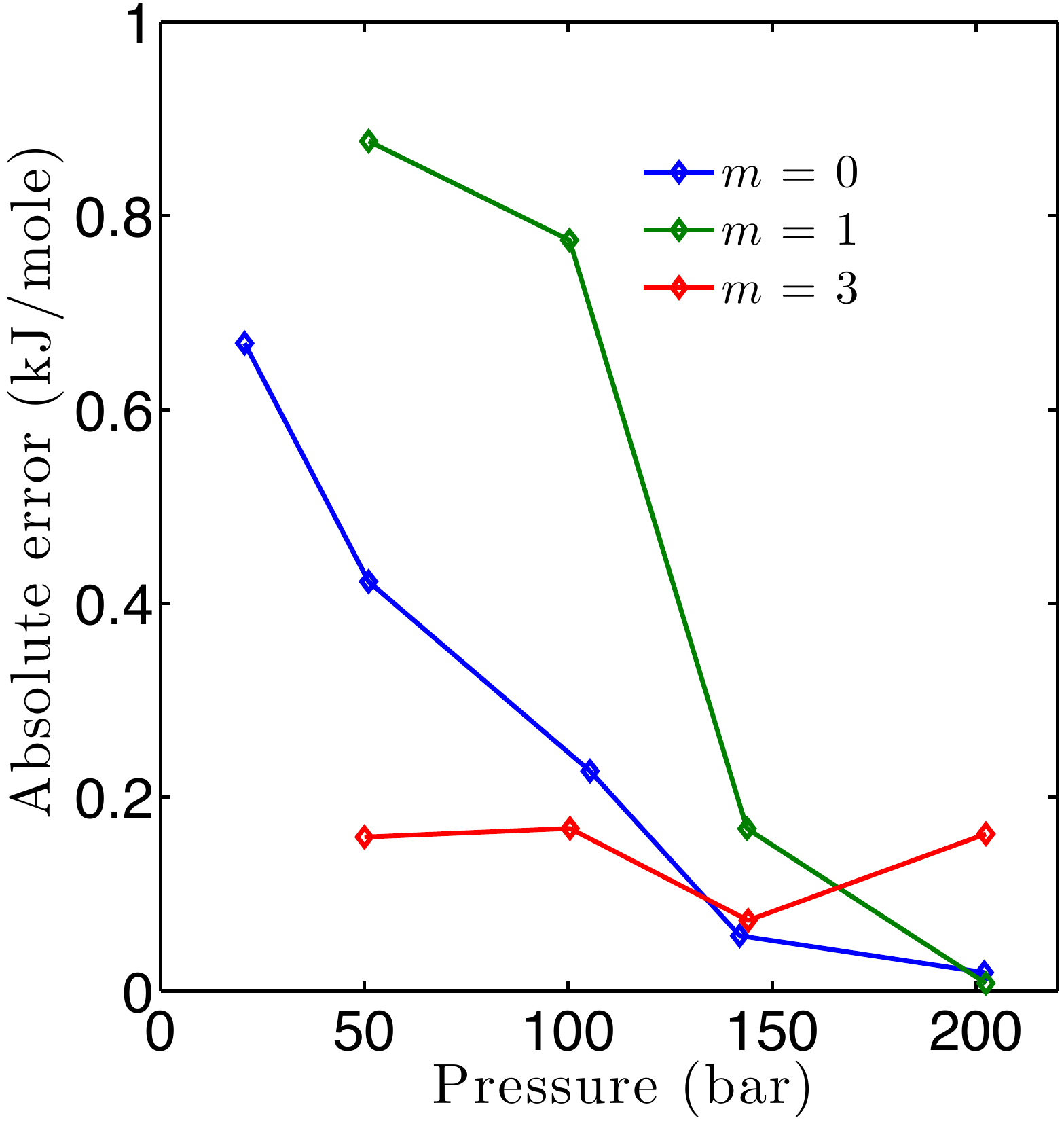} \\
\vspace{10pt} (c)\includegraphics[width=75mm,height=60mm]{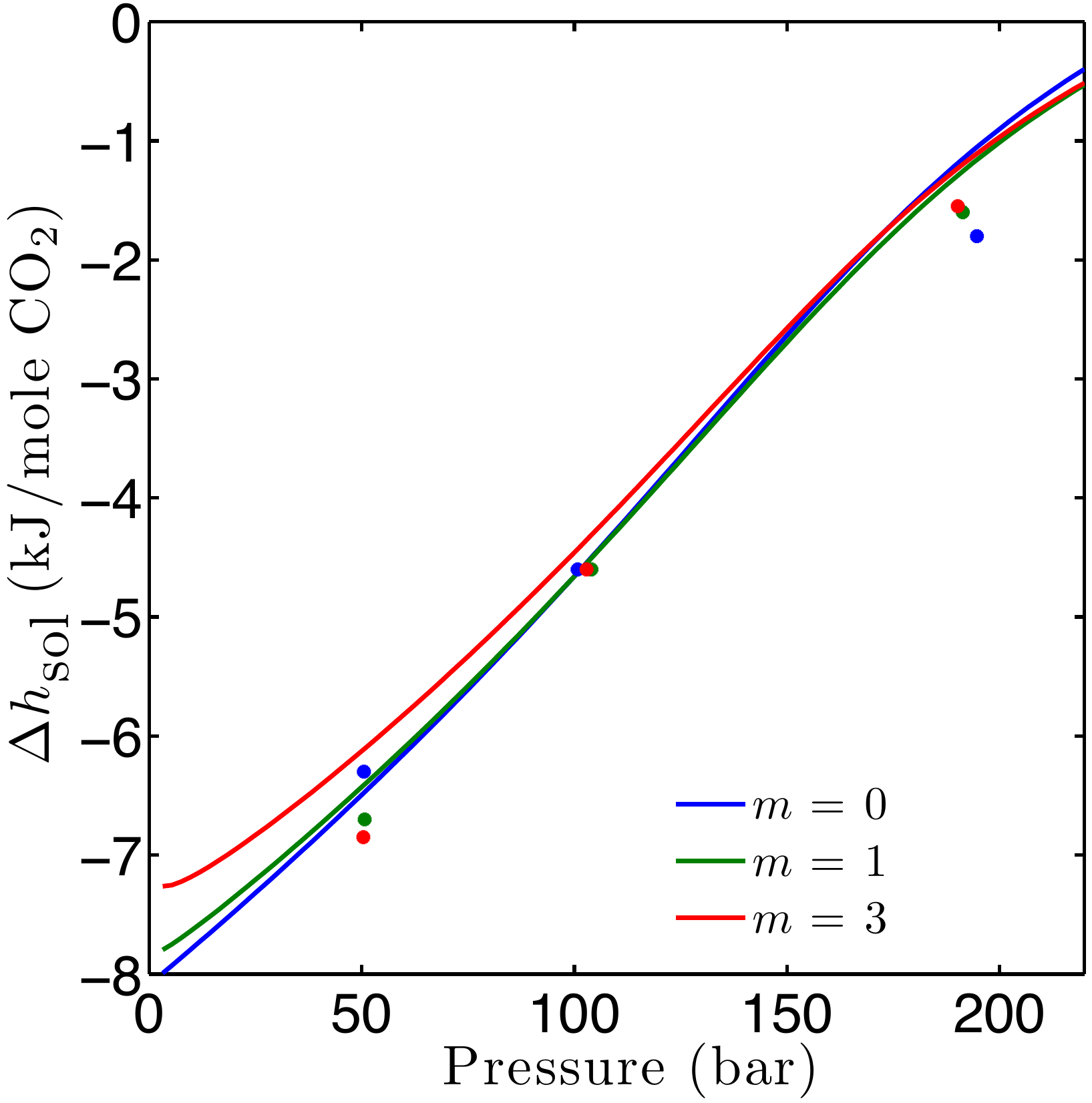} \quad
(d)\includegraphics[width=75mm,height=60mm]{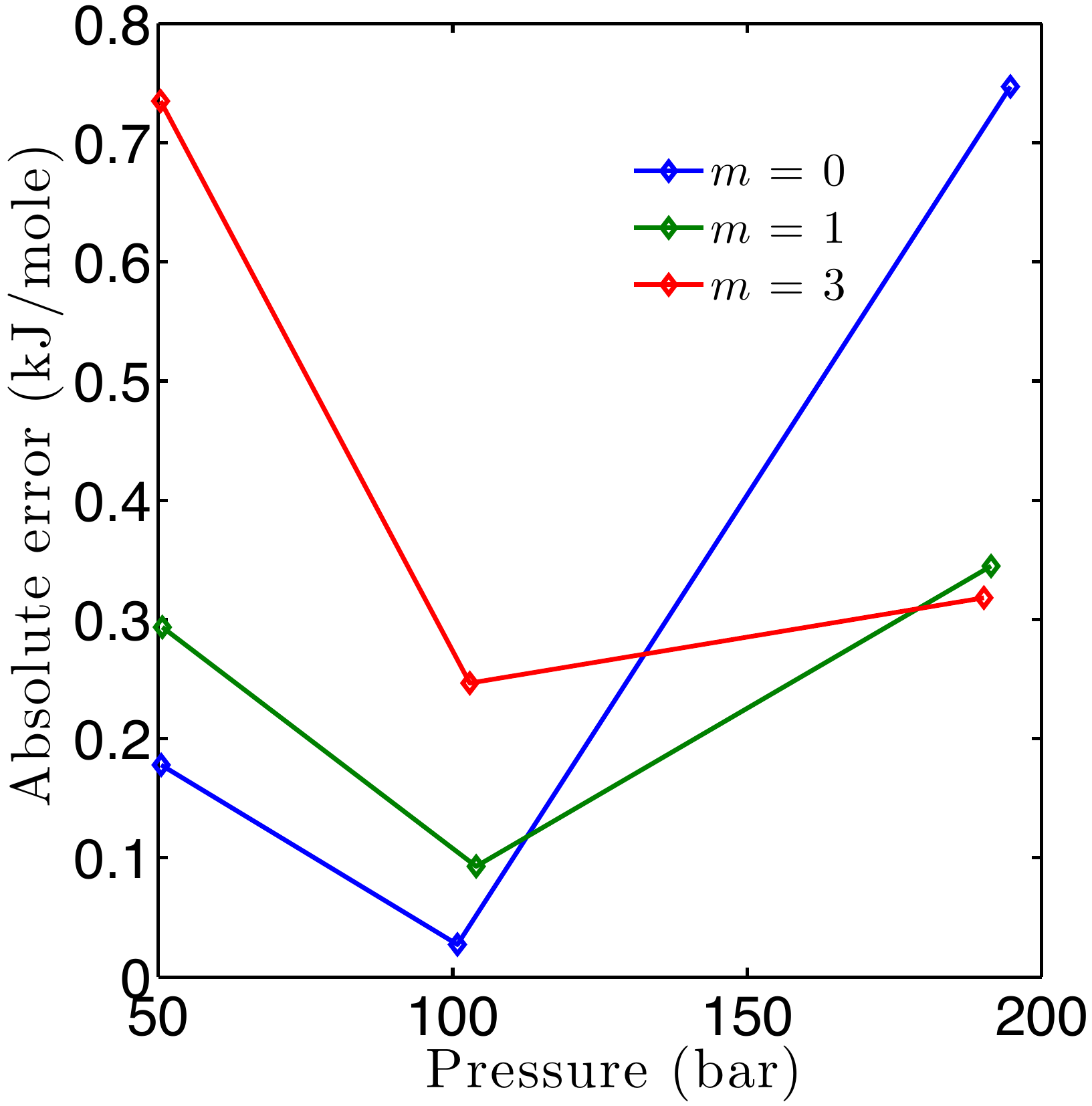} 
\caption{CO$_2$ molar enthalpy of solution derived from Duan and Sun's CO$_2$ activity coefficient model~\cite{Duan2003} vs.\ pressure at 323~K (a,b) and 373~K (c,d) and three different NaCl molalities~$m$. Experimental data are obtained from Koschel~\textit{et al}.~\cite{Koschel2006}.}
\label{fig:deltaHsolDuan}
\end{figure}

Duan and Sun developed their model for the purpose of performing CO$_2$ solubility calculations in pure water and brines. Appendix~\ref{sec:duan} describes how it can also be used to calculate the CO$_2$ partial molar enthalpy $\bar{H}_{\mathrm{CO}_2}(T, P, \mathbf{m})$ and~$\Delta h_\mathrm{sol}$. Here, $\mathbf{m}$ represents the molalities of all the solutes dissolved in the aqueous solvents. Figure~\ref{fig:deltaHsolDuan} compares~$\Delta h_\mathrm{sol}$ predictions from their model with data from Koschel~\textit{et al}. For mixtures in pure water ($m = 0$), there is marked improvement over the CPA EOS at 323~K. Since CO$_2$ is even less soluble in brines than in pure water, we follow the same reasoning in the preceding paragraph to argue that the CO$_2$ contribution to the error in the molar enthalpy $h(T, P, \mathbf{m})$ of an H$_2$O-CO$_2$-NaCl aqueous mixture is relatively small as long as $\bar{H}_{\mathrm{CO}_2}(T, P, \mathbf{m})$ is predicted to within a few~kJ/mole CO$_2$. The accuracy of Duan and Sun's model is well within this limit, at least for the conditions where experimental data are available (temperatures, pressures, and NaCl concentrations up to 373~K, 200~bar, and 3~molal, respectively).

In order to obtain the molar enthalpy $h(T, P, \mathbf{m})$, we must calculate the contributions from water and NaCl. Rather than treating water and NaCl separately, one approach is to treat them together as a single pseudocomponent, which we label as brine. In this approach, the H$_2$O-CO$_2$-NaCl mixture may be thought of as a binary brine-CO$_2$ mixture. A correlation for the specific enthalpy~$h_\mathrm{brine}$ of sodium chloride solutions, in units of~kJ/kg brine, has been presented by Michaelides~\cite{Michaelides1981}. This correlation contains an error in one of the coefficients that was corrected by a later study~\cite{Gudmundsson1989}. For every kilogram of water in the brine-CO$_2$ solution, the total moles are

\begin{equation*}
\frac{1000}{M_{\mathrm{H}_2\mathrm{O}}} + m_{\mathrm{NaCl}} + m_{\mathrm{CO}_2},
\end{equation*}
and the total mass of brine (H$_2$O + NaCl), in units of kg brine/kg H$_2$O, is

\begin{equation*}
1 + m_{\mathrm{NaCl}} \frac{M_{\mathrm{NaCl}}}{1000},
\end{equation*}
where $M_{\mathrm{NaCl}} = 58.44$~g/mole is the molecular weight of the salt. As a result, the enthalpy of the solution per kilogram of water is approximately given by

\begin{equation*}
m_{\mathrm{CO}_2} \bar{H}_{\mathrm{CO}_2}(T, P, \mathbf{m}) + \left( 1 + m_{\mathrm{NaCl}} \frac{M_{\mathrm{NaCl}}}{1000} \right) h_\mathrm{brine}.
\end{equation*}
The molar enthalpy is therefore

\begin{equation}
\label{eq:enthalpyAqueous}
h(T, P, \mathbf{m}) = \left[ \dfrac{m_{\mathrm{CO}_2}}{\dfrac{1000}{M_{\mathrm{H}_2\mathrm{O}}} + m_{\mathrm{NaCl}} + m_{\mathrm{CO}_2}} \right]  \bar{H}_{\mathrm{CO}_2}(T, P, \mathbf{m}) + \left[ \dfrac{\left( 1 + m_{\mathrm{NaCl}} \dfrac{M_{\mathrm{NaCl}}}{1000} \right)}{\dfrac{1000}{M_{\mathrm{H}_2\mathrm{O}}} + m_{\mathrm{NaCl}} + m_{\mathrm{CO}_2}} \right] h_\mathrm{brine}.
\end{equation}
Equation~(\ref{eq:enthalpyAqueous}) is an approximation because $h(T, P, \mathbf{m})$ should involve the specific enthalpy of brine that contains dissolved CO$_2$, and not the specific enthalpy of brine by itself. However, since the mixture is predominantly brine, the two quantities are not expected to be very different, just like how~$\bar{H}_{\mathrm{H}_2\mathrm{O}}$ and~$h_{\mathrm{H}_2\mathrm{O}}$ are similar. Michaelides states~$h_\mathrm{brine}$ is accurate to within 3~\%. This value is comparable to the error of the liquid water molar enthalpy in Figure~\ref{fig:enthalpyH2O}. Since Duan and Sun's model can compute~$\bar{H}_{\mathrm{CO}_2}$ more accurately than the CPA EOS for most of the conditions, it is reasonable to expect~$h(T, P, \mathbf{m})$ predictions to deviate from experimental data by no more than 0.40~kJ/kg (the maximum estimated error for the CPA EOS) over a wide range of conditions.

\section{Excess enthalpy}
\label{sec:excess}

\begin{figure}
\centering
(a)\includegraphics[width=75mm,height=70mm]{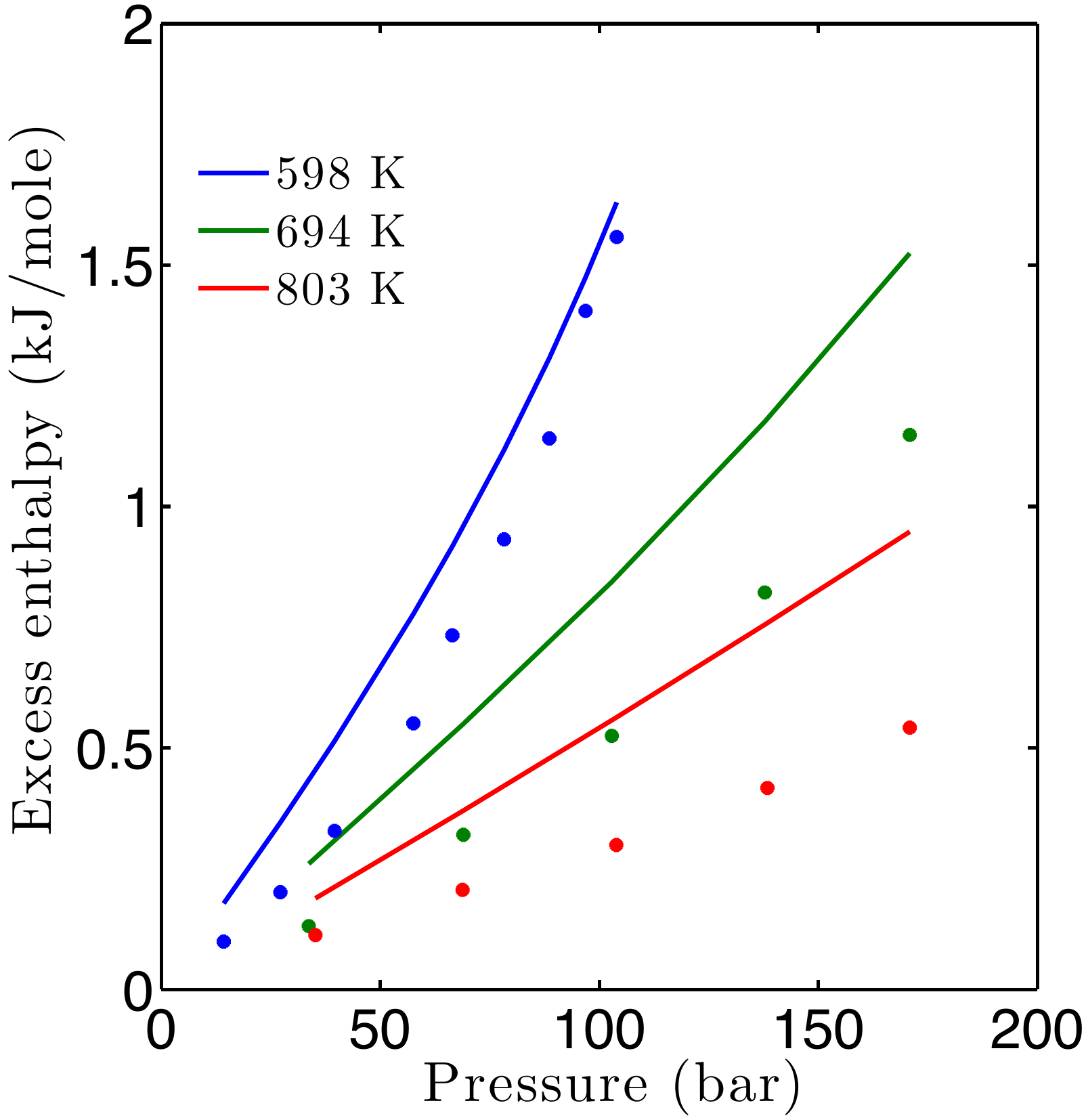} \quad
(b)\includegraphics[width=75mm,height=70mm]{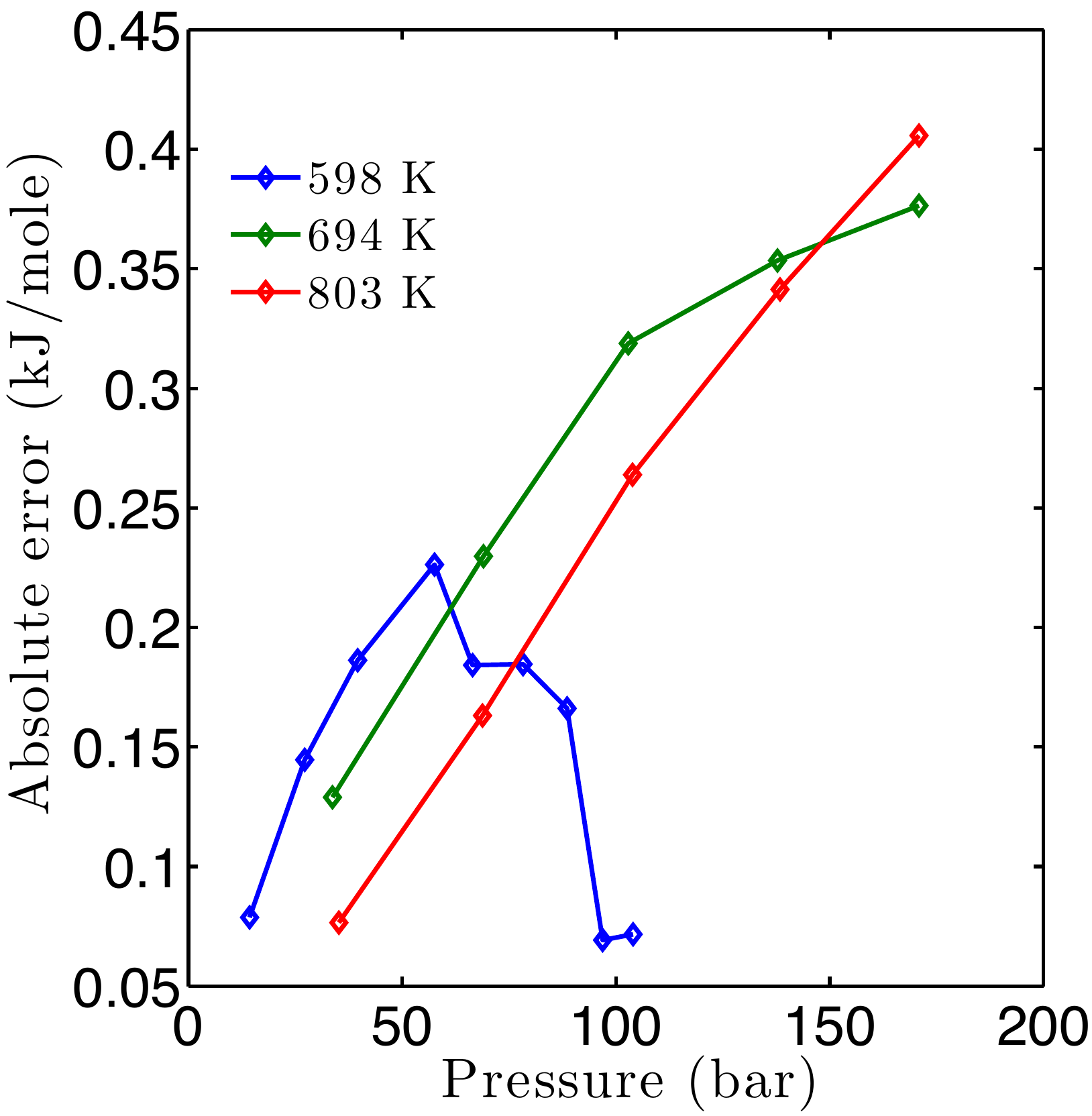} \\
\caption{CPA EOS-predicted molar excess enthalpies of gaseous CO$_2$-H$_2$O mixtures vs.\ pressure at three different~$T$. The experimental data at 598~K are from~\cite{Wormald1986} and represent 1:1 mole ratio mixtures of CO$_2$ to H$_2$O. Data at 694~K and 803~K are from~\cite{Wilson1983}, where the composition ranges between 47-51 mole~\% of CO$_2$.~(b) shows the absolute error of the results in (a).}
\label{fig:enthalpy_mixtureCO2H2O}
\end{figure}

This section explains how relaxing two of the assumptions in Li and Firoozabadi's version of the CPA EOS may allow for better agreement with experimental data for the molar excess enthalpy. This quantity is distinct from the molar enthalpy that we have so far considered. From the general definition in~(\ref{eq:excessProperty}), the molar excess enthalpy is 

\begin{equation*}
h^\textrm{excess}(T, P, \mathbf{z}) = h(T, P, \mathbf{z}) - h^\textrm{im}(T, P, \mathbf{z}),
\end{equation*}
where~$h^\textrm{im}(T, P, \mathbf{z})$ is the molar enthalpy of the corresponding ideal mixture at the conditions $(T, P, \mathbf{z})$. Since the molar enthalpy of an ideal mixture is calculated from the pure component enthalpies (see Section~\ref{sec:enthalpyDensity}),~$\Delta h_\mathrm{sol}$ in~(\ref{eq:deltaHsol}) is equivalent to the partial molar excess enthalpy of CO$_2$ in an infinitely dilute mixture. Therefore, the generalized CPA EOS proposed in this section is expected to also improve the agreement depicted in Figure~\ref{fig:deltaHsolCPA}. Bottini and Saville have measured the molar excess enthalpy of gaseous CO$_2$-H$_2$O mixtures in the temperature range 520 -- 620~K, pressure range 14 -- 45~bar, and CO$_2$ mole fractions between 20 -- 78 \%~\cite{Bottini1985}. For these mixtures, the CPA EOS yields an average absolute error of 1.1~kJ/kg (this corresponds to a 13.3~\% error) and a maximum error of 5.6~kJ/kg. Figure~\ref{fig:enthalpy_mixtureCO2H2O} compares CPA predictions of the molar excess enthalpy of gaseous CO$_2$-H$_2$O mixtures with data from two other studies. The absolute error can be a significant fraction of the excess enthalpy, especially at the higher temperatures. Nevertheless, these results are an improvement over those from Spycher and Pruess, who state that their average absolute error for 1:1 gaseous mixtures is about 0.30~kJ/mole (10~kJ/kg), and their average error for the Bottini and Saville study is 2.4~kJ/kg~\cite{Spycher2011}. We note that an earlier study of H$_2$O-CO$_2$ mixtures has achieved good agreement with excess enthalpy data up to 598~K, although their model also exhibits significant errors at higher temperatures~\cite{Gallagher1993}. This study uses optimization techniques to fit parameters to many different sets of data. The generalized corresponding states principle is employed to make the parameter space more manageable. Their model is targeted towards only the aqueous phase, however. It is not applicable to pure CO$_2$ or to CO$_2$-rich mixtures with water vapor, like the 90~mole~\% CO$_2$ mixtures in Figure~\ref{fig:enthalpyCO2rich}(a).

We may explain the behavior in Figure~\ref{fig:enthalpy_mixtureCO2H2O} by considering the nature of intermolecular forces, particularly hydrogen bonding, that exist in mixtures of CO$_2$ and H$_2$O (see Sections~\ref{sec:if_departure} and~\ref{sec:Helmholtz}). Breaking hydrogen bonds requires an energy input on the order of 10 kJ/mole~\cite{Burrows2013}, which is at least an order of magnitude larger than the excess enthalpy values. The large disparity suggests that the excess enthalpies are rather small in magnitude, so that even small discrepancies between theory and experiment appear as large relative errors in Figure~\ref{fig:enthalpy_mixtureCO2H2O}(b). The discrepancy may be largely due to the two assumptions listed in Section~\ref{sec:Li}. The assumptions state that the $\alpha$ (hydrogen-bond donor) and $\beta$ (hydrogen-bond acceptor) sites are equally numerous on a molecule of component $i$ and associate symmetrically so that their site fractions $\chi_{i\alpha}$ and $\chi_{i\beta}$ can be described by a single function $\chi_i$. In order to see the consequences of these assumptions, let us examine, as a representative example, a 1:1 CO$_2$-H$_2$O gaseous mixture at 598 K and 66.5 bar. The relevant values are:

\begin{itemize}
\item Experimental molar excess enthalpy = 0.73 kJ/mole.
\item CPA-predicted molar excess enthalpy = $h(T, P, \mathbf{z}) - h^\textrm{im}(T, P, \mathbf{z})$ = 0.92~kJ/mole.
\item CPA-predicted molar enthalpy departure function = $h(T, P, \mathbf{z}) - h^\textrm{ig}(T, P, \mathbf{z}) = -1.46$~kJ/mole.
\end{itemize}
Qualitatively, the signs and relative magnitudes of the excess enthalpy and departure function are as expected. The departure function should be negative because the intermolecular interactions present in the real (non-ideal) mixture reduce the energy of the mixture compared to the non-interacting ideal gas mixture. The reason why the excess enthalpy is positive can be understood by considering the process of mixing CO$_2$ with water. Mixing involves three processes:

\begin{enumerate}
\item Breaking CO$_2$-CO$_2$ intermolecular forces, primarily London dispersion forces.
\item Breaking H$_2$O-H$_2$O intermolecular forces, the most prominent being hydrogen bonds.
\item Forming CO$_2$-H$_2$O intermolecular forces, the most prominent being hydrogen bonds.
\end{enumerate}
The first two require energy input while the last one releases energy. The CO$_2$-H$_2$O hydrogen bonds are not taken into account in the ideal mixture, since the ideal mixture enthalpy is calculated from the pure component enthalpies. These hydrogen bonds are weaker than H$_2$O-H$_2$O hydrogen bonds because C is more electronegative than H, so the lone pairs of electrons on the oxygens in CO$_2$ are less effective hydrogen-bond acceptors than the lone pairs on the oxygen in H$_2$O. As a result, the real mixture should be higher in energy than the ideal mixture, meaning a positive excess enthalpy.  

A molecule of CO$_2$ has four $\beta$ sites, representing the 4 lone pairs on the two oxygens, and no $\alpha$ sites (hydrogens). Treating the $\alpha$ and $\beta$ sites as symmetric and equally numerous per molecule will overpredict the number of CO$_2$-H$_2$O hydrogen bonds formed since CO$_2$ does not have any $\alpha$~sites. Consequently, overprediction of the number of CO$_2$-H$_2$O hydrogen bonds will lead to an underprediction of the number of H$_2$O-H$_2$O hydrogen bonds in the mixture. This is because if a H$_2$O molecule forms a hydrogen bond with CO$_2$, it has fewer sites available to form a hydrogen bond with another H$_2$O. As a result, the theoretically-predicted excess enthalpy will be more positive than the experimental value, as we have observed. The small magnitude of the excess enthalpy is also consistent with our discussion above. We have reasoned that $h^\textrm{excess}$ is positive because CO$_2$-H$_2$O hydrogen bonds are weaker than H$_2$O-H$_2$O hydrogen bonds. This is due to the fact that carbon is more electronegative than hydrogen. However, since carbon is only slightly more electronegative than hydrogen, we would expect the excess enthalpy to be quite small, certainly much less than the energy required to break a hydrogen bond, which is $\mathcal{O}$(10~kJ/mole).

In order to achieve better agreement with excess enthalpy measurements, Li and Firoozabadi's CPA EOS must be modified so that it distinguishes between $\alpha$ and $\beta$ sites, treating them asymmetrically. We suggest relaxing the two assumptions in Section~\ref{sec:Li} so that equations in Sections~\ref{sec:fugacityCPA} and~\ref{sec:pressureCPA} for the fugacity coefficients~$\phi_i$ (on which all EOS predictions of the density, solubility, and enthalpy are based), pressure, and compressibility factor~$Z$ are solved instead. These equations are derived from the departure function in~(\ref{eq:FdepartCPAZ}) or equivalently, the one in~(\ref{eq:FdepartCPA}). From a practical perspective, the asymmetric treatment will introduce additional cross-association parameters (see Section~\ref{sec:FdepartAssoc}) that can be fit to excess enthalpy data. The pure component parameters that exist in the current model, such as~$\epsilon$ and~$\kappa$ described in Section~\ref{sec:FdepartAssoc}, do not need to be modified.

\section{Conclusions and future work}
\label{sec:future}

We have shown that the CPA EOS accurately calculates the molar enthalpy of pure CO$_2$, pure water, CO$_2$-rich mixtures, and aqueous mixtures for the conditions relevant to the applications discussed in Section~\ref{sec:intro}. Since the CPA EOS has previously been demonstrated to provide good agreement with density and solubility data~\cite{Li2009}, it can therefore serve as a single, unified model for thermodynamic property predictions of all of these fluids. Compared to the Spycher and Pruess model~\cite{Spycher2011}, the CPA EOS more accurately calculates the vapor-phase enthalpies of pure water at temperatures above 550~K along the vapor-liquid equilibrium curve (Figure~\ref{fig:HvapH2O}). For CO$_2$-H$_2$O mixtures, the CPA is more accurate at lower temperatures, higher pressures, and higher water concentrations~(although both models predict the molar enthalpy data in Figure \ref{fig:enthalpyCO2rich} well). For practical purposes, the most important difference between the two models is that the CPA EOS is not limited only to temperatures above 373~K for pure water or water-containing mixtures.

The CPA EOS is not applicable to aqueous H$_2$O-CO$_2$-NaCl mixtures, however. For these brine mixtures, we suggest calculating the molar enthalpy with~(\ref{eq:enthalpyAqueous}). We have argued that~(\ref{eq:enthalpyAqueous}) is expected to be accurate to within 0.40~kJ/mole, at least for the conditions where enthalpy data are available~\cite{Koschel2006} (temperatures, pressures, and NaCl concentrations up to 373~K, 200~bar, and 3~molal, respectively). Equation~(\ref{eq:enthalpyAqueous}) involves the partial molar enthalpy of CO$_2$, which we obtain by evaluating temperature derivatives of Duan and Sun's CO$_2$ activity coefficient model~\cite{Duan2003}. In a similar fashion, one can obtain the partial molar volume of CO$_2$ by evaluating pressure derivatives. The density of the brine can then be calculated from this partial molar volume. However, we find that the CO$_2$ partial molar volume derived in this way does not closely agree with experimental data~\cite{Yan2011}. Density calculations require finer precision than enthalpy calculations because the density increase from CO$_2$ dissolution into water or brine is small. It is less than 2~\% even at pressures where the CO$_2$ solubility is high (3 mole~\%). As a result, we recommend the density of H$_2$O-CO$_2$-NaCl mixtures be calculated using the models presented in more recent studies~\cite{Duan2008, Nomeli2014}. 

The CPA EOS may be combined with Duan and Sun's model to form a $\gamma$-$\phi$ model for H$_2$O-CO$_2$-NaCl mixtures. In this approach, the CPA EOS is used to compute the density, enthalpy, and composition of the CO$_2$-rich phase, which we treat as being composed only of CO$_2$ and water. For the aqueous phase, Duan and Sun's model is used to calculate the CO$_2$ solubility and the CO$_2$ partial molar enthalpy. The partial molar enthalpy may be combined with the specific enthalpy of brine (computed from the studies~\cite{Michaelides1981, Gudmundsson1989} described in Section~\ref{sec:aqueous}) to obtain the molar enthalpy~(\ref{eq:enthalpyAqueous}) of the mixture as whole. The density of the aqueous phase may be calculated from one of the aforementioned models~\cite{Duan2008, Nomeli2014}. A more theoretically robust alternative to a $\gamma$-$\phi$ model is a $\phi$-$\phi$ model, in which the same EOS is used for both phases. In our future work, we plan to develop a $\phi$-$\phi$ model by modifying the CPA EOS so that it can accurately predict the thermodynamic properties of both phases. The first step is to generalize the CPA EOS as described in Section~\ref{sec:excess} to further improve the agreement with CO$_2$-H$_2$O mixture enthalpy data. After this step, we may extend the applicability of the CPA EOS to H$_2$O-CO$_2$-NaCl mixtures by adding more terms to the departure function~(\ref{eq:Fdepart}) to account for intermolecular forces such as ion-ion and ion-dipole interactions. This work will involve many challenges, some of which arise due to the lack of experimental molar enthalpy data for aqueous mixtures. Nonetheless, a systematic and rigorous approach that builds on previous studies of electrolytic solutions~\cite{Harvey1989, Myers2002, Lin2007} may prove to be successful. 

\section*{Acknowledgments}
This work was performed under the auspices of the U.S. Department of Energy by Lawrence Livermore National Laboratory under Contract DE-AC52-07NA27344. We gratefully acknowledge financial support from Lawrence Livermore National Laboratory and the member institutions of the Reservoir Engineering Research Institute. 

\appendix

\section{Basic concepts in the thermodynamics of fluid mixtures}
\label{sec:thermo}

\subsection{Chemical potential, fugacity, fugacity coefficient}
\label{sec:fugacity}
Many fundamental thermodynamic relations for real fluid mixtures are inspired by the corresponding relations for ideal gas mixtures. One such example is the relation between the chemical potential $\mu_i^\textrm{ig} (T, P, \mathbf{n})$ of component~$i$ in an ideal gas mixture and the pure component ideal gas chemical potential $\mu_i^{\textrm{ig}} (T, P)$. Here,~$T$ is the temperature,~$P$ is the pressure,~$\mathbf{n}$ represents the mole numbers of the mixture's components. The difference between $\mu_i^\textrm{ig} (T, P, \mathbf{n})$ and $\mu_i^{\textrm{ig}} (T, P)$ is

\begin{equation}
\label{eq:IGmixing}
\mu_i^\textrm{ig} (T, P, \mathbf{n}) - \mu_i^{\textrm{ig}} (T, P) = RT \ln z_i,
\end{equation}
where~$R$ is the gas constant, and~$z_i = n_i / N$ is the mole fraction of component~$i$. Equation~(\ref{eq:IGmixing}) can be derived formally using statistical thermodynamics, but it can also be derived in a more informal manner as follows~\cite{Sandler2006}. Suppose a pure component exists as an ideal gas at pressure $P$. If this pure component is mixed isobarically with other ideal gases to form a mixture whose pressure is also $P$, the pressure of the component in the mixture becomes the partial pressure $z_i P$ so that

\begin{equation*}
\mu_i^\textrm{ig} (T, P, \mathbf{n}) - \mu_i^\textrm{ig} (T, P) = \int_{P}^{z_i P} \left. \left( \frac{\partial \mu_i^\textrm{ig}}{\partial P'} \right) \right|_{T,\mathbf{n}} \mathrm{d} P'.
\end{equation*}
In our study, we consider only systems where external fields like gravity are negligible and there is only pressure-volume work. For such systems,

\begin{equation}
\label{eq:dG}
\mathrm{d} G = - S \mathrm{d} T + V \mathrm{d} P + \sum_{i = 1}^c \mu_i \mathrm{d} n_i,
\end{equation}
where $G$ is the Gibbs energy, $S$ is the entropy, $V$ is the volume, and $c$ is the number of components. Applying the equality of mixed partial derivatives to~(\ref{eq:dG}), we have

\begin{equation}
\label{eq:MaxwellVmu}
\left. \left( \frac{\partial V}{\partial n_i} \right) \right|_{T, P,\mathbf{n}_i} = \left. \left( \frac{\partial \mu_i}{\partial P} \right) \right|_{T,\mathbf{n}},
\end{equation}
where the subscript $\mathbf{n}_i$ in the derivative of $V$ means that all mole numbers besides $n_i$ are held fixed. Using~(\ref{eq:MaxwellVmu}) and the volume $V^\textrm{ig} = NRT/P$ of an ideal gas, we obtain

\begin{equation*}
\int_{P}^{z_i P} \left. \left( \frac{\partial \mu_i^\textrm{ig}}{\partial P'} \right) \right|_{T,\mathbf{n}} \mathrm{d} P'= \int_{P}^{z_i P} \left. \left( \frac{\partial V^\textrm{ig}}{\partial n_i} \right) \right|_{T, P',\mathbf{n}_i} \mathrm{d} P'= RT \int_{P}^{z_i P} \frac{\mathrm{d} P'}{P'} = RT \ln z_i, 
\end{equation*}
which leads to~(\ref{eq:IGmixing}). Similarly, the chemical potential $\mu_i^\textrm{ig} (T, P, \mathbf{n})$ is related to $\mu_i^\textrm{ig} (T, P', \mathbf{n})$ at a different pressure $P'$ (but same temperature and composition) by

\begin{equation}
\label{eq:IGPressures}
\mu_i^\textrm{ig} (T, P, \mathbf{n}) - \mu_i^\textrm{ig} (T, P', \mathbf{n}) =  \int_{P'}^{P} \left. \left( \frac{\partial \mu_i^\textrm{ig}}{\partial P''} \right) \right|_{T,\mathbf{n}} \mathrm{d} P''= \int_{P'}^{P} \left. \left( \frac{\partial V^\textrm{ig}}{\partial n_i} \right) \right|_{T, P'',\mathbf{n}_i} \mathrm{d} P'' = RT \ln \frac{P}{P'}.
\end{equation}
Combining~(\ref{eq:IGmixing}) and~(\ref{eq:IGPressures}), the difference between $\mu_i^\textrm{ig} (T, P, \mathbf{n})$ and $\mu_i^\textrm{ig} (T, P', \mathbf{n}')$, which is at a different pressure and composition but same temperature, is

\begin{equation}
\label{eq:mu1mu2IG}
\mu_i^\textrm{ig} (T, P, \mathbf{n}) - \mu_i^\textrm{ig} (T, P', \mathbf{n}') = \left[ \mu_i^\textrm{ig} (T, P) - \mu_i^\textrm{ig} (T, P') \right] + RT \ln \frac{z_i}{z_i'} = RT \ln \frac{z_i P}{z_i' P'}.
\end{equation}

In analogy to~(\ref{eq:mu1mu2IG}), for any fluid (not just ideal gases) each component $i$ is assigned a fugacity $f_i = f_i(T, P, \mathbf{n})$ so that the difference between $\mu_i (T, P, \mathbf{n})$ and $\mu_i (T, P', \mathbf{n}')$ is 

\begin{equation}
\label{eq:mu1mu2}
\mu_i (T, P, \mathbf{n}) - \mu_i (T, P', \mathbf{n}') = RT \ln \frac{f_i (T, P, \mathbf{n})}{f_i(T, P', \mathbf{n}')}.
\end{equation}
In order for~(\ref{eq:mu1mu2}) to be consistent with~(\ref{eq:mu1mu2IG}), the fugacity of component $i$ in an ideal gas mixture must be equal to its partial pressure

\begin{equation}
\label{eq:fugacityIG}
f_i^\textrm{ig}(T, P, \mathbf{n}) = z_i P.
\end{equation}
The difference in the chemical potential of component $i$ in a real fluid and the chemical potential of that same component in an ideal gas mixture at the same $(T, P, \mathbf{n})$ is

\begin{equation}
\label{eq:mu1Realmu2IG}
\mu_i (T, P, \mathbf{n}) - \mu_i^{\textrm{ig}} (T, P, \mathbf{n}) = RT \ln \left(\frac{f_i}{z_i P} \right) = RT \ln \phi_i,
\end{equation}
where 

\begin{equation}
\label{eq:phiDefinition}
\phi_i(T, P, \mathbf{n}) = \frac{f_i(T, P, \mathbf{n})}{z_i P},
\end{equation}
is the fugacity coefficient of component $i$. Since any fluid mixture behaves like its corresponding ideal gas mixture as the pressure $P$ decreases to zero, $f_i$ becomes $f_i^\textrm{ig}(T, P, \mathbf{n}) = z_i P$ and $\phi_i$ becomes unity in this limit:

\begin{equation*}
\lim_{P \rightarrow 0} f_i(T, P, \mathbf{n}) = z_i P,
\end{equation*}
\begin{equation*}
\lim_{P \rightarrow 0} \phi_i(T, P, \mathbf{n}) = 1. 
\end{equation*}
If the fluid in question is a pure component fluid,

\begin{equation*}
\lim_{P \rightarrow 0} f(T, P) = P,
\end{equation*}
\begin{equation*}
\lim_{P \rightarrow 0} \phi(T,P) = 1. 
\end{equation*}
For a multiphase system of $p$ phases to be in equilibrium, a fundamental thermodynamic requirement~\cite{Prausnitz1999} is the equality of chemical potentials

\begin{equation}
\label{eq:phaseEquilibrium}
\mu_i^1 = \mu_i^2 = \dots = \mu_i^p \quad \textrm{for all $i$}.
\end{equation}
Using~(\ref{eq:mu1mu2}), the condition for phase equilibrium in~(\ref{eq:phaseEquilibrium}) can be equivalently stated as the equality of fugacities

\begin{equation}
\label{eq:phaseEquilibriumFugacity}
f_i^1 = f_i^2 = \dots = f_i^p \quad \textrm{for all $i$}.
\end{equation}
The fugacity~$f_i$ can be easily calculated from the fugacity coefficient~$\phi_i$ and vice-versa using~(\ref{eq:phiDefinition}). If we know how the fugacity of each component varies as a function of temperature, pressure, and composition,~(\ref{eq:phaseEquilibriumFugacity}) can be solved to determine the equilibrium composition (i.e., the solubility of each component) in each phase. Furthermore, we will see in Section~(\ref{sec:enthalpyDensity}) that other thermodynamic functions, like the density and the enthalpy, can in principle be derived from the fugacities or the fugacity coefficients. In essence, the purpose of an equation of state is to provide a relation from which we can derive the functional form of the fugacities for all components. Consequently, the equation of state can be used to derive other thermodynamic functions of the mixture as well. In this sense, the equation of state provides a complete thermodynamic specification of the fluids to which it is applicable. 

\subsection{Activity coefficient, activity, excess property, partial molar property}
\label{sec:activity}

In analogy to~(\ref{eq:IGmixing}), an ideal mixture (also called an ideal solution) is defined to be one in which
\begin{equation}
\label{eq:IMmixing}
\mu_i^\textrm{im} (T, P, \mathbf{n}) - \mu_i(T, P) = RT \ln z_i.
\end{equation}
Here,~$\mu_i(T, P)$ is the chemical potential of pure component~$i$ in the real fluid state, as opposed to the hypothetical ideal gas state in~(\ref{eq:IGmixing}). From~(\ref{eq:mu1mu2}) and~(\ref{eq:IMmixing}), the fugacity $f_i^\textrm{im}(T, P, \mathbf{n})$ in an ideal mixture and the fugacity $f_i(T, P)$ of the pure component are related by

\begin{equation}
\label{eq:fugacityIM}
f_i^\textrm{im}(T, P, \mathbf{n}) = z_i f_i(T, P).
\end{equation}
An ideal gas mixture can be thought of as an ideal mixture where the pure component fugacity $f_i(T, P) = P$ for all conditions such that the fugacity $f_i^\textrm{im}(T, P, \mathbf{n})$ in the mixture equals the partial pressure $z_i P$. In other words, an ideal gas mixture is an ideal mixture where~(\ref{eq:fugacityIG}) is satisfied. Equation~(\ref{eq:fugacityIM}) is a simple relation between a mixture property $f_i^\textrm{im}(T, P, \mathbf{n})$ and a pure component property $f_i(T, P)$; it is a key feature of ideal mixtures. A similar relation can be formed for a real (non-ideal) fluid mixture by introducing an activity coefficient $\gamma_i = \gamma_i(T, P, \mathbf{n})$ defined such that

\begin{equation}
\label{eq:fugacityActivity}
f_i(T, P, \mathbf{n}) = z_i \gamma_i(T, P, \mathbf{n}) f_i(T, P),
\end{equation}
where 

\begin{equation}
\label{eq:activityLimit}
\lim_{z_i \rightarrow 1} \gamma_i = 1.
\end{equation}
The activity~$\mathsf{a}_i$ of component~$i$ is defined as

\begin{equation}
\label{eq:activity}
\mathsf{a}_i = \gamma_i z_i = \frac{f_i(T, P, \mathbf{n})}{f_i(T, P)},
\end{equation}
so that for any fluid,

\begin{equation}
\label{eq:muActivity}
\mu_i (T, P, \mathbf{n}) - \mu_i(T, P) = RT \ln \mathsf{a}_i.
\end{equation}
With these definitions, one can view an ideal mixture as a real fluid in which $\gamma_i = 1$ (or equivalently,~$\mathsf{a}_i = z_i$) for all components over all conditions. An excess property is defined as 

\begin{equation}
\label{eq:excessProperty}
E^\textrm{excess}(T, P, \mathbf{n}) = E(T, P, \mathbf{n}) - E^\textrm{im}(T, P, \mathbf{n}),
\end{equation}
where $E$ is an extensive property (e.g., Gibbs energy) of the mixture. 

A related concept to the excess property is the partial molar property. The partial molar property of $E$ with respect to component $i$ is denoted as $\bar{E}_i$, and is defined as
\begin{equation}
\label{eq:Ebar_i}
\bar{E}_i = \left. \left( \frac{\partial E}{\partial n_i} \right) \right|_{T, P,\mathbf{n}_i}.
\end{equation}
One example that we have seen previously in~(\ref{eq:MaxwellVmu}) is the partial molar volume

\begin{equation*}
\bar{V}_i = \left. \left( \frac{\partial V}{\partial n_i} \right) \right|_{T, P,\mathbf{n}_i}.
\end{equation*}
Another example is the chemical potential in~(\ref{eq:dG}), which is the partial molar Gibbs energy

\begin{equation*}
\mu_i = \bar{G}_i = \left. \left( \frac{\partial G}{\partial n_i} \right) \right|_{T, P,\mathbf{n}_i}.
\end{equation*}
There exists a very important relation between $E$ and its partial molar properties $\bar{E}_i$. It can be shown~\cite{Sandler2006, Michelsen2007, Firoozabadi2015} that

\begin{equation*}
E(T, P, \mathbf{n}) = \sum_{i=1}^c n_i \bar{E}_i,
\end{equation*}
or on a molar basis, if $e(T, P, \mathbf{z}) = E(T, P, \mathbf{n})/N$, 

\begin{equation}
\label{eq:molarEbar}
e(T, P, \mathbf{z}) = \sum_{i=1}^c z_i \bar{E}_i.
\end{equation}
We denote the set of all mole fractions as $\mathbf{z} = \mathbf{n}/N$. The molar Gibbs energy is

\begin{equation}
\label{eq:molarG}
g(T, P, \mathbf{z}) = \sum_{i=1}^c z_i \bar{G}_i = \sum_{i=1}^c z_i \mu_i.
\end{equation}
Due to complexities arising from intermolecular interactions, the partial molar property of a mixture is in general not equal to the molar property $e_i(T,P) = E_i(T,P)/n_i$ of pure component~$i$. That is, $\bar{E}_i \neq e_i(T,P)$ except in special cases. One such special case is the partial molar enthalpy of an ideal mixture. Section~\ref{sec:enthalpyDensity} shows that for an ideal mixture, $\bar{H}_i^\textrm{im}(T,P,\mathbf{n}) = h_i(T,P)$. We will later use this relation to calculate the enthalpy of real fluid mixtures. It is important to note that $(T, P,\mathbf{n}_i)$ must be held fixed in the derivative for it to be considered a partial molar property. For example, we will see in Section~\ref{sec:fugacityCPA} that $\mu_i = \left. \left( \partial F/\partial n_i \right) \right|_{T, V,\mathbf{n}_i}$, where~$F$ is the Helmholtz energy of a fluid. Since it is the volume, and not the pressure, that is held fixed we conclude that $\mu_i \neq \bar{F}_i$. Instead, the partial molar Helmholtz energy is defined as

\begin{equation*}
\bar{F}_i = \left. \left( \frac{\partial F}{\partial n_i} \right) \right|_{T, P,\mathbf{n}_i}.
\end{equation*}

We can also define partial molar properties of excess properties. For example,

\begin{equation*}
\bar{G}_i^\textrm{excess} = \left. \left( \frac{\partial G^\textrm{excess} }{\partial n_i} \right) \right|_{T, P,\mathbf{n}_i} = \mu_i(T, P, \mathbf{n}) - \mu_i^\textrm{im} (T, P, \mathbf{n}).
\end{equation*}
Using this relation along with~(\ref{eq:mu1mu2}) and~(\ref{eq:fugacityIM})--(\ref{eq:fugacityActivity}), we have

\begin{equation*}
\bar{G}_i^\textrm{excess} = \mu_i(T, P, \mathbf{n}) - \mu_i^\textrm{im} (T, P, \mathbf{n}) = RT \ln \frac{f_i(T, P, \mathbf{n})}{ f_i^\textrm{im}(T, P, \mathbf{n})} = RT \ln \frac{z_i \gamma_i f_i(T, P)}{ z_i f_i(T, P)} = RT \ln \gamma_i.
\end{equation*}
This result shows that if we have a smooth function (such as a polynomial) that describes the variation of the excess Gibbs energy with the composition, we can take composition derivatives (i.e., evaluate the partial molar excess Gibbs energy $\bar{G}_i^\textrm{excess}$) to obtain the activity coefficient $\gamma_i$. In turn, we can use $\gamma_i$ to compute the real fluid mixture fugacity $f_i(T,P, \mathbf{n})$ from the pure component fugacity~$f_i(T,P)$ using~(\ref{eq:fugacityActivity}). The smooth function is called the activity coefficient model. It is essentially a correlation which provides a continuous fit to a discrete set of excess Gibbs energy data, which are obtained through experimental measurements at different compositions. See~\cite{Prausnitz1999, Elliott1999} for extensive discussions on activity coefficient models for various types of mixtures. Duan and Sun have presented a model for the CO$_2$ activity coefficient in mixtures where CO$_2$ is dissolved in aqueous sodium chloride solutions~\cite{Duan2003}. This model is partly based on the activity coefficient model of Pitzer, which is widely used to compute the excess Gibbs energy of electrolytic solutions~\cite{Pitzer1973, Prausnitz1999}. The Duan and Sun model is the focus of Section~\ref{sec:aqueous} and Appendix~\ref{sec:duan}.

\subsection{Enthalpy and density}
\label{sec:enthalpyDensity}

We stated in Section~\ref{sec:fugacity} that the enthalpy of a fluid mixture can be derived from the fugacity coefficients~$\phi_i$. To show this, we use the relation $G = H - TS$ between the Gibbs energy~$G$ and the enthalpy~$H$ so that the partial molar properties are related by

\begin{equation*}
\bar{G}_i = \mu_i = \bar{H}_i - T\bar{S}_i.
\end{equation*}
From~(\ref{eq:dG}), we have

\begin{equation*}
\bar{S}_i = \left. \left( \frac{\partial S}{\partial n_i} \right) \right|_{T, P,\mathbf{n}_i} = -\left. \left( \frac{\partial \mu_i}{\partial T} \right) \right|_{P,\mathbf{n}},
\end{equation*}
which leads to

\begin{equation*}
\mu_i = \bar{H}_i + T \left. \left( \frac{\partial \mu_i}{\partial T} \right) \right|_{P,\mathbf{n}}.
\end{equation*}
This result can be rearranged to

\begin{equation}
\label{eq:muHbar}
\frac{\partial }{\partial T} \left. \left( \frac{\mu_i}{T} \right) \right|_{P,\mathbf{n}} = -\frac{\bar{H}_i}{T^2}.
\end{equation}
Since an ideal mixture is defined by~(\ref{eq:IMmixing}),

\begin{equation}
\label{eq:HbarIdeal}
-\frac{\bar{H}_i^\textrm{im}}{T^2} = \frac{\partial }{\partial T} \left. \left( \frac{\mu_i^\textrm{im}}{T} \right) \right|_{P,\mathbf{n}} = \frac{\partial }{\partial T} \left. \left( \frac{\mu_i(T,P) + RT \ln z_i}{T} \right) \right|_{P,\mathbf{n}} = \frac{\partial }{\partial T} \left. \left( \frac{\mu_i(T,P)}{T} \right) \right|_{P,\mathbf{n}} =  -\frac{h_i(T,P)}{T^2}.
\end{equation}
This is the result we alluded to in Section~\ref{sec:activity}; the partial molar enthalpy $\bar{H}_i^\textrm{im}$ is equal to the molar enthalpy $h_i(T,P)$ of the pure component. The enthalpy of ideal mixtures, which includes ideal gas mixtures, can thus be computed from the pure component enthalpies. In other words, ideal mixtures are characterized by zero enthalpies of mixing:

\begin{equation*}
\Delta h^\textrm{im}_\textrm{mix} = h^\textrm{im}(T, P, \mathbf{z}) - \sum_{i = 1}^c z_i h_i (T, P)  =  \sum_{i = 1}^c z_i \left(\bar{H}_i^\textrm{im} - h_i \right) = 0.
\end{equation*}
Applying~(\ref{eq:muHbar}) and~(\ref{eq:HbarIdeal}) to~(\ref{eq:mu1Realmu2IG}), we have

\begin{equation*}
\frac{\partial }{\partial T} \left. \left( \frac{\mu_i - \mu_i^\textrm{ig}}{T} \right) \right|_{P,\mathbf{n}} = R \left. \left( \frac{\partial \ln \phi_i(T, P, \mathbf{n})}{\partial T} \right) \right|_{P,\mathbf{n}} = -\frac{\bar{H}_i - \bar{H}_i^\textrm{ig}}{T^2} = -\frac{\bar{H}_i - h_i^\textrm{ig}}{T^2}.
\end{equation*}
Combining this result with~(\ref{eq:molarEbar}), the molar enthalpy of any fluid mixture is

\begin{equation}
\label{eq:molarEnthalpy}
h(T, P, \mathbf{z}) = \sum_{i=1}^c z_i \bar{H}_i = -RT^2 \sum_{i=1}^c z_i \left. \left( \frac{\partial \ln \phi_i(T, P, \mathbf{n})}{\partial T} \right) \right|_{P,\mathbf{n}} + \sum_{i=1}^c z_i h_i^\textrm{ig}.
\end{equation}
This is the relation we seek between the fugacity coefficients and the enthalpy of the mixture. We can calculate the enthalpies~$h_i^\textrm{ig}$ of the pure components in their ideal gas states using correlations found in the references specified in Section~\ref{sec:pure}.

The density can be derived from the fugacity coefficients as well. Using~(\ref{eq:MaxwellVmu}) along with~(\ref{eq:mu1Realmu2IG}) and~(\ref{eq:molarEbar}), the molar volume $v = V/N$ is computed from the partial molar volumes $\bar{V}_i$ according to

\begin{equation*}
v = \sum_{i = 1}^c z_i \bar{V}_i = \sum_{i = 1}^c z_i \left. \left( \frac{\partial \mu_i}{\partial P} \right) \right|_{T,\mathbf{n}} = \sum_{i = 1}^c z_i \left. \left( \frac{\partial \mu_i^\textrm{ig}}{\partial P} \right) \right|_{T,\mathbf{n}} + RT \sum_{i = 1}^c z_i \left. \left( \frac{\partial \ln \phi_i}{\partial P} \right) \right|_{T,\mathbf{n}}
\end{equation*}
For an ideal gas where $V^\textrm{ig} = NRT/P$

\begin{equation*}
\left. \left( \frac{\partial \mu_i^\textrm{ig}}{\partial P} \right) \right|_{T,\mathbf{n}} =  \bar{V}_i^\textrm{ig} = \left. \left( \frac{\partial V^\textrm{ig}}{\partial n_i} \right) \right|_{T,P,\mathbf{n}_i}  = \frac{RT}{P}.
\end{equation*}
Therefore,

\begin{equation*}
v = RT \sum_{i = 1}^c z_i \left[\frac{1}{P} + \left. \left( \frac{\partial \ln \phi_i}{\partial P} \right) \right|_{T,\mathbf{n}} \right],
\end{equation*}
and the molar density $\rho$ is

\begin{equation*}
\rho = \frac{1}{v} = \frac{1}{RT \sum_{i = 1}^c z_i \left[\dfrac{1}{P} + \left. \left( \dfrac{\partial \ln \phi_i}{\partial P} \right) \right|_{T,\mathbf{n}} \right]}.
\end{equation*}
The mass density is

\begin{equation*}
\rho_\textrm{mass} = \rho \sum_{i = 1}^c z_i M_i,
\end{equation*}
where $M_i$ is the molecular weight of component $i$.

\section{Cubic-plus-association (CPA) equation of state}
\label{sec:CPA}

\subsection{Intermolecular forces and departure functions}
\label{sec:if_departure}

It was stated at the end of Section~\ref{sec:fugacity} that the fundamental role of an equation of state (EOS) is to provide an expression from which all other thermodynamic functions of a fluid can be derived. For equations of state that model complicated fluids like mixtures of CO$_2$ and water, this expression is often in the form of a departure function~\cite{Elliott1999, Sandler2006}, which is sometimes also referred to as a residual or a residual function~\cite{Prausnitz1999, Michelsen2007}. A departure function is defined as the difference in some property (e.g.,~Helmholtz energy) of a real fluid and that same property of the fluid if it were to exist as an ideal gas mixture. This definition resembles the definition of an excess property in~(\ref{eq:excessProperty}), but there are two important differences. First, the reference fluid in~(\ref{eq:excessProperty}) is an ideal mixture, while it is an ideal gas mixture in a departure function. Second, excess properties are always defined as differences in some property $E$ for two different fluids at the same temperature, pressure, and composition $(T, P, \mathbf{n})$. In contrast, departure functions can involve two different fluids at the same $(T, P, \mathbf{n})$, or two fluids at the same~$(T, V, \mathbf{n})$. The two types of departure functions are not necessarily the same. An ideal gas is by definition composed of non-interacting molecules represented by volumeless entities (points). In a real fluid, the departure from ideality arises because the molecules have a non-zero volume and interact with each other through intermolecular forces. The volume of the molecules is itself a reflection of repulsive intermolecular forces that have their origins in electrostatic forces and quantum mechanical effects (i.e., the Pauli exclusion principle)~\cite{Widom2002}. 

In order to understand the physical behavior represented by the CPA EOS, and indeed equations of state in general, we briefly review some basic concepts regarding intermolecular forces. The discussion here will also motivate the future work described in Sections~\ref{sec:excess} and~\ref{sec:future}. Intermolecular interactions can generally be divided into two sets: physical and chemical/quasi-chemical. Physical interactions include London dispersion forces (which can be thought of as instantaneous dipole-induced dipole interactions), dipole-dipole interactions (also called Keesom forces), and dipole-induced dipole interactions (Debye forces). They are sometimes collectively referred to as van der Waals forces~\cite{Burrows2013}. On a per-mole basis, the van der Waals forces are weaker than other types of interactions. However, because they form rather easily, especially London dispersion forces, they are relatively numerous in a given fluid and can significantly influence the properties of a fluid.

Chemical interactions include hydrogen bonds, which act between a molecule that contains the electronegative atoms oxygen (O), nitrogen (N), or fluorine (F) bound to hydrogen (H) and another molecule that contains O, N, or F. The hydrogen develops a partial positive charge, while the electronegative atom to which it is covalently bonded develops partial negative charge. Hydrogen bonding allows molecules of a fluid to associate to form polymers. For example, methanol consists of CH$_3$OH monomers, some of which polymerize to form CH$_3$OH dimers, trimers, or longer-chained polymers. The monomer units are linked together (associated) by hydrogen bonds between the oxygen on a methanol molecule and the hydrogen in the hydroxyl group of another methanol. Hydrogen bonds are classified as chemical interactions because they are quite strong and resemble chemical (i.e., covalent) bonds in that they involve partial overlap of the electron clouds of the atoms involved in the interaction~\cite{Prausnitz1999}. In contrast, physical interactions do not involve significant overlap of electron clouds. 

\begin{figure}
\centering
\centerline{\includegraphics[width=50mm,height=50mm]{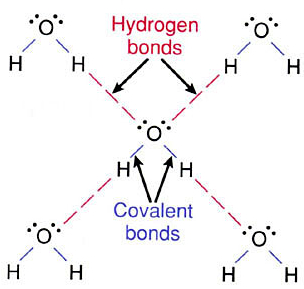}}
\caption{Hydrogen bonding in water. Each molecule can form up to four hydrogen bonds since it has four associating sites: 2 hydrogen-bond donor (also called electron acceptor) sites representing the two hydrogens, and 2 hydrogen-bond acceptor (electron donor) sites representing the two electron lone pairs on the oxygen. We will refer to these two site types as $\alpha$ and $\beta$ sites, respectively.}
\label{fig:waterHBond}
\end{figure}

Hydrogen bonding plays an essential role in determining water's unusual properties. Ice consists of a regular lattice of water molecules, with each molecule being bound to four other molecules through hydrogen bonds. Each molecule is said to have four associating sites (Figure~\ref{fig:waterHBond}), which come in two varieties: 2 hydrogen-bond donor ($\alpha$) sites representing the two hydrogens, and 2 hydrogen-bond acceptor ($\beta$) sites representing the two lone pairs of electrons on the oxygen. In liquid water, the molecules also form hydrogen bonds with each other, but these interactions are more transient; hydrogen bonds are constantly forming and breaking, and the molecules are able to move past each other. The regular, hydrogen-bonded lattice structure of ice limits the maximum packing achievable, which is why ice is less dense than liquid water at the normal melting point. 

In fluid mixtures, the components can be either self-associating or non-self-associating, based on whether the pure component forms hydrogen bonds. In mixtures with at least one self-associating component, there may be cross-association between the self-associating component and the other components. For example, CO$_2$ may cross-associate with water because the lone pairs on the oxygens of CO$_2$ can serve as hydrogen-bond acceptor sites. Molecules without associating sites, such as H$_2$S or alkanes, may also cross-associate with water. In this case, the cross-association provides a simple way to implicitly account for the complicated physical and chemical processes that occur during the solvation of these molecules in water.

\subsection{Helmholtz energy departure function}
\label{sec:Helmholtz}

\subsubsection{Overview of Peng-Robinson and CPA EOS}

The Peng-Robinson (PR) EOS belongs to the family of cubic equations of state, which are all based on the van der Waals equation of state~\cite{Sandler2006}. The PR EOS is widely used in the oil and gas industry to model fluid mixtures containing only non-self-associating components, such as hydrocarbons. It, along with other cubic EOS variants like the Soave-Redlich-Kwong (SRK) EOS, involves an interaction parameter $a$, which accounts for van der Waals interactions, and a co-volume parameter $b$, which accounts for the non-zero volume of the fluid molecules. If $a = b = 0$, the PR EOS reduces to the ideal gas EOS. Since $a$ accounts for only van der Waals interactions, the PR EOS in its original form cannot model fluid mixtures with associating components, such as water.

One approach to model mixtures with associating components is to append the PR EOS, or a different cubic EOS, with an additional term that accounts for the association. This approach forms the basis of the cubic-plus-association (CPA) EOS. The CPA EOS in its modern form was introduced by Kontogeorgis \textit{et al}.~\cite{Kontogeorgis1996}, although earlier authors have developed a similar EOS~\cite{Elliott1990}. Other models for associating fluids besides CPA are described in various studies~\cite{Economou1997, Kontogeorgis2010}, and will not be considered further in this report. The cubic part of the CPA EOS by Kontogeorgis \textit{et al}.\ comes from the SRK EOS. The association (hydrogen bonding) part is based on the thermodynamic perturbation theory of Wertheim~\cite{Wertheim1984, Wertheim1984a, Wertheim1986, Wertheim1987}. In fluids without associating components, the CPA EOS of Kontogeorgis \textit{et al}.\ reduces to the SRK EOS.

Li and Firoozabadi have presented a version of the CPA EOS where they introduce a novel scheme to model the cross-association between molecules of different components~\cite{Li2009}. Their study concludes that cross-association is necessary to faithfully reproduce experimental vapor-liquid equilibria (VLE) data, especially along the vapor phase, for mixtures containing water and one or more of the following components: CO$_2$, H$_2$S, and hydrocarbons (both aliphatic chains and aromatics). This conclusion is in agreement with earlier studies, who also report that cross-association is important in water/alkane mixtures and water/CO$_2$ mixtures~\cite{Kontogeorgis2006, Kontogeorgis2006a}. Li and Firoozabadi's original version of the CPA is targeted toward mixtures where water is the only self-associating component~\cite{Li2009}. They later modified their EOS to model mixtures where asphaltenes serve as the self-associating component~\cite{Li2010, Li2010a}. The Li and Firoozabadi CPA EOS with water as the only self-associating component will be the focus of the present study.

The starting point is to consider the CPA EOS's expression for the departure function~$F^\textrm{depart}$ of the Helmholtz energy. Since the Helmholtz energy~$F$ is a natural function of~($T, V, \mathbf{n}$),~$F^\textrm{depart}$ is defined as the difference between the Helmholtz energy of the real fluid and the Helmholtz energy of the corresponding ideal gas at the same~($T, V, \mathbf{n}$):

\begin{equation}
\label{eq:FdepartDefinition}
F^\textrm{depart} (T, V, \mathbf{n}) = F(T, V, \mathbf{n}) - F^\textrm{ig}(T, V, \mathbf{n}).
\end{equation}
Ideal gases by definition consist of non-interacting molecules of zero volume. The departure function is therefore a measure of the contribution to~$F(T, V, \mathbf{n})$ from intermolecular forces plus the non-zero volume taken up by the molecules. These contributions can be divided into two categories

\begin{equation}
\label{eq:Fdepart}
F^\textrm{depart} = F^\textrm{depart}_\textrm{phys}  + F^\textrm{depart}_\textrm{assoc},
\end{equation}
where $F^\textrm{depart}_\textrm{phys}$ is the contribution from physical (van der Waals) interactions and non-zero molecular volumes, and $F^\textrm{depart}_\textrm{assoc}$ is the contribution from association (hydrogen bonding). Note that because intermolecular forces like ion-dipole interactions and ion-ion interactions are not accounted for in the departure function, the CPA EOS cannot model mixtures involving brines, such as H$_2$O-CO$_2$-NaCl mixtures. It can model mixtures with CO$_2$ and H$_2$O only.

\subsubsection{Physical contribution to the Helmholtz energy departure function}
\label{sec:FdepartPhys}
The physical contribution~$F^\textrm{depart}_\textrm{phys}$ is modeled by the cubic part of the EOS. Unlike the original CPA introduced by Kontogeorgis \textit{et al}., Li and Firoozabadi use the PR EOS, not the SRK EOS, for their cubic part. The expression for $F^\textrm{depart}_\textrm{phys}$ given by the PR EOS is

\begin{equation}
\label{eq:FdepartPhys}
\frac{F^\textrm{depart}_\textrm{phys}}{N R T} = -\ln(1 - b \rho) - \frac{a}{2 \sqrt{2} b R T} \ln \left[ \frac{1 + (1 + \sqrt{2} ) b \rho}{1 + (1 - \sqrt{2}) b \rho} \right],
\end{equation}
where~$a = a(T, \mathbf{z})$ is the temperature-dependent parameter that accounts for the physical interactions, and $b = b(\mathbf{z})$ is the temperature-independent parameter that accounts for non-zero molecular volumes. In mixtures, $a$ and $b$ are computed from the pure component properties $a_i$ and $b_i$ using the standard van der Waals mixing rules:

\begin{equation}
\label{eq:a_ij}
a_{ij} = (1 - k_{i j}) a_i^{1/2} a_j^{1/2},
\end{equation}
\begin{equation*}
a(T, \mathbf{z}) = \sum_{i = 1}^c \sum_{j = 1}^c z_i z_j a_{ij} ,
\end{equation*}
\begin{equation}
\label{eq:b_mix}
b(\mathbf{z}) = \sum_{i = 1}^c z_i b_i,
\end{equation}
where $k_{i j} = k_{i j}(T)$ is the binary interaction coefficient between component $i$ and $j$. This coefficient is fitted to experimental VLE data. Fits to binary water/CO$_2$ mixture data give

\begin{equation*}
k_{i j}  = 0.5994 (T / T_{c,{\textrm{CO}_2}}) - 0.5088,
\end{equation*}
for $k_{i j}$ between water and CO$_2$ in which $T_{c,{\textrm{CO}_2}}$ is the CO$_2$ critical temperature. Correlations for $k_{i j}$ of non-self-associating components may be found in~\cite{Firoozabadi1999}. The pure component properties $a_i$ and $b_i$ in~(\ref{eq:a_ij}) and~(\ref{eq:b_mix}) are computed from three parameters if it is a non-self-associating: the reduced temperature~$T_{r,i}$; the reduced pressure~$P_{r,i}$; and the acentric factor~$\omega_i$. They are defined as

\begin{equation*}
T_{r,i} = T / T_{c,i},
\end{equation*}
\begin{equation*}
P_{r,i} = P / P_{c,i},
\end{equation*}
\begin{equation*}
\omega_i = - \log_{10} \left. \left(\frac{P^\textrm{vap}_i}{P_{c,i}} \right) \right|_{T_{r,i} = 0.7} - 1,
\end{equation*}
where $T_{c,i}$ is the critical temperature of component~$i$, $P_{c,i}$ is its critical pressure, and $P^\textrm{vap}_i$ is its vapor pressure. For the acentric factor, the vapor pressure is evaluated at a temperature corresponding to $T_{r,i} = 0.7$. For a non-self-associating component~$i$~\cite{Firoozabadi2015}, its properties $a_i$ and $b_i$ are

\begin{equation*}
a_i(T) = a_c (T_{c,i}) \alpha (\omega_i, T_{r,i}),
\end{equation*}
\begin{equation*}
a_c (T_{c,i}) = 0.45724 \frac{R^2 T_{c,i}^2}{P_{c,i}},
\end{equation*}
\begin{equation*}
\alpha (\omega_i, T_{r,i}) = \left[ 1 + m_i \left( 1 - \sqrt{T_{r,i}} \right) \right]^2,
\end{equation*}
\begin{equation*}
m_i = \begin{dcases}  
0.3796 + 1.485 \omega_i - 0.1644 \omega_i^2 + 0.01667 \omega_i^3, \quad 0.1 < \omega_i < 2.0, \\
0.37464 + 1.54226 \omega_i - 0.26992 \omega_i^2, \quad \omega_i < 0.1, 
\end{dcases}
\end{equation*}
\begin{equation*}
b_i = 0.0778 \frac{R T_{c,i}}{P_{c,i}}.
\end{equation*}
For water (the self-associating component), $a_{\textrm{H}_2\textrm{O}}$ is computed from the correlation suggested by Mathias \textit{et al}.~\cite{Mathias1989}:

\begin{equation*}
a_{\textrm{H}_2\textrm{O}} (T) = a_0 \left[1 + c_1 \left(1 -  \sqrt{T_{r,{\textrm{H}_2\textrm{O}}}} \right) + c_2 \left(1 -  \sqrt{T_{r,{\textrm{H}_2\textrm{O}}}} \right)^2 + c_3 \left(1 -  \sqrt{T_{r,{\textrm{H}_2\textrm{O}}}} \right)^3 \right], 
\end{equation*}
where $a_0$, $c_1$, $ c_2$, $c_3$ are constants. Li and Firoozabadi have found that choosing $c_1 = 1.7557$, $c_2 = 0.003518$, $c_3 = -0.2746$, and $a_0 = 0.9627$~L$^2 \cdot$bar/mole$^2$ provides a good match with experimental data. The volume parameter of water is $b_{\textrm{H}_2\textrm{O}} = 0.01458$ L/mole. Equation~(\ref{eq:FdepartPhys}) is commonly expressed in terms of the compressibility factor $Z$

\begin{equation}
\label{eq:Zcompress}
Z = \frac{PV}{NRT} = \frac{P}{\rho RT},
\end{equation}
by defining new variables $A$ and $B$ that are analogues to $a$ and $b$, respectively:

\begin{equation}
\label{eq:A_mix}
A(T, P, \mathbf{z}) = \frac{a P}{R^2 T^2},
\end{equation}
\begin{equation}
\label{eq:B_mix}
B(T, P, \mathbf{z}) = \frac{b P}{R T}.
\end{equation}
Substituting~(\ref{eq:Zcompress})--(\ref{eq:B_mix}) into~(\ref{eq:FdepartPhys}) yields

\begin{equation}
\label{eq:FdepartPhysZ}
\frac{F^\textrm{depart}_\textrm{phys}}{N R T} = \ln \left( \frac{Z}{Z - B} \right ) - \frac{A}{2 \sqrt{2} B} \ln \left[ \frac{Z + (1 + \sqrt{2} ) B}{Z + (1 - \sqrt{2}) B} \right].
\end{equation}

\subsubsection{Association contribution to the Helmholtz energy departure function}
\label{sec:FdepartAssoc} 
As mentioned in Section~\ref{sec:if_departure}, there are two types of association sites: hydrogen-bond donor sites, which we will denote as $\alpha$ sites, and hydrogen-bond acceptor sites, which we will denote as $\beta$ sites. Hydrogen bonds form only between sites of opposite types; $\alpha$ ($\beta$) sites on a molecule can associate with only the $\beta$ ($\alpha$) sites on another molecule. The general expression for $F^\textrm{depart}_\textrm{assoc}$ is 

\begin{equation}
\label{eq:FdepartAssoc}
\frac{F^\textrm{depart}_\textrm{assoc}}{NRT} = \sum_{i =1}^c z_i \sum_{j=\alpha,\beta} \eta_{ij} \left( \ln  \chi_{ij} - \frac{\chi_{ij} - 1}{2} \right),
\end{equation}
where $\eta_{ij}$ is the number of $j$ sites ($j = \alpha$ or $\beta$) on a molecule of component $i$, and $\chi_{ij}$ is the fraction of component~$i$ molecules whose $j$ sites are not associated (i.e., $j$ sites that are still free to form hydrogen bonds)~\cite{Kontogeorgis2010}. We make the following key assumption: water is the only self-associating component in the mixture. Other components may cross-associate with water, but they do not self-associate nor do they cross-associate with each other. This means that in addition to being inapplicable to electrolytic solutions [see discussion below Equation~(\ref{eq:Fdepart})], the CPA EOS that we consider is also limited to mixtures where water is the only self-associating component. It would not be applicable to aqueous mixtures of alcohols or amines, for example. Nonetheless, this assumption is sufficient for CO$_2$-H$_2$O mixtures, since CO$_2$ lacks the hydrogen atoms that are necessary for self-association. The site fractions for water (denoted by the subscript `w'), are obtained by solving equations suggested by Kontogeorgis \textit{et al}.~\cite{Kontogeorgis1996}:

\begin{equation}
\label{eq:chi_walpha}
\chi_{w\alpha} = \frac{1}{1 + \rho \sum_{i =1}^c z_i \eta_{i\beta} \chi_{i\beta} \delta_{w\alpha, i\beta}},
\end{equation}
\begin{equation}
\label{eq:chi_wbeta}
\chi_{w\beta} = \frac{1}{1 + \rho \sum_{i =1}^c z_i \eta_{i\alpha} \chi_{i\alpha} \delta_{w\beta, i\alpha}},
\end{equation}
in which $\delta_{w\alpha, i\beta}$ is the association strength between an $\alpha$ site on a water molecule and a $\beta$ site on a molecule of component $i$. The quantity $\delta_{w\beta, i\alpha}$ is defined similarly. Equations~(\ref{eq:chi_walpha}) and~(\ref{eq:chi_wbeta}) essentially state that the fraction of non-associated $\alpha$ ($\beta$) sites on water, as expressed by $\chi_{w\alpha}$ ($\chi_{w\beta}$), decreases with the density of available, meaning non-associated, $\beta$ ($\alpha$) sites on each component~$i$, which is given by $\rho z_i \eta_{i\beta} \chi_{i\beta}$ ($\rho z_i \eta_{i\alpha} \chi_{i\alpha}$), and with the association strength of $\beta$ ($\alpha$) sites on component~$i$. The association strengths are defined according to

\begin{equation}
\label{eq:deltaWater}
\delta_{w\alpha, w\beta} = \delta_{w\beta, w\alpha} = g \kappa \left[ \exp \left(\frac{\epsilon}{k_B T} \right) -1 \right],
\end{equation}
\begin{equation}
\label{eq:deltaWaterAlphaiBeta}
\delta_{w\alpha, i\beta} = s_{w\alpha, i\beta} \delta_{w\alpha, w\beta} \quad i \neq w,
\end{equation}
\begin{equation}
\label{eq:deltaWaterBetaiAlpha}
\delta_{w\beta, i\alpha} = s_{w\beta, i\alpha} \delta_{w\beta, w\alpha} \quad i \neq w,
\end{equation}
where $s_{w\alpha, i\beta} = s_{w\alpha, i\beta}(T)$ is the parameter of cross-association between an $\alpha$ site on water with a $\beta$ site on component~$i$, and $s_{w\beta, i\alpha}$ is defined similarly. The cross-association parameters are expressed as second-order polynomials of the reduced temperature $T_{r,i}$ whose coefficients can be fitted to experimental data~\cite{Li2009}. The energy and volume parameters of water self-association are denoted by $\epsilon$ and $\kappa$, respectively. Their values are

\begin{equation}
\label{eq:epsilon}
\frac{\epsilon}{k_B} = 1738.4 \textrm{ K},
\end{equation}
\begin{equation*}
\kappa = 1.8015 \times 10^{-3} \textrm{ }\frac{\textrm{L}}{\textrm{mole}},
\end{equation*}
where $k_B$ is the Boltzmann constant. These parameters have been fitted to data for pure water. Finally, $g = g(\eta)$ in~(\ref{eq:deltaWater}) represents the contact value of the radial distribution function for water. We may denote this function with its argument~$\eta$ to distinguish it from the molar Gibbs energy~$g(T, P, \mathbf{z})$. Li and Firoozabadi~\cite{Li2009} use the expression for $g(\eta)$ from the Starling-Carnahan EOS, which states

\begin{equation}
\label{eq:gRadial}
g(\eta) = \frac{1 - 0.5 \eta}{(1 - \eta)^3},
\end{equation}
\begin{equation}
\label{eq:eta}
\eta = \frac{b \rho}{4} = \frac{B}{4 Z}.
\end{equation}
Studies have found expressions for $g(\eta)$ simpler than that in~(\ref{eq:gRadial}) also work well~\cite{Kontogeorgis2006}, but we continue to use~(\ref{eq:gRadial}) to be consistent with the work of Li and Firoozabadi. Since water is the only self-associating component and other components may cross-associate only with water, as stated at the beginning of this section, the site fractions for these components are

\begin{equation}
\label{eq:chi_ialpha}
\chi_{i\alpha} = \frac{1}{1 + \rho z_w \eta_{w\beta} \chi_{w\beta} \delta_{w\beta, i\alpha}},
\end{equation}
\begin{equation}
\label{eq:chi_ibeta}
\chi_{i\beta} = \frac{1}{1 + \rho z_w \eta_{w\alpha} \chi_{w\alpha} \delta_{w\alpha, i\beta}}.
\end{equation}
Note that in non-associating fluids, the association strengths are all zero so that from~(\ref{eq:chi_walpha})--(\ref{eq:chi_wbeta}) and~(\ref{eq:chi_ialpha})--(\ref{eq:chi_ibeta}), all site fractions become zero and the CPA EOS reduces to the PR EOS. In summary, the association contribution to the Helmholtz energy departure function is given by~(\ref{eq:FdepartAssoc}), where the site fractions are obtained by solving~(\ref{eq:chi_walpha})--(\ref{eq:chi_wbeta}) and~(\ref{eq:chi_ialpha})--(\ref{eq:chi_ibeta}). The association strengths that affect the site fractions are calculated using~(\ref{eq:deltaWater})--(\ref{eq:eta}). Substituting~(\ref{eq:FdepartPhys}) and~(\ref{eq:FdepartAssoc}) into~(\ref{eq:Fdepart}), we obtain the full expression for the Helmholtz energy departure function

\begin{align}
\label{eq:FdepartCPA}
\frac{F^\textrm{depart}}{N R T} = -\ln(1 - b \rho) & - \frac{a}{2 \sqrt{2} b R T} \ln \left[ \frac{1 + (1 + \sqrt{2} ) b \rho}{1 + (1 - \sqrt{2}) b \rho} \right]  \nonumber \\ 
& + \sum_{i =1}^c z_i \sum_{j=\alpha,\beta} \eta_{ij} \left( \ln  \chi_{ij} - \frac{\chi_{ij} - 1}{2} \right).
\end{align}

Section~\ref{sec:pressureCPA} will derive the expression for the pressure from~(\ref{eq:FdepartCPA}). This form of~$F^\textrm{depart}$ is useful for this purpose because $P$ does not appear explicitly. However, for the purposes of the fugacity coefficient $\phi_i$, which we derive in the Section~\ref{sec:fugacityCPA}, it is more convenient to express~$F^\textrm{depart}$ in terms of the compressibility factor $Z = P / \rho R T$. To do so, we define new expressions for the association strengths according to

\begin{equation}
\label{eq:Kappa}
\mathcal{K} = \frac{\kappa P}{RT},
\end{equation}
\begin{equation}
\label{eq:DeltaWaterZ}
\Delta_{w\alpha, w\beta} = \Delta_{w\beta, w\alpha} = g \mathcal{K} \left[ \exp \left(\frac{\epsilon}{k_B T} \right) -1 \right],
\end{equation}
\begin{equation}
\label{eq:DeltaWaterAlphaiBetaZ}
\Delta_{w\alpha, i\beta} = s_{w\alpha, i\beta} \Delta_{w\alpha, w\beta} \quad i \neq w,
\end{equation}
\begin{equation}
\label{eq:DeltaWaterBetaiAlphaZ}
\Delta_{w\beta, i\alpha} = s_{w\beta, i\alpha} \Delta_{w\beta, w\alpha} \quad i \neq w.
\end{equation}
Note that $\Delta_{ij, kl} = (P/RT) \delta_{ij, kl}$ for all $i,j,k,l$. The site fractions are given by

\begin{equation}
\label{eq:chi_walphaZ}
\chi_{w\alpha} = \frac{Z}{Z + \sum_{i =1}^c z_i \eta_{i\beta} \chi_{i\beta} \Delta_{w\alpha, i\beta}},
\end{equation}
\begin{equation}
\label{eq:chi_wbetaZ}
\chi_{w\beta} = \frac{Z}{Z + \sum_{i =1}^c z_i \eta_{i\alpha} \chi_{i\alpha} \Delta_{w\beta, i\alpha}},
\end{equation}
\begin{equation}
\label{eq:chi_ialphaZ}
\chi_{i\alpha} = \frac{Z}{Z + z_w \eta_{w\beta} \chi_{w\beta} \Delta_{w\beta, i\alpha}},
\end{equation}
\begin{equation}
\label{eq:chi_ibetaZ}
\chi_{i\beta} = \frac{Z}{Z + z_w \eta_{w\alpha} \chi_{w\alpha} \Delta_{w\alpha, i\beta}}.
\end{equation}
In terms of $Z$, the association contribution to the Helmholtz energy departure function is given by~(\ref{eq:FdepartAssoc}), where the site fractions are obtained by solving~(\ref{eq:chi_walphaZ})--(\ref{eq:chi_ibetaZ}). The association strengths that affect the site fractions are calculated using~(\ref{eq:Kappa})--(\ref{eq:DeltaWaterBetaiAlphaZ}) and~(\ref{eq:epsilon})--(\ref{eq:eta}). Substituting~(\ref{eq:FdepartPhysZ}) and~(\ref{eq:FdepartAssoc}) into~(\ref{eq:Fdepart}), we obtain the full expression for the Helmholtz energy departure function

\begin{equation}
\label{eq:FdepartCPAZ}
\frac{F^\textrm{depart}}{N R T} = \ln \left( \frac{Z}{Z - B} \right ) - \frac{A}{2 \sqrt{2} B} \ln \left[ \frac{Z + (1 + \sqrt{2} ) B}{Z + (1 - \sqrt{2}) B} \right] + \sum_{i =1}^c z_i \sum_{j=\alpha,\beta} \eta_{ij} \left( \ln  \chi_{ij} - \frac{\chi_{ij} - 1}{2} \right).
\end{equation}

\subsection{Fugacity coefficient, enthalpy, density}
\label{sec:fugacityCPA}

In this section, we derive the expression for the fugacity coefficient~$\phi_i$ from the Helmholtz energy departure function~(\ref{eq:FdepartCPAZ}). The fugacity coefficient is an important quantity in phase equilibria (solubility) calculations, as discussed in Section~\ref{sec:fugacity}. The fugacity coefficient can also be used to compute the enthalpy and the density, as described in Section~\ref{sec:enthalpyDensity}. We start with the expression for the differential of the Helmholtz energy~$F$. For systems where external fields like gravity are negligible and there is only pressure-volume work,

\begin{equation}
\label{eq:dF}
\mathrm{d} F = - S \mathrm{d} T - P \mathrm{d} V  + \sum_{i = 1}^c \mu_i \mathrm{d} n_i,
\end{equation}
from which it follows that

\begin{equation*}
\mu_i  = \left. \left( \frac{\partial F}{\partial n_i} \right) \right|_{T, V,\mathbf{n}_i}.
\end{equation*}
This expression, when combined with the definition of $F^\textrm{depart}$ in~(\ref{eq:FdepartDefinition}), leads to

\begin{equation}
\label{eq:muDepartureTV}
\frac{\mu_i(T, V, \mathbf{n}) - \mu_i^{\textrm{ig}}(T, V, \mathbf{n})}{RT}  = \frac{1}{RT} \left. \left( \frac{\partial F^\textrm{depart}}{\partial n_i} \right) \right|_{T, V,\mathbf{n}_i}.
\end{equation}
The left-hand side of~(\ref{eq:muDepartureTV}) is similar to the left-hand side of~(\ref{eq:mu1Realmu2IG}), except that the ideal gas in~(\ref{eq:mu1Realmu2IG}) is at the same pressure (but not same volume) as the real fluid, while the ideal gas in~(\ref{eq:muDepartureTV}) is at the same volume (but not same pressure) as the real fluid. We can denote the volume of the ideal gas in~(\ref{eq:mu1Realmu2IG}) as $V^{\textrm{ig}} = NRT/P$ so that its chemical potential is $\mu_i^{\textrm{ig}}(T, P, \mathbf{n}) = \mu_i^{\textrm{ig}}(T, V^{\textrm{ig}}, \mathbf{n})$. Similarly, the pressure of the ideal gas in~(\ref{eq:muDepartureTV}) is $P^\textrm{ig} = NRT/V$ so that its chemical potential is $\mu_i^{\textrm{ig}}(T, V, \mathbf{n}) = \mu_i^{\textrm{ig}}(T, P^{\textrm{ig}}, \mathbf{n})$. The two different ideal gas states are related by

\begin{equation*}
\mu_i^{\textrm{ig}}(T, V, \mathbf{n})  - \mu_i^{\textrm{ig}}(T, V^{\textrm{ig}}, \mathbf{n}) = \int_{V^{\textrm{ig}} }^V \left. \left( \frac{\partial \mu_i^{\textrm{ig}}}{\partial V'} \right) \right|_{T,\mathbf{n}} \mathrm{d} V'.
\end{equation*}
From the equality of mixed derivatives in~(\ref{eq:dF}), we readily obtain 

\begin{equation*}
-\left. \left( \frac{\partial P}{\partial n_i} \right) \right|_{T, V,\mathbf{n}_i} = \left. \left( \frac{\partial \mu_i}{\partial V} \right) \right|_{T,\mathbf{n}},
\end{equation*}
which we apply to the integral 

\begin{equation*}
 \int_{V^{\textrm{ig}} }^V \left. \left( \frac{\partial \mu_i^{\textrm{ig}}}{\partial V'} \right) \right|_{T,\mathbf{n}} \mathrm{d} V' =  -\int_{V^{\textrm{ig}} }^V \left. \left( \frac{\partial P^{\textrm{ig}}}{\partial n_i} \right) \right|_{T,V',\mathbf{n}_i} \mathrm{d} V' = RT \ln \left(\frac{V^{\textrm{ig}}}{V} \right) = - RT \ln Z.
\end{equation*}
Thus,

\begin{equation*}
\mu_i(T, V, \mathbf{n}) - \mu_i^{\textrm{ig}}(T, V, \mathbf{n}) = \mu_i(T, P, \mathbf{n}) - \left[ \mu_i^{\textrm{ig}}(T, P, \mathbf{n}) - RT \ln Z \right],
\end{equation*}
so that combining with~(\ref{eq:mu1Realmu2IG}) and~(\ref{eq:muDepartureTV}), we obtain

\begin{equation}
\label{eq:phiHelmholtzDeparture}
\ln \phi_i = \frac{1}{RT} \left. \left( \frac{\partial F^\textrm{depart}}{\partial n_i} \right) \right|_{T, V,\mathbf{n}_i} - \ln Z.
\end{equation}

Since the departure function can be divided into a physical contribution and an association contribution [see Equation~(\ref{eq:Fdepart})], we can also divide the fugacity coefficient into a physical contribution and an association contribution according to

\begin{equation*}
\ln \phi_i^\textrm{phys} = \frac{1}{RT} \left. \left( \frac{\partial F^\textrm{depart}_\textrm{phys}}{\partial n_i} \right) \right|_{T, V,\mathbf{n}_i} - \ln Z,
\end{equation*}
and 
\begin{equation}
\label{eq:lnphiAssocDeparture}
\ln \phi_i^\textrm{assoc} = \frac{1}{RT} \left. \left( \frac{\partial F^\textrm{depart}_\textrm{assoc}}{\partial n_i} \right) \right|_{T, V,\mathbf{n}_i}.
\end{equation}
It can be shown (see for example,~\cite{Elliott1999}) that the fugacity coefficient of the PR EOS is

\begin{align}
\label{eq:lnphiPhysZ}
\ln \phi_i^\textrm{phys} = -\ln \left( Z - B \right ) & + \frac{B_i}{B} \left(\frac{B}{Z - B} - \frac{A Z}{Z^2 + 2 B Z - B^2} \right) \nonumber \\ 
& - \frac{A}{2 \sqrt{2} B} \left(\frac{2 \sum_{j = 1}^c z_j A_{ij}}{A} - \frac{B_i}{B} \right) \ln \left[ \frac{Z + (1 + \sqrt{2} ) B}{Z + (1 - \sqrt{2}) B} \right],
\end{align}
where 

\begin{equation}
\label{eq:B_i}
B_i = \frac{b_i P}{R T},
\end{equation}
\begin{equation}
\label{eq:A_ij}
A_{ij} = \frac{a_{ij} P}{R^2 T^2} = \frac{(1 - k_{i j}) a_i^{1/2} a_j^{1/2} P}{R^2 T^2}.
\end{equation}

Equation~(\ref{eq:lnphiAssocDeparture}) is difficult to evaluate because the site fractions $\boldsymbol{\chi} = (\chi_{1\alpha}, \chi_{1\beta}, \chi_{2\alpha}, \dots, \chi_{c\beta})$ are implicit functions of each other~[see~(\ref{eq:chi_walphaZ})--\ref{eq:chi_ibetaZ})], and thus, it is not immediately clear how to evaluate their derivatives. We avoid having to evaluate the site-fraction derivatives by employing the method developed by Michelsen and Hendriks~\cite{Michelsen2001}. Their method involves constructing a function $Q = Q(T, V, \mathbf{n}, \boldsymbol{\chi}$) such that $Q$ is equal to $F^\textrm{depart}_\textrm{assoc}/RT$ at a stationary point (labeled `sp') with respect to the site fractions $\boldsymbol{\chi}$. That is, $Q = Q_\textrm{sp} = F^\textrm{depart}_\textrm{assoc}/RT$ at a point where 

\begin{equation}
\label{eq:dQspdChi}
\left. \left( \frac{\partial Q_\textrm{sp}}{\partial \chi_{ij}} \right) \right|_{(T, V, \mathbf{n}, \boldsymbol{\chi}_{ij})} = 0 \quad \textrm{for all }\chi_{ij}. 
\end{equation}
The subscript $\boldsymbol{\chi}_{ij}$ in the derivative above means that all site fractions besides $\chi_{ij}$ are held fixed. Therefore,

\begin{equation*}
\ln \phi_i^\textrm{assoc} = \frac{1}{RT} \left. \left( \frac{\partial F^\textrm{depart}_\textrm{assoc}}{\partial n_i} \right) \right|_{T, V,\mathbf{n}_i} = \left. \left( \frac{\partial Q_\textrm{sp}}{\partial n_i} \right) \right|_{T, V,\mathbf{n}_i}.
\end{equation*}
Applying the chain rule,

\begin{equation}
\label{eq:dQdni}
\left. \left( \frac{\partial Q_\textrm{sp}}{\partial n_i} \right) \right|_{T, V,\mathbf{n}_i} = \left. \left( \frac{\partial Q_\textrm{sp}}{\partial n_i} \right) \right|_{T, V,\mathbf{n}_i, \boldsymbol{\chi}} + \sum_{i=1}^c \sum_{j=\alpha, \beta} \left. \left( \frac{\partial Q_\textrm{sp}}{\partial \chi_{ij}} \right) \right|_{(T, V, \mathbf{n}, \boldsymbol{\chi}_{ij})} \left. \left( \frac{\partial \chi_{ij}}{\partial n_i} \right) \right|_{T, V,\mathbf{n}_i}.
\end{equation}
By definition of the stationary point in~(\ref{eq:dQspdChi}), Equation~(\ref{eq:dQdni}) simplifies to

\begin{equation}
\label{eq:dQspdni}
\left. \left( \frac{\partial Q_\textrm{sp}}{\partial n_i} \right) \right|_{T, V,\mathbf{n}_i} = \left. \left( \frac{\partial Q_\textrm{sp}}{\partial n_i} \right) \right|_{T, V,\mathbf{n}_i, \boldsymbol{\chi}}.
\end{equation}
The site fractions $\boldsymbol{\chi}$ are held constant in~(\ref{eq:dQspdni}). The result of all this is that the troublesome derivatives $\left. \left( \partial \chi_{ij} / \partial n_i \right) \right|_{T, V,\mathbf{n}_i}$ do not need to be evaluated. The task remains to select a good choice for the function $Q$. One such choice is

\begin{align}
\label{eq:Q}
Q(T, V, \mathbf{n}, \boldsymbol{\chi}) = & \sum_{i =1}^c n_i \sum_{j=\alpha,\beta} \eta_{ij} \left( \ln  \chi_{ij} - \chi_{ij} + 1 \right) \nonumber \\
& - \frac{1}{2 V} \sum_{i =1}^c n_i \sum_{j=\alpha,\beta} \eta_{ij} \chi_{ij} \sum_{k =1}^c n_k \sum_{l=\alpha,\beta} \eta_{kl} \chi_{kl} \delta_{ij, kl}.
\end{align}
 
It must first be shown that $Q_\textrm{sp} = F^\textrm{depart}_\textrm{assoc}/RT$. At a stationary point,

\begin{equation*}
\left. \left( \frac{\partial Q_\textrm{sp}}{\partial \chi_{ij}} \right) \right|_{(T, V, \mathbf{n}, \boldsymbol{\chi}_{ij})} = n_i \eta_{ij} \left(\frac{1}{\chi_{ij}} - 1 \right) - \frac{n_i \eta_{ij}}{V} \sum_{k =1}^c n_k \sum_{l=\alpha,\beta} \eta_{kl} \chi_{kl} \delta_{ij, kl} = 0.
\end{equation*}
Rearranging, we have

\begin{equation}
\label{eq:Qsp}
\frac{1}{V} \sum_{k =1}^c n_k \sum_{l=\alpha,\beta} \eta_{kl} \chi_{kl} \delta_{ij, kl} = \frac{1}{\chi_{ij}} - 1.
\end{equation}
Substituting this result into~(\ref{eq:Q}) and comparing with~(\ref{eq:FdepartAssoc}), we see (since $z_i = n_i / N$) that

\begin{equation*}
Q_\textrm{sp} = \sum_{i =1}^c n_i \sum_{j=\alpha,\beta} \eta_{ij} \left( \ln  \chi_{ij} - \frac{\chi_{ij} - 1}{2} \right) = \frac{F^\textrm{depart}_\textrm{assoc}}{RT} .
\end{equation*}
Substituting~(\ref{eq:Q}) into~(\ref{eq:dQspdni}), we have

\begin{align*}
\left. \left( \frac{\partial Q_\textrm{sp}}{\partial n_m} \right) \right|_{T, V,\mathbf{n}_m, \boldsymbol{\chi}} = & \sum_{j=\alpha,\beta} \eta_{mj} \left( \ln  \chi_{mj} - \chi_{mj} + 1 \right) -\frac{\sum_{j=\alpha,\beta} \eta_{mj} \chi_{mj} }{V} \sum_{k =1}^c n_k \sum_{l=\alpha,\beta} \eta_{kl} \chi_{kl} \delta_{mj, kl} \nonumber \\
& - \frac{1}{2 V} \sum_{i =1}^c n_i \sum_{j=\alpha,\beta} \eta_{ij} \chi_{ij} \sum_{k =1}^c n_k \sum_{l=\alpha,\beta} \eta_{kl} \chi_{kl} \left. \left( \frac{\partial \delta_{ij, kl}}{\partial n_m} \right) \right|_{T, V,\mathbf{n}_m, \boldsymbol{\chi}}.
\end{align*}
Using~(\ref{eq:Qsp}), this simplifies to

\begin{align*}
\left. \left( \frac{\partial Q_\textrm{sp}}{\partial n_m} \right) \right|_{T, V,\mathbf{n}_m, \boldsymbol{\chi}} = & \sum_{j=\alpha,\beta} \eta_{mj} \ln  \chi_{mj} \nonumber \\
& - \frac{1}{2 V} \sum_{i =1}^c n_i \sum_{j=\alpha,\beta} \eta_{ij} \chi_{ij} \sum_{k =1}^c n_k \sum_{l=\alpha,\beta} \eta_{kl} \chi_{kl} \left. \left( \frac{\partial \delta_{ij, kl}}{\partial n_m} \right) \right|_{T, V,\mathbf{n}_m, \boldsymbol{\chi}}.
\end{align*}
From~(\ref{eq:deltaWater}), we see that the only part of $\delta_{ij, kl}$ that depends on the mole numbers is the contact value of the radial distribution function~$g(\eta)$. Therefore, using~(\ref{eq:eta}) yields

\begin{equation*}
\left. \left( \frac{\partial \delta_{ij, kl}}{\partial n_m} \right) \right|_{T, V,\mathbf{n}_m, \boldsymbol{\chi}} = \frac{\delta_{ij, kl}}{g}\frac{\mathrm{d} g}{\mathrm{d} \eta} \left. \left( \frac{\partial \eta}{\partial n_m} \right) \right|_{T, V,\mathbf{n}_m, \boldsymbol{\chi}} = \frac{\delta_{ij, kl}}{g}\frac{\mathrm{d} g}{\mathrm{d} \eta} \frac{b_m}{4V} = \frac{\delta_{ij, kl}}{g}\frac{\mathrm{d} g}{\mathrm{d} \eta} \frac{B_m}{4Z} \frac{1}{N},
\end{equation*}
where 

\begin{equation*}
\frac{\mathrm{d} g}{\mathrm{d} \eta} = \frac{2.5 - \eta}{(1- \eta)^4}.
\end{equation*}
Substituting into the above, we have

\begin{align*}
\left. \left( \frac{\partial Q_\textrm{sp}}{\partial n_m} \right) \right|_{T, V,\mathbf{n}_m, \boldsymbol{\chi}} = & \sum_{j=\alpha,\beta} \eta_{mj} \ln  \chi_{mj} \nonumber \\
& - \frac{B_m}{8 g Z} \frac{\mathrm{d} g}{\mathrm{d} \eta} \sum_{i =1}^c z_i \sum_{j=\alpha,\beta} \eta_{ij} \chi_{ij} \left( \frac{1}{V} \sum_{k =1}^c n_k \sum_{l=\alpha,\beta} \eta_{kl} \chi_{kl} \delta_{ij, kl} \right). 
\end{align*}
Substituting~(\ref{eq:Qsp}) gives

\begin{equation}
\label{eq:lnphiAssocZ}
\ln \phi_i^\textrm{assoc} = \sum_{j=\alpha,\beta} \eta_{ij} \ln  \chi_{ij} + \frac{B_i}{8 g Z} \frac{\mathrm{d} g}{\mathrm{d} \eta} \sum_{k =1}^c z_k \sum_{j=\alpha,\beta} \eta_{kj} (\chi_{kj} - 1).
\end{equation}
Combining~(\ref{eq:lnphiPhysZ}) and~(\ref{eq:lnphiAssocZ}), we obtain the full expression for the fugacity coefficient

\begin{align}
\label{eq:lnphiZ}
\ln \phi_i = -\ln \left( Z - B \right ) & + \frac{B_i}{B} \left(\frac{B}{Z - B} - \frac{A Z}{Z^2 + 2 B Z - B^2} \right) \nonumber \\ 
& - \frac{A}{2 \sqrt{2} B} \left(\frac{2 \sum_{j = 1}^c z_j A_{ij}}{A} - \frac{B_i}{B} \right) \ln \left[ \frac{Z + (1 + \sqrt{2} ) B}{Z + (1 - \sqrt{2}) B} \right] \nonumber \\
& + \sum_{j=\alpha,\beta} \eta_{ij} \ln  \chi_{ij} + \frac{B_i}{8 g Z} \frac{\mathrm{d} g}{\mathrm{d} \eta} \sum_{k =1}^c z_k \sum_{j=\alpha,\beta} \eta_{kj} (\chi_{kj} - 1).
\end{align}
For completeness, we also present the formula for $\ln \phi_i$ in terms of the molar density $\rho$, which can be obtained from~(\ref{eq:lnphiZ}) using the definitions~(\ref{eq:Zcompress})--(\ref{eq:B_mix}) and~(\ref{eq:B_i})--(\ref{eq:A_ij})

\begin{align}
\label{eq:lnphi}
\ln \phi_i = -\ln \left( 1 - b \rho \right ) -\ln Z & + \frac{b_i}{b} \left(\frac{b \rho}{1 - b \rho} - \frac{1}{R T} \frac{a \rho}{1 + 2 b \rho - (b \rho)^2} \right) \nonumber \\ 
& - \frac{a}{2 \sqrt{2} b R T} \left(\frac{2 \sum_{j = 1}^c z_j a_{ij}}{a} - \frac{b_i}{b} \right) \ln \left[ \frac{1 + (1 + \sqrt{2} ) b \rho }{1 + (1 - \sqrt{2}) b \rho} \right] \nonumber \\
& + \sum_{j=\alpha,\beta} \eta_{ij} \ln  \chi_{ij} + \frac{b_i \rho}{8 g} \frac{\mathrm{d} g}{\mathrm{d} \eta} \sum_{k =1}^c z_k \sum_{j=\alpha,\beta} \eta_{kj} (\chi_{kj} - 1).
\end{align}

\subsection{Pressure and compressibility factor}
\label{sec:pressureCPA}

The main purpose of this section is to derive the expression for the pressure from the Helmholtz energy departure function. From~(\ref{eq:dF}), the pressure is
\begin{equation*}
P  = -\left. \left( \frac{\partial F}{\partial V} \right) \right|_{T, \mathbf{n}}, 
\end{equation*}
so that if $P^\textrm{ig} = NRT/V$ is the pressure of the ideal gas at ($T, V, \mathbf{n}$),

\begin{equation*}
P  - P^\textrm{ig} = - \left. \left( \frac{\partial F^\textrm{depart}}{\partial V} \right) \right|_{T, \mathbf{n}},
\end{equation*}
As mentioned in Section~\ref{sec:FdepartAssoc}, these derivatives are difficult to evaluate because $F^\textrm{depart}$ in~(\ref{eq:FdepartCPAZ}) depends explicitly on pressure, and so the right-hand side will contain explicit volume terms as well as terms which depend on the volume implicitly through the pressure. We can overcome this problem by noting that since all of the mole numbers in these derivatives are held constant, the total number of moles $N$ is also constant. The condition that all mole numbers $\mathbf{n}$ be held constant is equivalent to the condition that all mole fractions $\mathbf{z}$ be held constant. (The converse is not true, however.) We can therefore write the pressure in terms of the molar density $\rho = N/V$ as

\begin{equation*}
P  = \frac{\rho^2}{N} \left. \left( \frac{\partial F}{\partial \rho} \right) \right|_{T, \mathbf{z}}. 
\end{equation*}
The pressure and the compressibility factor $Z = P/\rho R T$ are related to the Helmholtz energy departure function according to

\begin{equation}
\label{eq:pressureHelmholtzDeparture}
P  - P^\textrm{ig} = \frac{\rho^2}{N} \left. \left( \frac{\partial F^\textrm{depart}}{\partial \rho} \right) \right|_{T, \mathbf{z}},
\end{equation}
\begin{equation}
\label{eq:ZHelmholtzDeparture}
\frac{Z  - 1}{\rho} =  \frac{1}{N R T} \left. \left( \frac{\partial F^\textrm{depart}}{\partial \rho} \right) \right|_{T, \mathbf{z}}.
\end{equation}
These expressions are useful because we can use~(\ref{eq:FdepartCPA}) for $F^\textrm{depart}$, which is the form of $F^\textrm{depart}$ in which $P$ does not appear explicitly.

The pressure is derived from $F^\textrm{depart}$ using~(\ref{eq:pressureHelmholtzDeparture}). Since $F^\textrm{depart} = F^\textrm{depart}_\textrm{phys} + F^\textrm{depart}_\textrm{assoc}$, we can divide the pressure into a physical contribution (from the PR EOS) and an association contribution, just like we did for the fugacity coefficient in the previous section. The two contributions to $P$ are  

\begin{equation*}
P^\textrm{phys} = P^\textrm{ig} + \frac{\rho^2}{N} \left. \left( \frac{\partial F^\textrm{depart}_\textrm{phys}}{\partial \rho} \right) \right|_{T, \mathbf{z}} = \rho RT + \frac{\rho^2}{N} \left. \left( \frac{\partial F^\textrm{depart}_\textrm{phys}}{\partial \rho} \right) \right|_{T, \mathbf{z}},
\end{equation*}
\begin{equation}
\label{eq:PAssocDeparture}
P^\textrm{assoc} = \frac{\rho^2}{N} \left. \left( \frac{\partial F^\textrm{depart}_\textrm{assoc}}{\partial \rho} \right) \right|_{T, \mathbf{z}}.
\end{equation}
The derivative of $F^\textrm{depart}_\textrm{phys}$ is

\begin{align*}
\left. \left( \frac{\partial F^\textrm{depart}_\textrm{phys}}{\partial \rho} \right) \right|_{T, \mathbf{n}} & = NRT \frac{\partial}{\partial \rho} \left. \left( - \ln(1 - b \rho) - \frac{a}{2 \sqrt{2} b R T} \ln \left[ \frac{1 + (1 + \sqrt{2} ) b \rho}{1 + (1 - \sqrt{2}) b \rho} \right] \right) \right|_{T, \mathbf{z}} \nonumber \\
& = NRT \left[ \frac{b}{1- b \rho} - \frac{a}{2 \sqrt{2} b R T} \left(\frac{(1 + \sqrt{2} )b }{1 + (1 + \sqrt{2} ) b \rho} - \frac{(1 - \sqrt{2} )b }{1 + (1 - \sqrt{2} ) b \rho} \right) \right] \nonumber \\
& = NRT \left[ \frac{b}{1- b \rho} - \frac{1}{R T} \frac{a}{1 + 2 b \rho - (b \rho)^2} \right],
\end{align*}
so that $P^\textrm{phys}$ is

\begin{align}
\label{eq:Pphysrho}
P^\textrm{phys} & = \rho RT + \rho^2 R T \left[ \frac{b}{1- b \rho} - \frac{1}{R T} \frac{a}{1 + 2 b \rho - (b \rho)^2} \right] \nonumber \\
& = \frac{\rho R T}{1 - b \rho} - \frac{a \rho^2}{1 + 2 b \rho - (b \rho)^2}.
\end{align}

For the derivative in~(\ref{eq:PAssocDeparture}), we encounter the same issue as was encountered with the fugacity coefficient. Namely, we have to evaluate derivatives of the site fractions. We use the same method from Michelsen and Hendriks~\cite{Michelsen2001} to overcome this problem. The only difference is that $Q$ depends on density instead of volume and on mole fractions instead of mole numbers. That is, $Q = Q(T, \rho, \mathbf{z}, \boldsymbol{\chi})$. Let us choose $Q$ to be

\begin{align}
\label{eq:Qrho}
Q(T, \rho, \mathbf{z}, \boldsymbol{\chi}) = & \sum_{i =1}^c z_i \sum_{j=\alpha,\beta} \eta_{ij} \left( \ln  \chi_{ij} - \chi_{ij} + 1 \right) \nonumber \\
& - \frac{\rho}{2} \sum_{i =1}^c z_i \sum_{j=\alpha,\beta} \eta_{ij} \chi_{ij} \sum_{k =1}^c z_k \sum_{l=\alpha,\beta} \eta_{kl} \chi_{kl} \delta_{ij, kl}.
\end{align}
It must first be shown that $Q_\textrm{sp} = F^\textrm{depart}_\textrm{assoc}/NRT$. At a stationary point,

\begin{equation*}
\left. \left( \frac{\partial Q_\textrm{sp}}{\partial \chi_{ij}} \right) \right|_{(T, \rho, \mathbf{z}, \boldsymbol{\chi}_{ij})} = z_i \eta_{ij} \left(\frac{1}{\chi_{ij}} - 1 \right) - z_i \eta_{ij} \left( \rho \sum_{k =1}^c z_k \sum_{l=\alpha,\beta} \eta_{kl} \chi_{kl} \delta_{ij, kl} \right) = 0.
\end{equation*}
Rearranging, we have

\begin{equation}
\label{eq:Qsprho}
\rho \sum_{k =1}^c z_k \sum_{l=\alpha,\beta} \eta_{kl} \chi_{kl} \delta_{ij, kl} = \frac{1}{\chi_{ij}} - 1.
\end{equation}
Substituting this result into~(\ref{eq:Qrho}) and comparing with~(\ref{eq:FdepartAssoc}), we see that

\begin{equation*}
Q_\textrm{sp} = \sum_{i =1}^c z_i \sum_{j=\alpha,\beta} \eta_{ij} \left( \ln  \chi_{ij} - \frac{\chi_{ij} - 1}{2} \right) = \frac{F^\textrm{depart}_\textrm{assoc}}{NRT} .
\end{equation*}
Taking the derivative of~(\ref{eq:Qrho}) and using~(\ref{eq:Qsprho}), we have

\begin{align*}
& \frac{1}{NRT} \left. \left( \frac{\partial F^\textrm{depart}_\textrm{assoc}}{\partial \rho} \right) \right|_{T, \mathbf{z}} = \left. \left( \frac{\partial Q_\textrm{sp}}{\partial \rho} \right) \right|_{T,\mathbf{z}, \boldsymbol{\chi}} \nonumber \\
& = \frac{1}{2 \rho} \sum_{i =1}^c z_i \sum_{j=\alpha,\beta} \eta_{ij} (\chi_{ij} - 1) - \frac{\rho}{2} \sum_{i =1}^c z_i \sum_{j=\alpha,\beta} \eta_{ij} \chi_{ij} \sum_{k =1}^c z_k \sum_{l=\alpha,\beta} \eta_{kl} \chi_{kl} \left. \left( \frac{\partial \delta_{ij, kl}}{\partial \rho} \right) \right|_{T, \mathbf{z}, \boldsymbol{\chi}}.
\end{align*}
Equation~(\ref{eq:deltaWater}) shows that the only part of $\delta_{ij, kl}$ that depends on the density is~$g$. Therefore, using~(\ref{eq:eta}) yields

\begin{equation*}
\left. \left( \frac{\partial \delta_{ij, kl}}{\partial \rho} \right) \right|_{T, \mathbf{z}, \boldsymbol{\chi}} = \frac{\delta_{ij, kl}}{g}\frac{\mathrm{d} g}{\mathrm{d} \eta} \left. \left( \frac{\partial \eta}{\partial \rho} \right) \right|_{T, \mathbf{z}, \boldsymbol{\chi}} = \frac{\delta_{ij, kl}}{g}\frac{\mathrm{d} g}{\mathrm{d} \eta} \frac{b}{4}.
\end{equation*}
Substituting into the above, we have

\begin{align*}
& \frac{1}{NRT} \left. \left( \frac{\partial F^\textrm{depart}_\textrm{assoc}}{\partial \rho} \right) \right|_{T, \mathbf{z}} = \left. \left( \frac{\partial Q_\textrm{sp}}{\partial \rho} \right) \right|_{T,\mathbf{z}, \boldsymbol{\chi}} \nonumber \\
& = \frac{1}{2 \rho} \sum_{i =1}^c z_i \sum_{j=\alpha,\beta} \eta_{ij} (\chi_{ij} - 1) - \frac{b}{8 g} \frac{\mathrm{d} g}{\mathrm{d} \eta} \sum_{i =1}^c z_i \sum_{j=\alpha,\beta} \eta_{ij} \chi_{ij} \left( \rho \sum_{k =1}^c z_k \sum_{l=\alpha,\beta} \eta_{kl} \chi_{kl} \delta_{ij, kl} \right).
\end{align*}
Substituting~(\ref{eq:eta}) and~(\ref{eq:Qsprho}) gives

\begin{align*}
& \frac{1}{NRT} \left. \left( \frac{\partial F^\textrm{depart}_\textrm{assoc}}{\partial \rho} \right) \right|_{T, \mathbf{z}} = \left. \left( \frac{\partial Q_\textrm{sp}}{\partial \rho} \right) \right|_{T,\mathbf{z}, \boldsymbol{\chi}} = \frac{1}{2 \rho} \left( 1 + \frac{\eta }{g} \frac{\mathrm{d} g}{\mathrm{d} \eta} \right) \sum_{i =1}^c z_i \sum_{j=\alpha,\beta} \eta_{ij} (\chi_{ij} - 1).
\end{align*}
Multiplying both sides by~$\rho^2 RT$, we get the association contribution to the pressure

\begin{equation}
\label{eq:Passocrho}
P^\textrm{assoc} = \frac{\rho^2}{N} \left. \left( \frac{\partial F^\textrm{depart}_\textrm{assoc}}{\partial \rho} \right) \right|_{T, \mathbf{z}} = \frac{\rho R T}{2} \left( 1 + \frac{\eta}{g} \frac{\mathrm{d} g}{\mathrm{d} \eta} \right) \sum_{i =1}^c z_i \sum_{j=\alpha,\beta} \eta_{ij} (\chi_{ij} - 1).
\end{equation}
We combine~(\ref{eq:Pphysrho}) and~(\ref{eq:Passocrho}) to write the complete expression for the pressure

\begin{equation}
\label{eq:pressureCPA}
P = \frac{\rho R T}{1 - b \rho} - \frac{a \rho^2}{1 + 2 b \rho - (b \rho)^2} + \frac{\rho R T}{2} \left( 1 + \frac{\eta}{g} \frac{\mathrm{d} g}{\mathrm{d} \eta} \right) \sum_{i =1}^c z_i \sum_{j=\alpha,\beta} \eta_{ij} (\chi_{ij} - 1).
\end{equation}
Dividing both sides by $\rho R T$, we obtain the compressibility factor

\begin{equation*}
Z = \frac{1}{1 - b \rho} - \frac{1}{RT} \frac{a \rho}{1 + 2 b \rho - (b \rho)^2} + \frac{1}{2} \left( 1 + \frac{\eta}{g} \frac{\mathrm{d} g}{\mathrm{d} \eta} \right) \sum_{i =1}^c z_i \sum_{j=\alpha,\beta} \eta_{ij} (\chi_{ij} - 1).
\end{equation*}
Using~(\ref{eq:A_mix})--(\ref{eq:B_mix}), we can also express the above as

\begin{equation}
\label{eq:ZCPA}
Z = \frac{Z}{Z - B} - \frac{A Z}{Z^2 + 2 B Z - B^2} + \frac{1}{2} \left( 1 + \frac{\eta}{g} \frac{\mathrm{d} g}{\mathrm{d} \eta} \right) \sum_{i =1}^c z_i \sum_{j=\alpha,\beta} \eta_{ij} (\chi_{ij} - 1).
\end{equation}
In the absence of associating components, only the first two terms on the right-hand side of~(\ref{eq:ZCPA}) are retained. These terms come from the PR EOS. We obtain a cubic polynomial in $Z$ if we rearrange the subsequent equation to have all terms on one side. This result shows why the PR EOS is said to be a cubic equation of state. If the association terms are non-zero, solving for $Z$ becomes more difficult because we do not know \textit{a priori} the number of roots in~(\ref{eq:ZCPA}). As a result, iterative root-finding procedures (usually a combination of the bisection method with Newton's method) must be used over a large search interval such as $Z \in [ B + \delta, 1000B]$, where $\delta$ is a small number. The lower limit in this interval is only slightly larger than $B$ because $B = B(T, P, \mathbf{z})$ represents the closest packing (most compressed state achievable) for a fluid at the conditions $(T, P, \mathbf{z})$.

\subsection{Simplifications employed in the Li and Firoozabadi model}
\label{sec:Li}

In addition to the limitation that water is the only self-associating component in the mixture, which we have already discussed in Section~\ref{sec:FdepartAssoc}, Li and Firoozabadi~\cite{Li2009} make two more assumptions:

\begin{enumerate}
\item All components have a total of four association sites, with two of each type, so that $\eta_{i\alpha} = \eta_{i\beta} = 2$ for all components~$i$. This four-site model has been shown to work well for water~\cite{Kontogeorgis2006a}.
\item The energetics of association is symmetric between the two types of sites so that $\chi_{i\alpha} = \chi_{i\beta}$ for all $i$. We denote these quantities as $\chi_i = \chi_{i\alpha} = \chi_{i\beta}$.
\end{enumerate}
With these simplifications,~(\ref{eq:FdepartAssoc}) becomes

\begin{equation*}
\frac{F^\textrm{depart}_\textrm{assoc}}{NRT} = 4 \sum_{i =1}^c z_i \left( \ln  \chi_{i} - \frac{\chi_{i} - 1}{2} \right).
\end{equation*}
As a result, the expressions~(\ref{eq:FdepartCPA}) and~(\ref{eq:FdepartCPAZ}) for the Helmholtz energy departure function simplify to

\begin{equation*}
\frac{F^\textrm{depart}}{N R T} = -\ln(1 - b \rho) - \frac{a}{2 \sqrt{2} b R T} \ln \left[ \frac{1 + (1 + \sqrt{2} ) b \rho}{1 + (1 - \sqrt{2}) b \rho} \right] + 4 \sum_{i =1}^c z_i \left( \ln  \chi_{i} - \frac{\chi_{i} - 1}{2} \right),
\end{equation*}
and

\begin{equation*}
\frac{F^\textrm{depart}}{N R T} = \ln \left( \frac{Z}{Z - B} \right ) - \frac{A}{2 \sqrt{2} B} \ln \left[ \frac{Z + (1 + \sqrt{2} ) B}{Z + (1 - \sqrt{2}) B} \right] + 4 \sum_{i =1}^c z_i \left( \ln  \chi_{i} - \frac{\chi_{i} - 1}{2} \right),
\end{equation*}
respectively. The association strengths, cross-association parameters, and site fractions in~(\ref{eq:DeltaWaterZ})--(\ref{eq:chi_ibetaZ}) are now given by a more tractable set of expressions:

\begin{equation*}
\Delta_{w, w} = g \mathcal{K} \left[ \exp \left(\frac{\epsilon}{k_B T} \right) -1 \right],
\end{equation*}
\begin{equation*}
\Delta_{w, i} = s_{w, i} \Delta_{w, w} \quad i \neq w,
\end{equation*}
\begin{equation*}
\chi_{w} = \frac{Z}{Z + 2 \sum_{i =1}^c z_i \chi_{i} \Delta_{w, i}},
\end{equation*}
\begin{equation*}
\chi_{i} = \frac{Z}{Z + 2 z_w \chi_{w} \Delta_{w, i}}.
\end{equation*}
Li and Firoozabadi have found that fitting to binary water/CO$_2$ mixture VLE data gives

\begin{equation*}
s_{w,\textrm{CO}_2}  = 0.0529 T_{r,{\textrm{CO}_2}}^2 + 0.0404 T_{r,{\textrm{CO}_2}}-0.0693,
\end{equation*}
for the cross-association parameter $s_{w,\textrm{CO}_2}$ between water and CO$_2$. Equations~(\ref{eq:lnphiZ}) and~(\ref{eq:lnphi}) for the fugacity coefficient~$\phi_i$ become

\begin{align}
\label{eq:lnphiZLi}
\ln \phi_i = -\ln \left( Z - B \right ) & + \frac{B_i}{B} \left(\frac{B}{Z - B} - \frac{A Z}{Z^2 + 2 B Z - B^2} \right) \nonumber \\ 
& - \frac{A}{2 \sqrt{2} B} \left(\frac{2 \sum_{j = 1}^c z_j A_{ij}}{A} - \frac{B_i}{B} \right) \ln \left[ \frac{Z + (1 + \sqrt{2} ) B}{Z + (1 - \sqrt{2}) B} \right] \nonumber \\
& + 4 \ln \chi_{i} + \frac{B_i}{2 g Z} \frac{\mathrm{d} g}{\mathrm{d} \eta} \sum_{k =1}^c z_k (\chi_{k} - 1),
\end{align}
\begin{align*}
\ln \phi_i = -\ln \left( 1 - b \rho \right ) -\ln Z & + \frac{b_i}{b} \left(\frac{b \rho}{1 - b \rho} - \frac{1}{R T} \frac{a \rho}{1 + 2 b \rho - (b \rho)^2} \right) \nonumber \\ 
& - \frac{a}{2 \sqrt{2} b R T} \left(\frac{2 \sum_{j = 1}^c z_j a_{ij}}{a} - \frac{b_i}{b} \right) \ln \left[ \frac{1 + (1 + \sqrt{2} ) b \rho }{1 + (1 - \sqrt{2}) b \rho} \right] \nonumber \\
& + 4 \ln \chi_{i} + \frac{b_i \rho}{2 g} \frac{\mathrm{d} g}{\mathrm{d} \eta} \sum_{k =1}^c z_k  (\chi_{k} - 1).
\end{align*}
Equations~(\ref{eq:pressureCPA})--(\ref{eq:ZCPA}) simplify to

\begin{equation*}
P = \frac{\rho R T}{1 - b \rho} - \frac{a \rho^2}{1 + 2 b \rho - (b \rho)^2} + 2 \rho R T \left( 1 + \frac{\eta}{g} \frac{\mathrm{d} g}{\mathrm{d} \eta} \right) \sum_{i =1}^c z_i (\chi_{i} - 1),
\end{equation*}
\begin{equation*}
Z = \frac{1}{1 - b \rho} - \frac{1}{RT} \frac{a \rho}{1 + 2 b \rho - (b \rho)^2} + 2 \left( 1 + \frac{\eta}{g} \frac{\mathrm{d} g}{\mathrm{d} \eta} \right) \sum_{i =1}^c z_i (\chi_{i} - 1),
\end{equation*}
\begin{equation*}
Z = \frac{Z}{Z - B} - \frac{A Z}{Z^2 + 2 B Z - B^2} + 2 \left( 1 + \frac{\eta}{g} \frac{\mathrm{d} g}{\mathrm{d} \eta} \right) \sum_{i =1}^c z_i (\chi_{i} - 1).
\end{equation*}
The enthalpy results presented in Sections~\ref{sec:pure} and~\ref{sec:mixture} are calculated using the equations presented in this section. As we demonstrate in those sections, this simplified version of the CPA EOS is sufficient for molar enthalpy calculations of mixtures containing CO$_2$ and H$_2$O, as well as enthalpies of pure components. It has previously been shown to provide excellent agreement with VLE (solubility) and density data~\cite{Li2009}. It cannot, however, accurately calculate excess enthalpies of CO$_2$-H$_2$O mixtures. In Section~\ref{sec:excess}, we reason why it fails for excess enthalpy calculations and explain how relaxing the two assumptions stated in this section, so that the equations in Sections~\ref{sec:fugacityCPA} and~\ref{sec:pressureCPA} are followed instead, may allow for better agreement with experimental excess enthalpy data.

\section{Duan and Sun CO$_2$ activity coefficient model}
\label{sec:duan}

In electrolytic solutions, concentrations are usually expressed in terms of molality ($m$) rather than mole fractions, so that (\ref{eq:activityLimit})--(\ref{eq:muActivity}) must be recast in terms of molality. In the molality convention, a natural choice for the activity coefficient $\gamma_i$ of a solute species~$i$ is

\begin{equation*}
\lim_{m_i \rightarrow 1} \gamma_i = 1.
\end{equation*}
With~$\gamma_i$ defined in this way, the activity~$\mathsf{a}_i$ and chemical potential~$\mu_i$ of~$i$ are given by

\begin{equation*}
\mathsf{a}_i = \gamma_i m_i = \frac{f_i(T, P, m_i)}{f_i^\textrm{im}(T, P, m_i = 1)}.
\end{equation*}

\begin{equation}
\label{eq:muSolute}
\mu_i (T, P, m_i) - \mu_i^\textrm{im}(T, P, m_i = 1) = RT \ln \mathsf{a}_i.
\end{equation}
Here,~$f_i^\textrm{im}(T, P, m_i = 1)$ and~$\mu_i^\textrm{im}(T, P, m_i = 1)$ are the fugacity and chemical potential of~$i$, respectively, in an ideal mixture where~$m_i = 1$. These equations apply to the aqueous phase, which is the electrolytic phase. For the CO$_2$-rich phase, which is assumed to contain only CO$_2$ and water, Duan and Sun~\cite{Duan2003} choose the reference chemical potential to be that of the pure component in the ideal gas state at a pressure~$P'$~=~1~bar. Using~(\ref{eq:mu1mu2})--(\ref{eq:phiDefinition}), the chemical potential~$\mu_i (T, P, \mathbf{z})$ of component~$i$ in the CO$_2$-rich phase, with pressure expressed in units of bars, is

\begin{equation*}
\mu_i (T, P, \mathbf{z}) =  \mu_i^\textrm{ig} (T, P' = 1\mathrm{~bar}) + RT \ln z_i P + RT \ln \phi_i(T, P, \mathbf{z}).
\end{equation*}
For pure CO$_2$, this equation simplifies to

\begin{equation}
\label{eq:muCO2}
\mu_{\mathrm{CO}_2} (T, P) =  \mu_{\mathrm{CO}_2}^\textrm{ig} (T, P' = 1\mathrm{~bar}) + RT \ln P + RT \ln \phi_{\mathrm{CO}_2}(T, P).
\end{equation}
Substituting~(\ref{eq:muHbar}) and~(\ref{eq:HbarIdeal}) into the definition~(\ref{eq:deltaHsol}) of the CO$_2$ molar enthalpy of solution~$\Delta h_\mathrm{sol}$,

\begin{align*}
\Delta h_\mathrm{sol} = &  \bar{H}_{\mathrm{CO}_2}(T, P, m_{\mathrm{CO}_2} \rightarrow 0) - h_{\mathrm{CO}_2}(T, P) \nonumber \\
& = -T^2 \frac{\partial }{\partial T} \left[ \left. \left( \frac{\mu_{\mathrm{CO}_2} (T, P, m_{\mathrm{CO}_2} \rightarrow 0)}{T} \right) \right|_{P, \mathbf{m}} - \left. \left( \frac{\mu_{\mathrm{CO}_2} (T, P)}{T} \right) \right|_{P} \right].
\end{align*}
The subscript~$\mathbf{m}$ indicates that all species concentrations are held fixed. From~(\ref{eq:muSolute}) and~(\ref{eq:muCO2}),

\begin{align*}
\frac{\partial }{\partial T} \left. \left( \frac{\mu_{\mathrm{CO}_2} (T, P, m_{\mathrm{CO}_2} \rightarrow 0)}{T} \right) \right|_{P, \mathbf{m}} = & \frac{\partial }{\partial T} \left. \left( \frac{\mu_{\mathrm{CO}_2}^\textrm{im}(T, P, m_{\mathrm{CO}_2} = 1)}{T} \right) \right|_{P} \nonumber \\
& + R \left. \left( \frac{\partial \ln \gamma_{\mathrm{CO}_2}(T, P, m_{\mathrm{CO}_2} \rightarrow 0)}{\partial T}  \right) \right|_{P, \mathbf{m}},
\end{align*}

\begin{equation*}
\frac{\partial }{\partial T} \left. \left( \frac{\mu_{\mathrm{CO}_2} (T, P)}{T} \right) \right|_{P} = \frac{\partial }{\partial T} \left. \left( \frac{\mu_{\mathrm{CO}_2}^\textrm{ig} (T, P' = 1\mathrm{~bar})}{T} \right) \right|_{P} + R \left. \left( \frac{\partial \ln \phi_{\mathrm{CO}_2}(T, P)}{\partial T}  \right) \right|_{P}.
\end{equation*}

For a brine solution where the only electrolyte is NaCl, Duan and Sun's CO$_2$ activity coefficient model is

\begin{equation}
\label{eq:activityDuan}
\ln \gamma_{\mathrm{CO}_2} = 2 \lambda_{\mathrm{CO}_2-\mathrm{Na}} m_\mathrm{Na} + 2 \lambda_{\mathrm{CO}_2-\mathrm{Cl}} m_\mathrm{Cl} + \zeta_{\mathrm{CO}_2-\mathrm{Na}-\mathrm{Cl}} m_\mathrm{Na} m_\mathrm{Cl}.
\end{equation}
The coefficients~$\lambda_{\mathrm{CO}_2-\mathrm{Na}},~\lambda_{\mathrm{CO}_2-\mathrm{Cl}},~\zeta_{\mathrm{CO}_2-\mathrm{Na}-\mathrm{Cl}}$ are temperature- and pressure-dependent functions that represent intermolecular interactions between CO$_2$ and the dissolved ions. Equation~(\ref{eq:activityDuan}) is obtained by taking the derivative, with respect to the molality of CO$_2$, of a Pitzer-type excess Gibbs energy function~\cite{Pitzer1973}. Duan and Sun treat~$\mu_{\mathrm{CO}_2}^\textrm{ig} (T, P' = 1\mathrm{~bar})$ and~$ \lambda_{\mathrm{CO}_2-\mathrm{Cl}}$ as being identically equal to zero. Using many sources of CO$_2$ solubility data in pure water and NaCl solutions, they fit the functions $\mu_{\mathrm{CO}_2}^\textrm{im}(T, P, m_{\mathrm{CO}_2} = 1)/RT$, $\lambda_{\mathrm{CO}_2-\mathrm{Na}}$, and $\zeta_{\mathrm{CO}_2-\mathrm{Na}-\mathrm{Cl}}$ to expressions of the form

\begin{equation*}
c_1 + c_2 T + \frac{c_3}{T} + c_4 T^2 + \frac{ c_5}{630 - T} + c_6 P + c_7 P \ln T + \frac{c_8 P}{T} +\frac{ c_9 P}{630 - T} + \frac{c_{10} P^2}{(630 - T)^2} + c_{11} T \ln P,
\end{equation*}
where the $c$'s are constant coefficients. The fugacity coefficient~$\phi_{\mathrm{CO}_2}(T, P)$ is obtained by iteratively solving a nonlinear equation, as described in their paper. In a later study, Duan~\textit{et al}.\ present a series of piecewise expressions (curve fits) for $\phi_{\mathrm{CO}_2}(T, P)$ so that it can be calculated directly, without the need for iterative methods~\cite{Duan2006}. Although these expressions may increase the efficiency of CO$_2$ solubility computations, they cannot be used for enthalpy computations. We have found that for certain conditions, they lead to sharp jumps in the enthalpy. This unphysical behavior occurs because the enthalpy is obtained from the fugacity coefficient by taking a temperature derivative, so that it should be represented by a smooth function of temperature, rather than a series of piecewise expressions. We therefore compute~$\phi_{\mathrm{CO}_2}(T, P)$ as described in the earlier study.

\bibliographystyle{elsarticle-num}


\end{document}